\pdfoutput=1
\documentclass[fleqn,usenatbib,usedcolumn]{mnras}
\usepackage[british]{babel}             
\usepackage{amsmath}
\usepackage{amssymb}
\usepackage{txfonts}                    
\let\la=\lesssim 
\usepackage[T1]{fontenc}                
\usepackage{graphicx}                   
\usepackage{color}
\usepackage{multirow}
\usepackage{url}
\hypersetup{pdfauthor={Rumsey et al.},
            pdftitle={AMI observations of ten CLASH galaxy clusters:
                      SZ and X-ray data used together to determine cluster
                      dynamical states},
            pdfkeywords={galaxies: clusters: general -- methods:
                         observational -- techniques:
                         interferometric -- large-scale
                         structure of the Universe},
            bookmarksnumbered=true,
            breaklinks=true}
\volume{{\rm in press}}
\def\vpad{{\Large$\mathstrut$}}
\def\Msun{M$_{\odot}$}

\title[Cluster dynamical states from SZ and X-ray]{AMI observations of ten CLASH galaxy clusters: SZ and X-ray data used together to determine cluster dynamical states}
\author[Rumsey et~al.]{Clare Rumsey,$^{1}$\thanks
{Email:cr461@mrao.cam.ac.uk} Malak Olamaie$^{1}$, Yvette C. Perrott$^{1}$, Helen R. Russell$^{3}$
\newauthor Farhan Feroz$^{1}$, Keith J. B. Grainge$^{4,1}$, Will J. Handley$^{1,2}$, Michael P. Hobson$^1$,
\newauthor Richard D. E. Saunders$^{1,2}$, Michel P. Schammel$^{1}$ \vspace{0.3cm}\\
$^{1}$ Astrophysics Group, Cavendish Laboratory, 19 J. J.
Thomson Avenue, Cambridge, CB3 0HE\\
$^{2}$ Kavli Institute for Cosmology Cambridge, Madingley
Road,Cambridge, CB3 0HA\\
$^{3}$ Institute of Astronomy, Madingley Road, Cambridge CB3 0HA \\
$^{4}$ Jodrell Bank Centre for Astrophysics,
School of Physics and Astronomy,
Manchester, M13 9PL}

\date{Accepted ??????; Received ???????}

\pagerange{\pageref{firstpage}--\pageref{lastpage}}

\pubyear{2016}

\begin{document}
\label{firstpage}

\maketitle

\begin{abstract}
Using Arcminute Microkelvin Imager (AMI) SZ observations towards ten CLASH clusters we investigate the influence of cluster mergers on observational galaxy cluster studies.
Although selected to be largely relaxed, there is disagreement in the literature on the dynamical states of CLASH sample members. We analyse our AMI data in a fully Bayesian way to produce estimated cluster parameters and consider the intrinsic correlations in our NFW/GNFW-based model. Varying pressure profile shape parameters, illustrating an influence of mergers on scaling relations, induces small deviations from the canonical self-similar predictions -- in agreement with simulations of \citet{2007MNRAS.380..437P} who found that merger activity causes only small scatter perpendicular to the relations. We demonstrate this effect observationally using the different dependencies of SZ and X-ray signals to $n_{\rm e}$ that cause different sensitivities to the shocking and/or fractionation produced by mergers.
Plotting $Y_{\rm X}$--$M_{\rm gas}$ relations (where $Y_{\rm X}=M_{\rm gas}T$) derived from AMI SZ and from $Chandra$ X-ray gives ratios of AMI and $Chandra$ $Y_{\rm X}$ and $M_{\rm gas}$ estimates that indicate movement of clusters \textit{along} the scaling relation, as predicted by \citet{2007MNRAS.380..437P}.
Clusters that have moved most along the relation have the most discrepant $T_{\rm SZ}$ and $T_{\rm X}$ estimates: all the other clusters (apart from one) have SZ and X-ray estimates of $M_{\rm gas}$, $T$ and $Y_{\rm X}$ that agree within $r_{500}$.
We use SZ vs X-ray discrepancies in conjunction with $Chandra$ maps and $T_{\rm X}$ profiles, making comparisons with simulated cluster merger maps in \citet{2006MNRAS.373..881P}, to identify disturbed members of our sample and estimate merger stages.
\end{abstract}


\begin{table*}
\small
\begin{center}
\caption{Summary of assessments that we have found in the literature of the dynamical states of the AMI-CLASH sub-sample.}\label{tab:litrev}\setlength\extrarowheight{3pt}\vspace{1mm}
\begin{tabular}{lccl}\hline
Cluster       & Relaxed\vpad   & Unrelaxed\vpad  & Notes   \\
 \\\hline
A611\vpad              & \checkmark                  &                             & Widely agreed relaxed, see e.g. \citet{2008MNRAS.383..879A}, \citet{2011AA...528A..73D} \vspace{1mm}\\\hline
\multirow{2}{*}{A1423} & \multirow{2}{*}{\checkmark} &\multirow{2}{*}{\checkmark}  & Relaxed: e.g. classified `non-distorted' by \citet{2007AA...467..485H} and `regular' by \citet{2005MNRAS.359.1481B} \\
                       &                &                                          & Unrelaxed: due to centroid shift, cuspiness and cooling time -- \citet{2013MNRAS.433.2790L}\vspace{1mm}\\\hline
\multirow{2}{*}{A2261} & \multirow{2}{*}{\checkmark} &\multirow{2}{*}{\checkmark}  & Widely considered relaxed, see e.g. \citet{2013MNRAS.433.2790L}, \citet{2012MNRAS.420.2120M}\\
                       &                &                                          & Unrelaxed due to extra structure to the south-west, \citet{2009MNRAS.392.1509G}\vspace{1mm}\\\hline
\multirow{2}{*}{CLJ1226+3332}  & \multirow{2}{*}{\checkmark}    & \multirow{2}{*}{\checkmark}  & Classified relaxed e.g. \citet{2004MNRAS.351.1193M}, \citet{2008MNRAS.383..879A}\\
                       &                &                                          & Merger given X-ray morphology, \citet{2009ApJ...692.1033V}, and temperature map, \citet{2007ApJ...659.1125M}    \vspace{1mm}\\\hline
MAJ0647+7015  &                & \checkmark                                        & Strong-lens selected, highly unrelaxed: \citet{2013ApJ...762L..30Z} \vspace{1mm}\\\hline
MAJ0717+3745  &                & \checkmark                                        & Strong-lens selected, highly unrelaxed and complex merger: \citet{2013ApJ...777...43M} \vspace{1mm}\\\hline
\multirow{2}{*}{MAJ0744+3927}  & \multirow{2}{*}{\checkmark}    & \multirow{2}{*}{\checkmark} & In samples of large relaxed clusters \citet{2004MNRAS.353..457A}, \citet{2007MNRAS.379..209S}, and \cite{2008MNRAS.383..879A}\\
                       &                &                                          & Substructure identified, e.g. \citet{2007MNRAS.379..209S}, \citet{2008MNRAS.383..879A}. Disturbed, \citet{2013ApJ...768..177S} \vspace{1mm}\\\hline
MAJ1149+2223  &                & \checkmark                                        & Strong-lens selected, highly complex merger, see \citet{2009ApJ...707L.163S}\vspace{1mm}\\\hline
MAJ1423+2404  & \checkmark     &                                                   & Highly relaxed state, pronounced cool-core, e.g. \citet{2007ApJ...661L..33E} and \citet{2010MNRAS.405..777L} \vspace{1mm}\\\hline
\multirow{2}{*}{RXJ1532+3021}  & \multirow{2}{*}{\checkmark} &\multirow{2}{*}{\checkmark}  & Presence of a cool-core e.g. \citet{2008MNRAS.383..879A}, \citet{2013ApJ...777..163H} classifies it as relaxed\\
                       &                &                                          & Cold front possibly from low-level merger turbulence \citep{2013ApJ...777..163H}\vspace{1mm}\\
\hline
\end{tabular}
\end{center}
\end{table*}	

\section{Introduction}\label{sec:intro}

Physical parameters of glaxy clusters, such as total mass and gas mass, are commonly studied through scaling relations.
These relations assume that both growing and mature clusters are relaxed, self-similar systems such that relations between e.g. $L_{\rm X}$, $L_{\rm SZ}$, $M_{\rm tot}$, $M_{\rm gas}$, $T$, etc. are simple power laws (see e.g. \citealt{1986MNRAS.222..323K} and \citealt{2013SSRv..177..247G} for a recent review).
Deviations from hydrostatic equilibrium (HSE) (or from virialisation) and self-similarity during cluster mergers will cause scatter around the scaling relations.
Studies in the literature aim to use these relations to make accurate determinations of e.g. total cluster mass, and therefore often focus on minimising the scatter either by careful sample selection of low-redshift, relaxed clusters (e.g. \citealt{2008A&A...482..451Z}, \citealt{2009A&A...498..361P}, \citealt{2009ApJ...693.1142S}, \citealt{2009ApJ...692.1033V}), or by finding a particularly low-scatter mass proxy (e.g. \citealt{2006ApJ...650..128K}, \citealt{2007A&A...474L..37A}, \citealt{2008A&A...482..451Z}, \citealt{2013ApJ...767..116M}).
These approaches often produce low-scatter relations that agree with the self-similar predictions. However, \citet{2007MNRAS.380..437P}, using simulations of two-body cluster mergers to track the evolution of a merger (from a relaxed system before the start of the merger through to relaxation of the merged system) in the plane of a scaling relation, find large changes in cluster observables \textit{along} the relation with little perpendicular displacement.

Assessment of these cluster parameter values through calculation from Sunyaev--Zel'dovich (SZ, \citealt{1972CoASP...4..173S}) and X-ray observation provides a critical probe of the dynamical state of the cluster gas due to the difference in dependencies of the SZ and X-ray flux densities on the electron number density, $n_{\rm e}$.
The SZ effect is the inverse Compton scattering of CMB photons by hot cluster gas, and is $\propto\int{{n_{\rm e}}{T}\,{\rm d}l}$, where $T$ is the plasma temperature and ${\rm d}l$ the line element along the line of sight through the cluster. The X-ray Bremsstrahlung signal is $\propto\int{{n_{\rm e}}^{2} \Lambda({T})\,{\rm d}l}$, where $\Lambda$ is the cooling function ($\Lambda$$\propto$ $T^{\frac{1}{2}}$ for the clusters in this paper).
Parameter values estimated from measurement of SZ and X-ray signals will, therefore, also depend differently on $n_{\rm e}$ and $T$.

As cluster mergers are known to produce regions of higher density gas, through processes such as shocking, X-ray parameter estimation is likely more sensitive to dynamical state, and will produce larger displacements along scaling relations during a merger than SZ parameter values. This implies that merger activity can be identified by looking at discrepancies between SZ and X-ray measurements.

To test this observationally, we use the CLASH sample of well-studied clusters selected by \citet{2012ApJS..199...25P} to form a sample of massive clusters, most of which are classified in the literature as relaxed, plus a small number of clusters with pronounced strong gravitational lensing (see Section \ref{sec:sample}).
Here we discuss measurements of a sub-sample of CLASH clusters via the SZ effect using the Arcminute Microkelvin Imager (AMI, \citealt{2008MNRAS.391.1545Z}).

The SZ signal measures the Comptonization parameter, $y$, the line-of-sight integral of the number of collisions multiplied by the mean fractional energy change of the CMB photons per collision:
\begin{eqnarray}
 y &=& \frac{\sigma_{T}}{m_{\rm e}c^2} \int{n_{\rm e}k_
{\rm B}T\,{\rm d}l}\\
&=& \frac{\sigma_{T}}{m_{\rm e}c^2} \int{P_{\rm e}\,{\rm
d}l},\label{eq:ypar}
\end{eqnarray}
where $\sigma_{\rm T}$ is the Thomson scattering cross-section, $m_{\rm e}$ the electron mass, $c$ the speed of light.
Equation \ref{eq:ypar} shows that the SZ surface brightness is proportional to the electron pressure, $P_{\rm e}$, assuming an ideal gas law, integrated along the line of sight. Integrating $y$ over the solid angle $\Omega$ subtended by the cluster gives $Y_{\rm SZ}$, which quantifies the internal energy of the cluster gas, providing a proxy for total mass, given redshift information.

In X-ray studies $Y_{\rm X}$, found from $Y_{\rm X}=M_{\rm gas}T_{\rm X}$, is used as an analogue of $Y_{\rm SZ}$ which is proportional to the product of the gas mass and the mean temperature measured from SZ within a sphere (or a cylinder). \citet{2006ApJ...650..128K} find, using simulated data, that $Y_{\rm X}$ provides an equally good proxy for total mass as $Y_{\rm SZ}$.
The mean cluster temperature has also been widely used as a proxy for total cluster mass.
Cluster $T$ has traditionally been measured through X-ray spectroscopy; with good enough sensitivity and angular resolution, annular averaging gives temperature profiles out to, for some clusters, radii of $\approx$\,1\,Mpc (see e.g. ACCEPT Database, \citealt{2009ApJS..182...12C}, \citealt{2006ApJ...640..691V}, \citealt{2006A&A...446..429P}).
\citet{2012MNRAS.423.1534O} and \citet{2013MNRAS.430.1344O} show that a gas temperature profile can also be obtained via SZ observation, given assumed geometry and dynamical state, and given a prior on the gas fraction $f_{\rm gas}$ {\color{black}at $r_{200}$}.

In this study, cluster parameters are derived from our AMI SZ measurements in a fully Bayesian way using the model described in \citet{2012MNRAS.423.1534O} and (2013). This model uses a Navarro, Frenk and White (NFW) profile to describe the dark matter density, which is believed, from cosmological N-body simulations, to accurately model all dark matter halos \citep{1997ApJ...490..493N}. A generalised Navarro, Frenk and White (GNFW) profile is used to describe the gas pressure, shown to follow self-similarity more closely than the density or temperature at high radius \citep{2007ApJ...668....1N}. Further conditions of spherical symmetry, HSE, and a small $f_{{\rm gas},\,r_{200}}$ compared to unity, produces cluster properties as functions of radius.

Throughout, we assume $H_{0}$ = 70~km~$\rm s^{-1} Mpc^{-1}$ and a concordance $\Lambda$CDM cosmology with $\Omega_{m}$ = 0.3, $\Omega_{\Lambda}$ = 0.7, $\Omega_{k}$ = 0, $\Omega_{b}$ = 0.041, $\omega_{0}$ =$-$1, $\omega_{a}$ = 0 and $\sigma_{8}$ = 0.8. All cluster parameter values are at the redshift of the cluster. We emphasise that we denote $M_{\rm gas}T$ as $Y_{\rm X}$ for either SZ or X-ray.

\section{The AMI-CLASH sub-sample}\label{sec:sample}

The CLASH (Cluster Lensing and Supernova Survey with Hubble) sample
consists of 25 massive clusters, covering a large redshift range ($z$
from 0.213 to 0.888), selected for strong and weak lensing
observation with Hubble and \textit{Subaru} \citep{2012ApJS..199...25P}.
20 of these clusters were selected from $Chandra$ X-ray observations to
be dynamically relaxed. The remaining five clusters were selected solely
based on their high lensing-strength and consequently include some of
the most disturbed clusters known \citep{2012AA...547A..66R}. Eleven of
these clusters are visible to AMI which is currently restricted to a
declination range of 20$^\circ$ to 85$^\circ$.
Since the CLASH sample is
composed mainly of Abell and MACS clusters, all eleven clusters in the
sub-sample had in fact been observed (with varying sensitivities) by AMI
in 2009 to 2012. These observations showed that the field of view of
MAJ1720+3536 contains too bright a source to allow SZ mapping with AMI.
We discard this cluster leaving an AMI-CLASH sub-sample of ten. We have
searched the literature to find additional assessments of the dynamical
states and X-ray morphologies of the ten clusters; Table
\ref{tab:litrev} summarises these findings.

\begin{table}
\caption{Summary of AMI characteristics}\label{table_ami}\setlength\extrarowheight{0.5pt}
\centering
\begin{tabular}{{l}{c}{c}}
\hline
                           & SA                   & LA                      \\
\hline
Antenna diameter           & 3.7~m                 & 12.8~m                 \\
Number of antennas         & 10                    & 8                      \\
Baseline lengths (current) & 5--20~m               & 18--110~m              \\
Primary beam (15.7~GHz)    & 20.1 arcmin           & 5.5 arcmin             \\
Typical synthesized beam   &  3 arcmin             &  30 arcsec             \\
Flux sensitivity           & 30~mJy~s$^{1/2}$       & 3~mJy~s$^{1/2}$       \\
Observing frequency        & \multicolumn{2}{c}{13.9--18.2{~GHz}}           \\
Bandwidth                  & \multicolumn{2}{c}{4.3{~GHz}}                  \\
Channel bandwidth          & \multicolumn{2}{c}{0.72{~GHz}}                 \vspace{1mm}\\
\hline
\end{tabular}
\vspace{0.3cm}
\end{table}
\begin{table}
\centering
\caption{AMI actual observing time in hours for each cluster in the AMI-CLASH sub-sample.}\label{table:obs}\setlength\extrarowheight{0.5pt}
\begin{tabular}{{l}{c}{c}{c}{c}}
\hline
Cluster       & \multicolumn{2}{c}{SA}    & \multicolumn{2}{c}{LA}  \\\hline
              & 2009-12  & new   & 2009-12  & new   \\\hline
A611          & 18       & 16    & 24      & 14         \\
A1423         & 57       & --    & 18      & 8          \\
A2261         & 60       & 8     & 7       & 124        \\
CLJ1226+3332  & 77       & --    & 48      & --         \\
MAJ0647+7015  & 25       & 8     & 21      & 8          \\
MAJ0717+3745  & 49       & 9     & 30      & 38         \\
MAJ0744+3927  & 16       & 9     & 10      & 8          \\
MAJ1149+2223  & 38       & 7     & 18      & 16         \\
MAJ1423+2404  & 42       & 7     & 27      & 8          \\
RXJ1532+3021  & 63       & 77    & 25      & 33         \vspace{1mm}\\
\hline
\end{tabular}
\vspace{0.3cm}
\end{table}

\begin{table*}
\centering
\caption{Summary of the priors used in the analysis using the model described in Section \ref{sec:model}. $M_{{\rm tot},\,r_{200}}$ and $f_{{\rm gas},\,r_{200}}$ are the total mass internal to $r_{200}$ and gas fraction internal to $r_{200}$, respectively.}\label{table_priors}\setlength\extrarowheight{1pt}
\begin{tabular}{{l}{l}}
\hline
Parameter                          & Prior                      \\
\hline
Cluster position $(x_{\rm c},y_{\rm c})$       & Gaussian at SA pointing centre, $\sigma$ set at 60 arcsec regardless of precision of cluster position \\
Mass $(M_{{\rm tot},\,r_{200}}$/\Msun$)$        & Uniform in log space between 1$\times10^{14}$ and 6$\times10^{15}$                   \\
Gas fraction $(f_{{\rm gas},\,r_{200}})$        & Gaussian with $\mu$\,=\,0.13, $\sigma$\,=\, 0.02                                            \\
Shape parameters $(a, b, c, c_{500})$          & Delta-function with ``universal" values, see \citet{2010AA...517A..92A}                   \\
Source position $(x_{\rm s},y_{\rm s})$        & Delta-function at LA position                                                         \\
Source flux density $(S_0/Jy)$ $>$\,4$\sigma_{\rm SA}$    & Gaussian at LA value, with $\sigma$\,=\,40 per cent of LA flux density      \\
Source spectral index when $S_{0}$ $>$\,4$\sigma_{\rm SA}$  & Gaussian centred at LA-fitted spectral index with $\sigma$\,=\,LA error, or prior based on 10C, see Section \ref{sec:priors} \\
Source flux density $(S_0/Jy)$ $<$\,4$\sigma_{\rm SA}$    & Delta-function at LA value, unless close to cluster              \\
Source spectral index when $S_{0}$ $<$\,4$\sigma_{\rm SA}$  & Delta-function centred at LA-fitted spectral index, or based on 10C, see Section \ref{sec:priors}          \vspace{1mm}\\
\hline
\end{tabular}
\vspace{0.3cm}
\end{table*}


\section{AMI and Observations}
\label{sec:observations}

SZ observations were carried out with AMI, a dual array of
interferometers observing at centre-frequency 15.7\,GHz, located near
Cambridge. The arrangement of ten 3.7-m antennas with baselines of
5--20\,m in the Small Array (SA) and eight 12.8-m antennas with
baselines of 18--110\,m in the Large Array (LA) allows for study of the
large-scale SZ effect from clusters along with subtraction of otherwise
confusing radio sources. For a full description of the instrument see
\citet{2008MNRAS.391.1545Z}. Our pointing strategy for each cluster is
as follows. A single SA pointing centre is observed, centred on the
cluster X-ray centre. For the LA, the smaller primary beam size (see
Table \ref{table_ami}) requires that the same area is covered with a
mosaiced 61-point LA observation; the central 19 pointings of this are
observed 3$\times$ longer than the outer pointings to reach lower noise
levels near the cluster centre. Our sensitivity aim for each cluster has
been to achieve thermal noise levels below 100\,$\mu$Jy in the centre of
the SA maps and averaging below 80\,$\mu$Jy over the central
19-pointings of the LA maps. The 2009-2012 observations of CLJ1226+3332
on both arrays and the SA 2009-2012 observations of A1423 and
MAJ1423+2404 have the sensitivities that we require. For the rest we
have made new observations so that these, plus the 2009-2012
observations, achieve the sensitivity aim. A summary of the observations
carried out is presented in Table \ref{table:obs}. Note that, for
RXJ1532+3021, 2009-2012 observations were made with a somewhat erroneous
pointing centre, and we use only new, correctly centred, observations.

Data processing is carried out with our in-house software package
\textsc{reduce}, in which the raw data are calibrated and flagged for
interference and telescope errors. Note that all AMI data, whenever
taken, have been reduced with our latest reduction pipeline with
optimised interference flagging and calibration. Flux calibration is
carried out with observations of 3C\,48, 3C\,147 and 3C\,286, with
3C\,286 flux densities calibrated against Very Large Array (VLA)
measurements \citep{2013ApJS..204...19P}. Data are Fourier transformed
into frequency channels and then written out as $uv$-\textsc{fits}
files. Data $uv$-\textsc{fits} files from 2009-2012 and from new
observations are concatenated in $uv$-space to produce a single data set
in each array. For source-finding and for initial visual inspection of a
cluster field, the $uv$-data are imaged in
\textsc{aips}{\footnote{http://aips.nrao.edu/}} using automated
\textsc{clean} procedures with noise levels determined using
\textsc{imean}. The LA channel maps are passed through the
\textsc{Source-Find} algorithm \citep{2011MNRAS.415.2699A} which
estimates LA source positions, flux densities and spectral indices to
use as priors in \textsc{McAdam} \citep{2009MNRAS.398.2049F}, our
Bayesian analysis system (Section \ref{sec:analysis}). If sources in the
fields exhibit variability, concatenating $uv$-data taken a few years
apart could introduce inaccuracies in source flux estimates between the
arrays. If the proportion of old data and new data is different between
the arrays, the average SA flux may be significantly different from the
average LA flux that is being subtracted, leaving residuals. Only a few
sources in the maps show variability between old (2009-2012) and new
(2013-2014) observations at a level that would effect parameter
estimation if the LA flux was directly subtracted from the SA data. This
effect is accounted for in \textsc{McAdam} (Section \ref{sec:analysis}).

\section{Parameter Estimation}
\label{sec:analysis}

Analysis of AMI data is carried out in $uv$-space using our Bayesian
analysis package \textsc{McAdam} \citep{2009MNRAS.398.2049F} which uses
the fast sampler \textsc{multinest} (\citealt{2008MNRAS.384..449F},
\citealt{2009MNRAS.398.1601F} and \citealt{2013arXiv1306.2144F}) to
estimate cluster parameters given a cluster model; \textsc{McAdam}
simultaneously fits the radio source environment and takes full account
of the power spectrum of the primordial CMB structure as a function of
angular scale as seen by AMI, thermal noise in the $uv$-data, and
confusion noise, in the form of a generalised noise covariance matrix.
Concatenated SA visibilities must first be binned to reduce the size of
the data, easing memory demands; for minimal loss of information, bin
sizes are less than aperture illumination function FWHM.

\subsection{Modelling}\label{sec:model}

The model we employ for this analysis is described in
\citet{2012MNRAS.423.1534O} and \citet{2013MNRAS.430.1344O}, and is
based on a number of key assumptions. The first two are the functional
forms of the dark matter density profile and the gas pressure profile
within a spherical geometry. The non-baryonic matter distribution of a
cluster is described using a NFW model,

\begin{equation}\label{eq:DMdensity}
   \rho_{\rm {DM}}(r)=\frac{\rho_{\rm {s}}}{\left(\frac{r}{R_{\rm s}}\right)\left(1 + \frac{r}{R_{\rm s}}
   \right)^2},
\end{equation}
where $\rho_{\rm {s}}$ is an overall normalisation coefficient and the scale radius, $R_{\rm s}$, is the value of $r$ at which ${\rm d}\ln \rho(r)/{\rm d}\ln r=-2$. A GNFW model is used to describe the gas pressure profile
\begin{equation}\label{eq:GNFW}
  P_{\rm e}(r)=\frac{P_{\rm {ei}}}{\left(\frac{r}{r_{\rm p}}\right)^c\left(1+\left(\frac{r}{r_{\rm
  p}}\right)^{a}\right)^{(b-c)/ a}},
\end{equation}
where $P_{\rm {ei}}$ is an overall normalisation coefficient, and $r_{\rm p}$ is the scale radius. The parameters $a$, $b$ and $c$ describe the slopes of the pressure profile at $r\approx r_{\rm p}$, $r> r_{\rm p}$ and $r \ll r_{\rm p}$ respectively.

The third assumption is of hydrostatic equilibrium throughout the cluster. The final assumption is that the gas mass fraction $f_{{\rm gas},\,r_{200}}$ is small compared to unity, so that $M_{{\rm tot},\,r} \approx M_{{\rm DM},\,r}$. These assumptions allow the total mass within a radius $r$ to be derived as
\begin{eqnarray} \label{eq:DMmass}
M_{\rm {tot},\,r}&=& \int_{0}^{r}{\rho_{\rm {DM}}(r')(4\pi r^{'2}{\rm d}r')}   \nonumber\\
&=& 4\pi \rho_{\rm {s}}R^3_{\rm s} \left \{\ln\left(1 +\frac{r}{R_{\rm s}}\right)- \left(1+
\frac{R_{\rm s}}{r}\right)^{-1}\right\}.
\end{eqnarray}

Using the assumption of HSE, the electron number density profile $n_{\rm e}(r)$ can be found and, taking the gas to be ideal, the electron temperature profile is calculated by ${\rm k}_{\rm B} T(r)\,=\,P_{\rm e}(r)/n_{\rm e}(r)$,
\begin{eqnarray}\label{eq:Tgas}
{\rm k_{\rm B}}T(r) & = & (4\pi \mu {\rm G}\rho_{\rm {s}})(R^3_{\rm s})\times \nonumber\\
 &  &  \left [ \frac{\ln\left(1 +\frac{r}{R_{\rm s}}\right)- \left(1+\frac{R_{\rm s}}{r}\right)^{-1}}{r}  \right] \times
\nonumber\\
 &  &  \left [1 + \left(\frac{r}{r_{\rm p}}\right)^{ a} \right]\left[{b} \left(\frac{r}{r_{\rm p}}\right)^{a} + {c}
\right]^{-1}.
\end{eqnarray}

Parameter values are estimated at or within an overdensity radius $r_{\Delta}$, the cluster radius internal to which the mean density is $\Delta$ times the critical density at the cluster redshift. Throughout this paper we use parameters estimated from this model at or within $r_{200}$ and $r_{500}$.

\begin{table*}
\small
\begin{center}
\caption{Some \textsc{McAdam} derived large-scale cluster parameter values for the AMI-CLASH
sub-sample, internal to $r_{200}$. $T_{\rm SZ}(r_{200})$ is the gas temperature at $r_{200}$. Uncertainties are 68 per cent confidence uncertainties (1$\sigma$). ``/" means ``in units of". Due to the strong correlations between the parameters, their uncertainties are also correlated.}\label{tab:r200}\setlength\extrarowheight{1pt}
\begin{tabular}{lcccccc}\hline
Cluster    & $M_{{\rm tot},\,r_{200}}$\vpad  & $M_{{\rm gas},\,r_{200}}$\vpad  & $r_{200}$    & $T_{\rm SZ}(r_{200})$ &$Y_{r_{200}}$               & Redshift \\
              & /($ 10^{14}$\Msun)      & /($ 10^{14}$\Msun)       & /\,Mpc                   & /\,keV                &/($ 10^{-4}{\rm Mpc}^{-2}$) &      \vspace{1mm}\\\hline
A611\vpad     & $7.7\pm1.1$             & $0.99\pm0.10$            & $1.71\pm0.08$            & $3.3\pm0.3$           & $0.67\pm0.11$              & 0.288\\
A1423         & $4.5\pm1.1$             & $0.57\pm0.14$            & $1.46\pm0.12$            & $2.2\pm0.4$           & ${0.28}^{+0.10}_{-0.09}$   & 0.213\\
A2261         & $13.3\pm1.9$            & $1.69\pm0.08$            & $2.10\pm0.10$            & ${4.6}^{+0.5}_{-0.4}$ & ${1.58}^{+0.19}_{-0.17}$    & 0.224\\
CLJ1226+3332  & $4.9\pm1.1$             & $0.62\pm0.14$            & $1.15\pm0.09$            & $3.1\pm0.5$           & $0.36\pm0.13$              & 0.890\\
MAJ0647+7015  & ${11.4}^{+1.5}_{-1.4}$  & $1.41\pm0.11$            & $1.74\pm0.07$            & $4.8\pm0.4$           & $1.27\pm0.16$              & 0.584\\
MAJ0717+3745  & $12.9\pm1.8$            & $1.65\pm0.16$            & $1.84\pm0.09$            & $5.2\pm0.5$           & $1.60\pm0.25$              & 0.548\\
MAJ0744+3927  & $11.5\pm1.4$            & $1.50\pm0.12$            & $1.68\pm0.07$            & $5.1\pm0.4$           & $1.40\pm0.17$              & 0.686\\
MAJ1149+2223  & $15.7\pm1.7$            & $2.00\pm0.11$            & $1.97\pm0.07$            & $5.9\pm0.4$           & $2.20\pm0.20$              & 0.544\\
MAJ1423+2404  & ${7.7}^{+2.1}_{-2.3}$   & ${0.99}^{+0.26}_{-0.27}$ & $1.54\pm0.15$            & $3.6\pm0.7$           & ${0.72}^{+0.29}_{-0.30}$   & 0.545\\
RXJ1532+3021  & $5.6\pm1.8$             & ${0.73}^{+0.22}_{-0.23}$ & ${1.49}^{+0.16}_{-0.17}$ & $2.7\pm0.6$           & ${0.42}^{+0.20}_{-0.21}$   & 0.345\\
MAJ0717 RH    & $14.1\pm1.9$            & $1.80\pm0.16$            & $1.89\pm0.09$            & $5.5\pm0.5$           & $1.85\pm0.27$              & 0.548\vspace{1mm}\\\hline
\end{tabular}
\vspace{2mm}
\end{center}
\end{table*}
\begin{table*}
\small
\begin{center}
\caption{Same as Table \ref{tab:r200} for cluster parameter estimates at, or internal to, $r_{500}$.}\label{tab:r500}\setlength\extrarowheight{1pt}\vspace{1mm}
\begin{tabular}{lccccc}\hline
Cluster   & $M_{{\rm tot},\,r_{500}}$\vpad & $M_{{\rm gas},\,r_{500}}$\vpad & $r_{500}$    & $T_{\rm SZ}(r_{500})$  &$Y_{r_{500}}$               \\
              & /($ 10^{14}$\Msun)         & /($ 10^{14}$\Msun)           & /\,Mpc         & /\,keV                 &/($ 10^{-4}{\rm Mpc}^{-2}$) \vspace{1mm}\\\hline
A611\vpad     & $5.7\pm0.8$                & $0.69\pm0.07$                & $1.14\pm0.06$  & $4.0\pm0.4$            & $0.51\pm0.08$              \\
A1423         & $3.3\pm0.8$                & $0.39\pm0.09$                & $0.98\pm0.08$  & $2.8\pm0.5$            & $0.21\pm0.08$              \\
A2261         & $9.8\pm1.4$                & ${1.17}^{+0.05}_{-0.06}$     & $1.40\pm0.07$  & $5.6\pm0.5$            & ${1.20}^{+0.14}_{-0.13}$   \\
CLJ1226+3332  & $3.6\pm0.8$                & $0.45\pm0.10$                & $0.77\pm0.06$  & $3.6\pm0.6$            & $0.27\pm0.10$              \\
MAJ0647+7015  & $8.4\pm1.1$                & $1.00\pm0.08$                & $1.16\pm0.05$  & $5.6\pm0.5$            & $0.97\pm0.12$              \\
MAJ0717+3745  & $9.6\pm1.3$                & $1.17\pm0.11$                & $1.23\pm0.06$  & ${6.1}^{+0.5}_{-0.6}$  & $1.22\pm0.19$              \\
MAJ0744+3927  & $8.5\pm1.0$                & $1.07\pm0.09$                & $1.12\pm0.05$  & $5.9\pm0.5$            & $1.10\pm0.13$              \\
MAJ1149+2223  & $11.6\pm1.3$               & $1.42\pm0.08$                & $1.31\pm0.05$  & $6.9\pm0.5$            & $1.67\pm0.15$              \\
MAJ1423+2404  & $5.7\pm1.6$                & $0.70\pm0.19$                & $1.03\pm0.10$  & $4.3\pm0.8$            & ${0.55}^{+0.22}_{-0.23}$   \\
RXJ1532+3021  & ${4.2}^{+1.3}_{-1.4}$      & $0.51\pm0.16$                & $1.00\pm0.11$  & $3.3\pm0.7$            & $0.32\pm0.16$              \\
MAJ0717 RH    & $10.4\pm1.4$               & ${1.30}^{+0.12}_{-0.14}$     & $1.26\pm0.06$  & $6.4\pm0.6$            & $1.40\pm0.21$              \vspace{1mm}\\\hline
\end{tabular}
\vspace{2mm}
\end{center}
\vspace{1mm}
\end{table*}


\subsection{Priors}
\label{sec:priors}
Priors are summarised in Table \ref{table_priors}.
The prior on cluster total mass is based on {\color{black}statistics of cluster masses and} AMI observing capabilities. {\color{black}The prior on the gas fraction at $r_{200}$, denoted $f_{{\rm gas},\,r_{200}}$, is tight as SZ data do not contrain this property: our priors are based on X-ray studies and WMAP, see e.g. \citet{2006ApJ...640..691V} \citet{2009ApJ...692.1033V} and \citet{2011ApJS..192...18K}.}
The shape parameters $a$, $b$, $c$, and $c_{500}$ are given delta priors at the ``universal" values from \citet{2010AA...517A..92A}, who assign these parameters the values $a$\,=\,1.0510, $b$\,=\,5.4905, $c$\,=\,0.3081 and $c_{500}$\,=\,1.177.
Imposing a fixed pressure profile shape to all sample members, we expect the derived cluster parameters to be highly correlated; this is discussed further in Section \ref{sc:scalingrel}.

Priors on the radio point source parameters ($x_{\rm s}, y_{\rm s}, S_0, \alpha_{\rm s}$) are taken from source-finding in the LA maps, as described in Section \ref{sec:observations}, and are dependent on the detection significance of the source in the SA map. For all point sources, priors on position $(x_{\rm s}, y_{\rm s})$ are set as delta functions to the LA values due to the much higher resolution of the LA. Sources detected at $>$\,4$\sigma_{\rm SA}$ are given Gaussian priors on the LA flux estimate and the spectral index (either fitted from the LA channel data or based on the 10C survey.) These parameters are modeled simultaneously by \textsc{McAdam} with the cluster parameters to account for possible flux calibration errors between the arrays and source variability between observations, discussed earlier.
Sources detected at $<$\,4$\sigma_{\rm SA}$ are given delta-priors on LA flux and spectral index, but where the source position is close to the cluster centre, we replace delta priors with Gaussian priors {\color{black}as detailed in Table \ref{table_priors}}.

\section{SZ maps and physical parameters}
\label{sec:results}

\begin{figure*}
\includegraphics[trim= 0mm 16.7mm 0mm 12mm, clip,width=88mm]{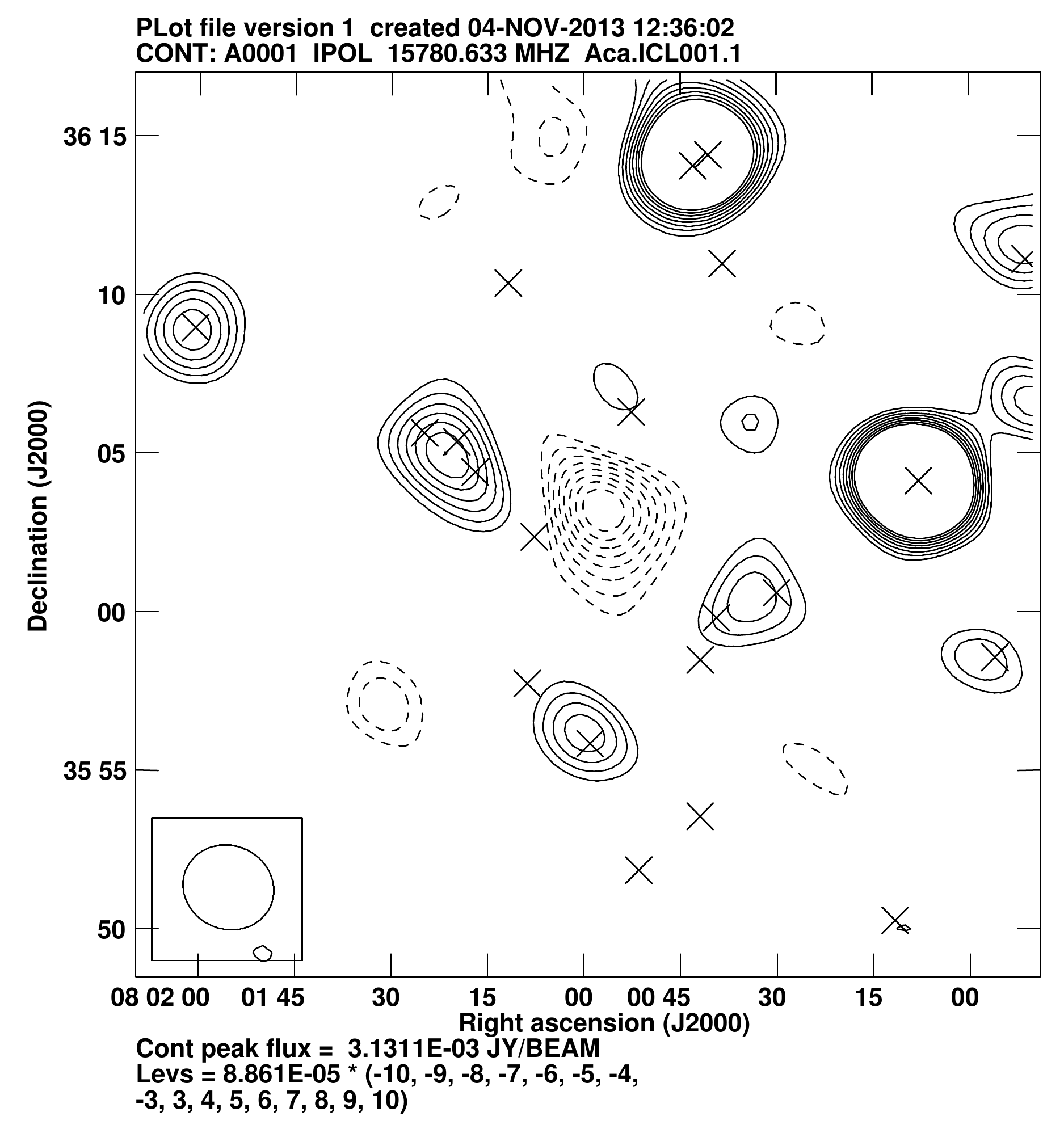}
\includegraphics[trim= 0mm 16.7mm 0mm 12mm, clip,width=88mm]{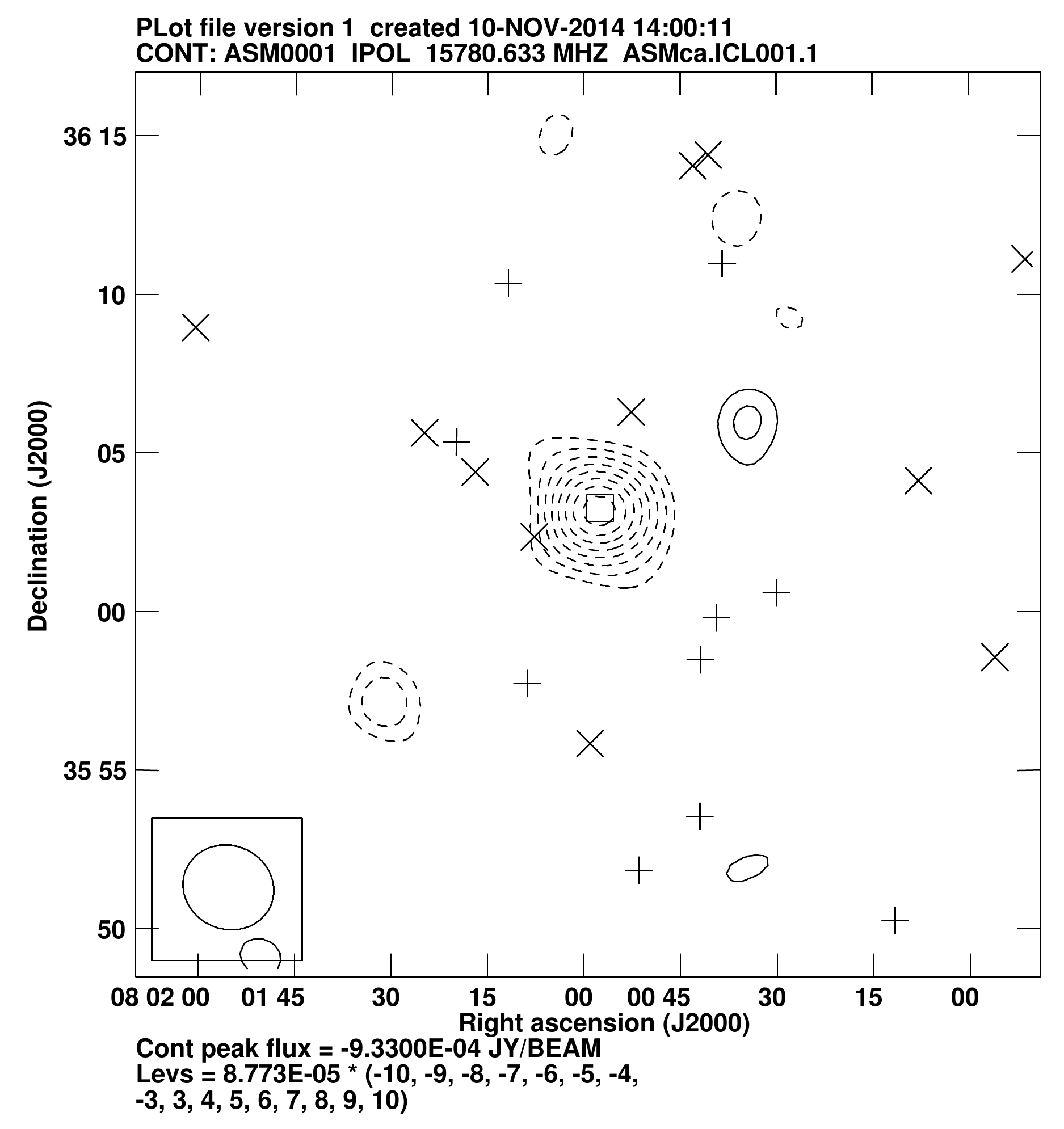}
\includegraphics[trim= 0mm 5mm 0mm -10mm, clip,width=100mm]{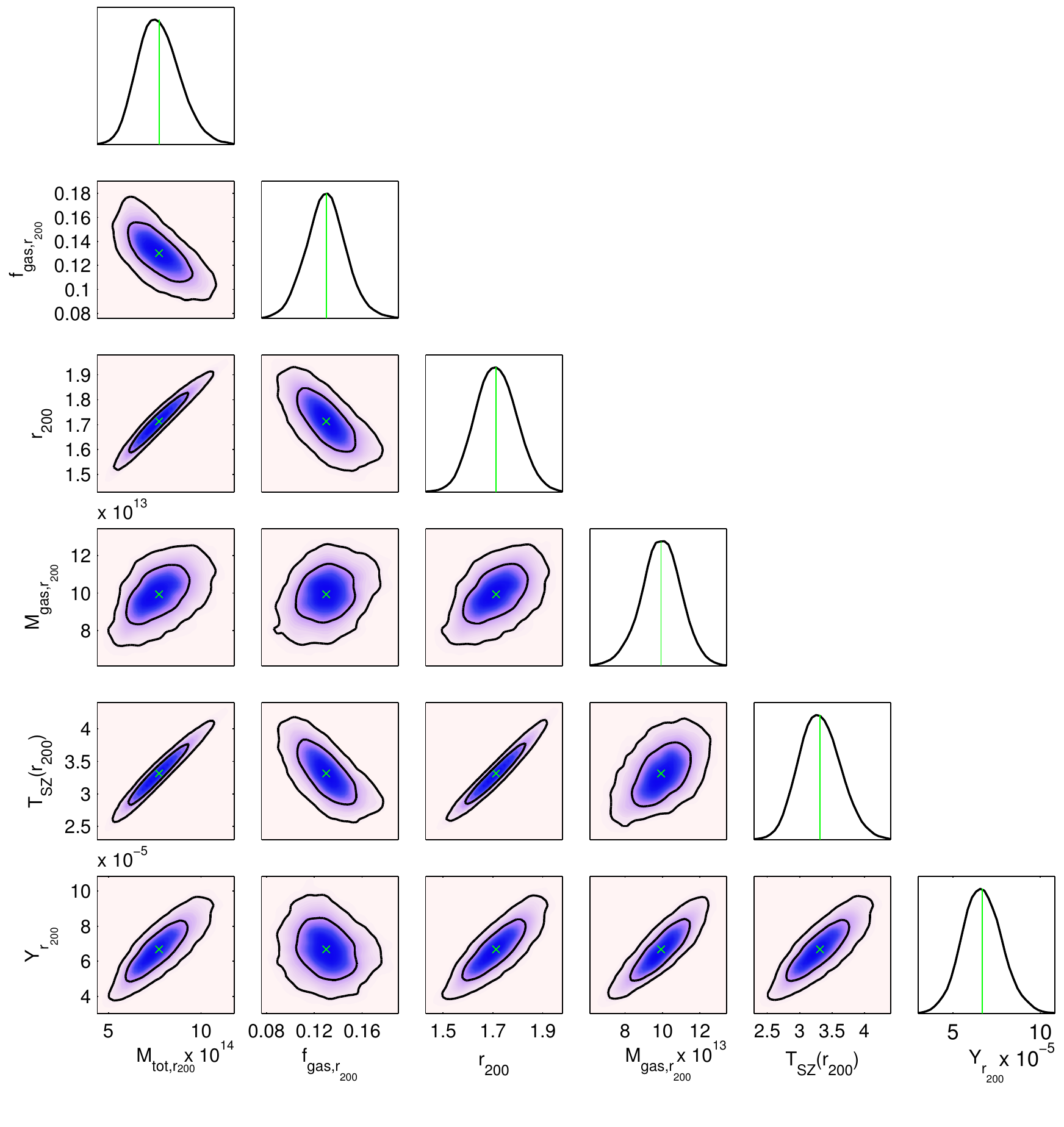}
\includegraphics[trim= -10mm 0mm 0mm -5mm, clip,width=71mm]{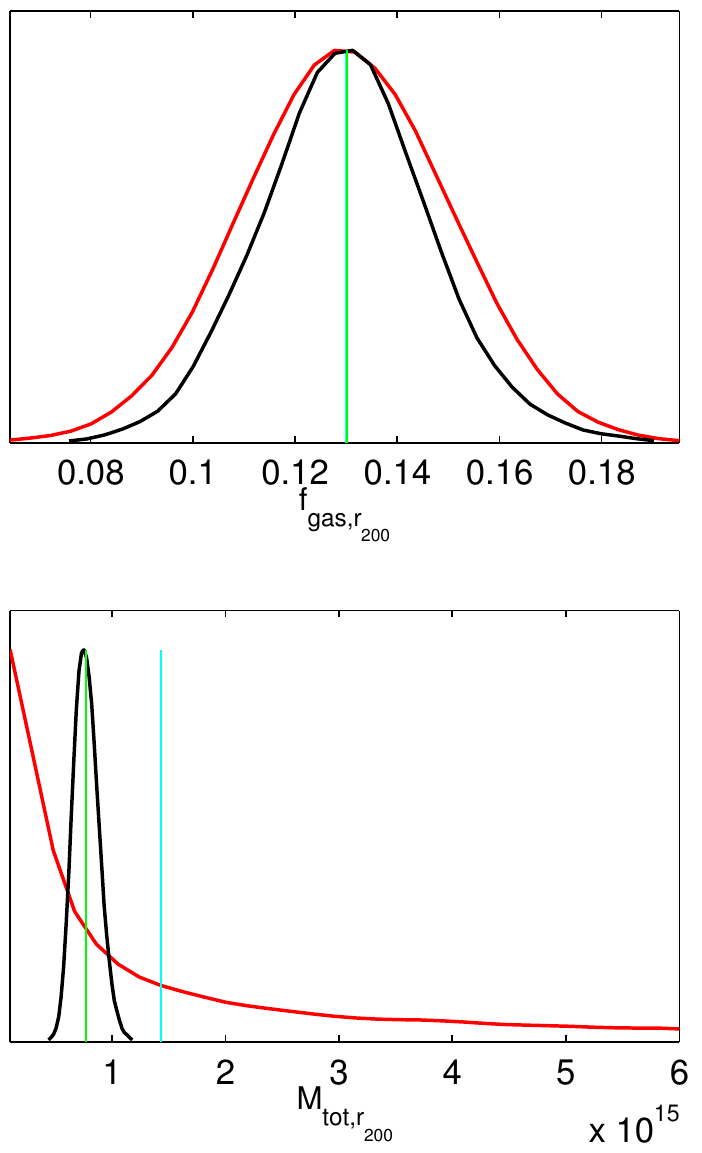}
\caption{\textbf{ A611}. Top two panels: SA contour maps -- left shows the non-source-subtracted map ($\sigma_{\rm SA} = 89\,\mu$Jy) with crosses showing the positions of LA sources, right shows the source subtracted map ($\sigma_{\rm SA} = 88\,\mu$Jy, {\color{black}11$\sigma_{\rm SA}$ decrement}) with crosses marking positions of sources that have been modelled by \textsc{McAdam} and plus signs showing those given delta priors on flux as well as position. Solid contours are positive emission, dashed are negative. Bottom left panel: \textsc{McAdam} fitted parameter probability distributions, the green lines and crosses show the mean and the contour levels represent 68 per cent and 95 per cent confidence limits. {\color{black}Bottom right panel: 1D marginalised posterior distributions for $f_{{\rm gas},\,r_{200}}$ (upper) and $M_{{\rm tot},\,r_{200}}$ (lower) (as shown in the bottom left panel) overlaid with the priors in red. Fitted mean values are shown in green and prior mean values in cyan.} \label{fig:A611}}
\end{figure*}

\begin{figure*}
\includegraphics[trim= 0mm 16.7mm 0mm 12mm, clip,width=88mm]{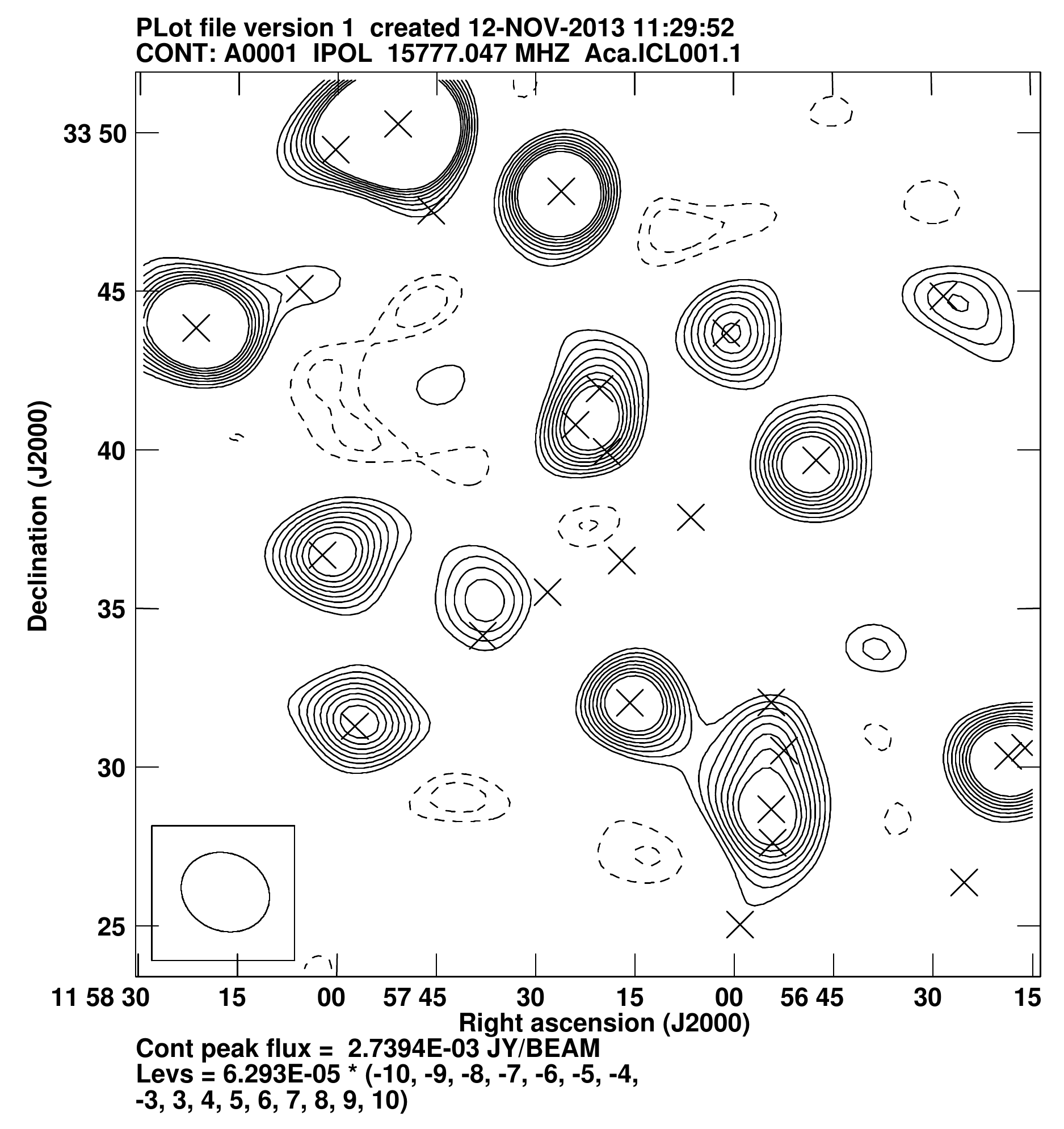}
\includegraphics[trim= 0mm 16.7mm 0mm 12mm, clip,width=88mm]{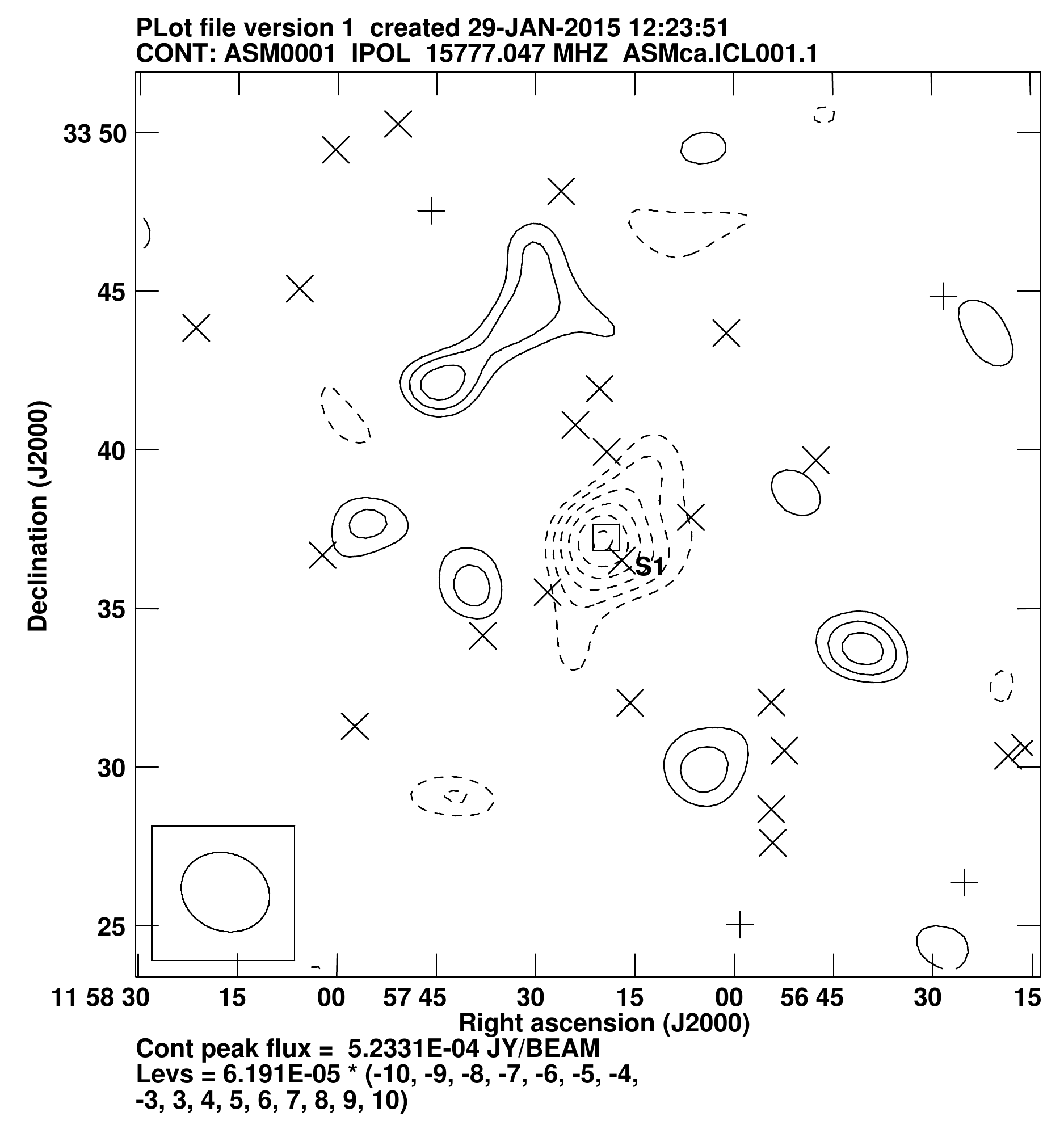}
\includegraphics[trim= 0mm 0mm 0mm -8mm, clip,width=88mm]{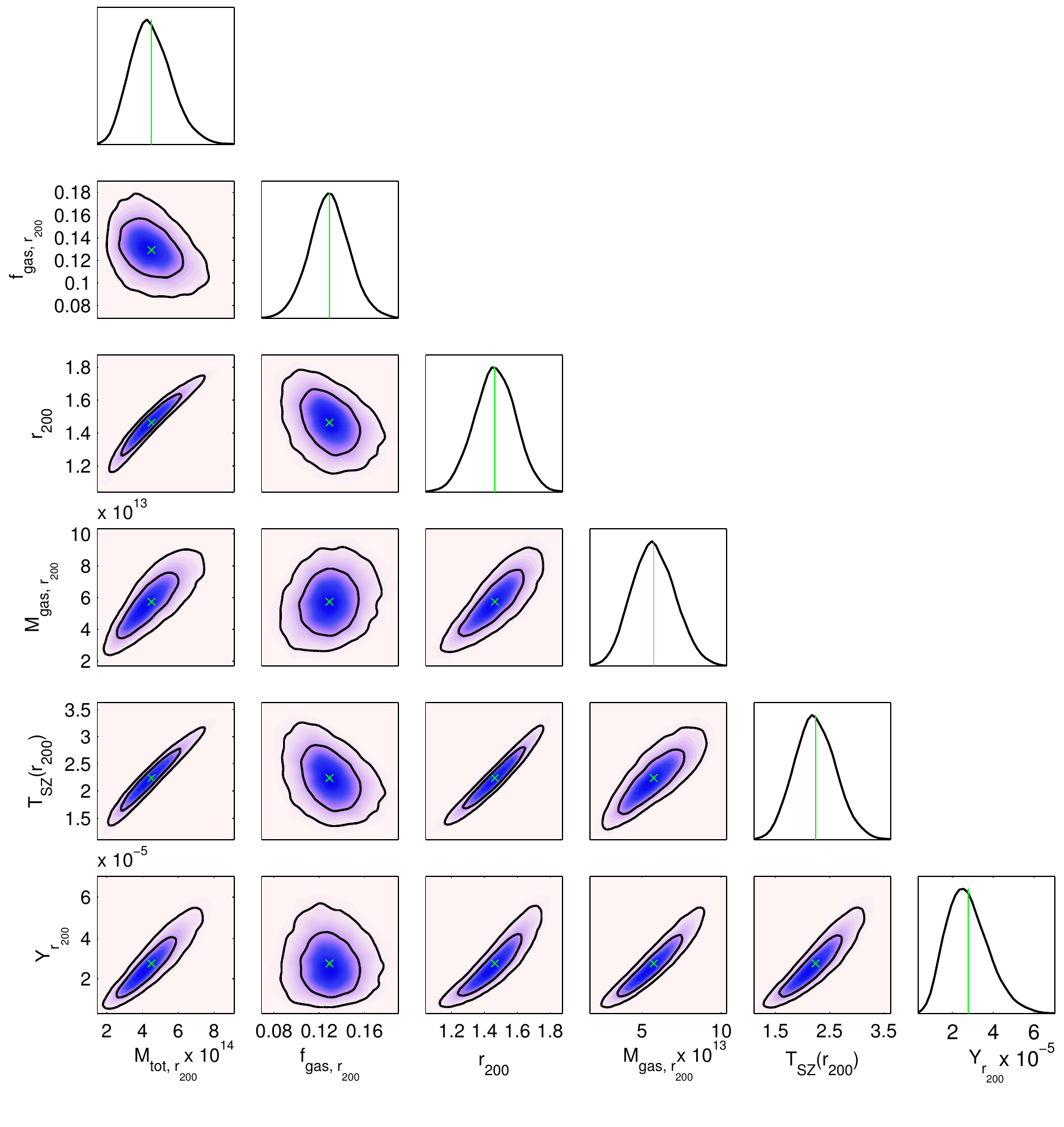}
\includegraphics[trim= 0mm 0mm 0mm -8mm, clip,width=88mm]{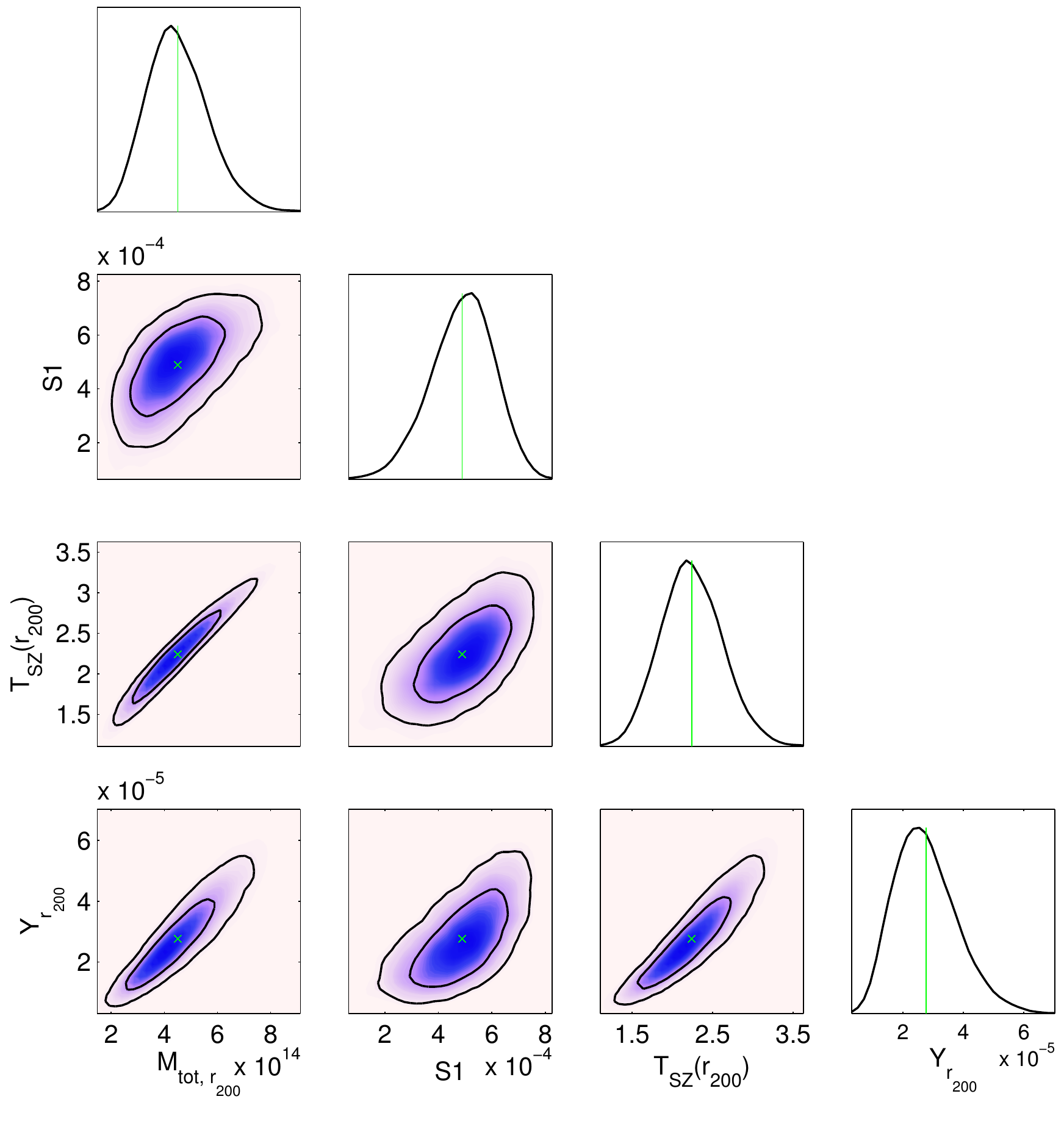}
\caption{\textbf{ A1423}. Top two panels: SA contour maps -- left shows the non-source-subtracted map ($\sigma_{\rm SA} = 63\,\mu$Jy), right shows the source subtracted map ($\sigma_{\rm SA} = 62\,\mu$Jy, {\color{black}8$\sigma_{\rm SA}$ decrement}). See Figure \ref{fig:A611} caption for more details. Bottom two panels: \textsc{McAdam} fitted probability distributions -- left shows parameter probability distributions, right shows degeneracies between the fitted cluster parameter values and the flux of source $\rm S1$, labelled on the map in the top right panel. The green lines and crosses show the mean and the contour levels represent 68 per cent and 95 per cent confidence limits. \label{fig:A1423}}
\end{figure*}

\begin{figure*}
\includegraphics[trim= 0mm 16.7mm 0mm 12mm, clip,width=88mm]{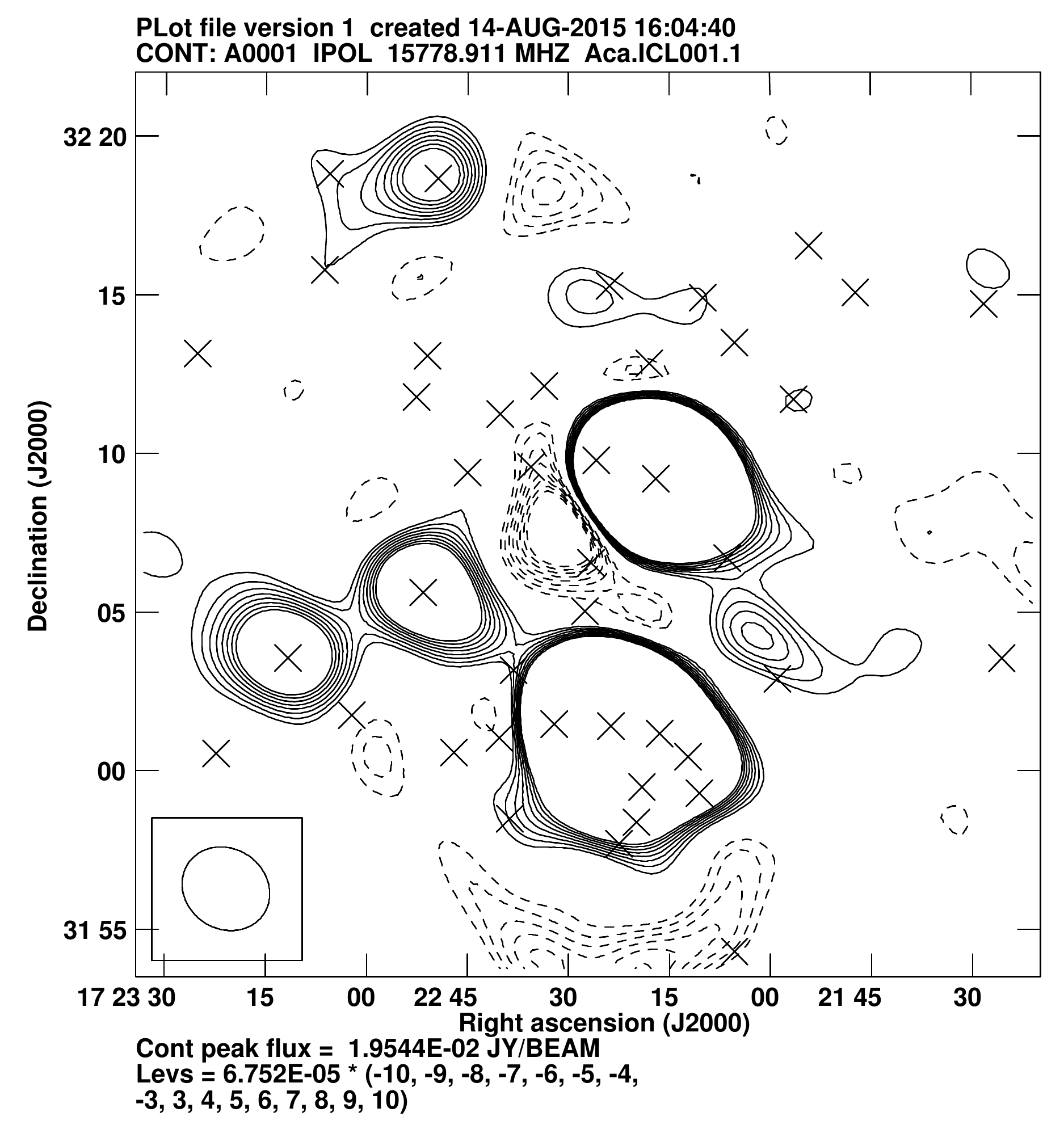}
\includegraphics[trim= 0mm 16.7mm 0mm 12mm, clip,width=88mm]{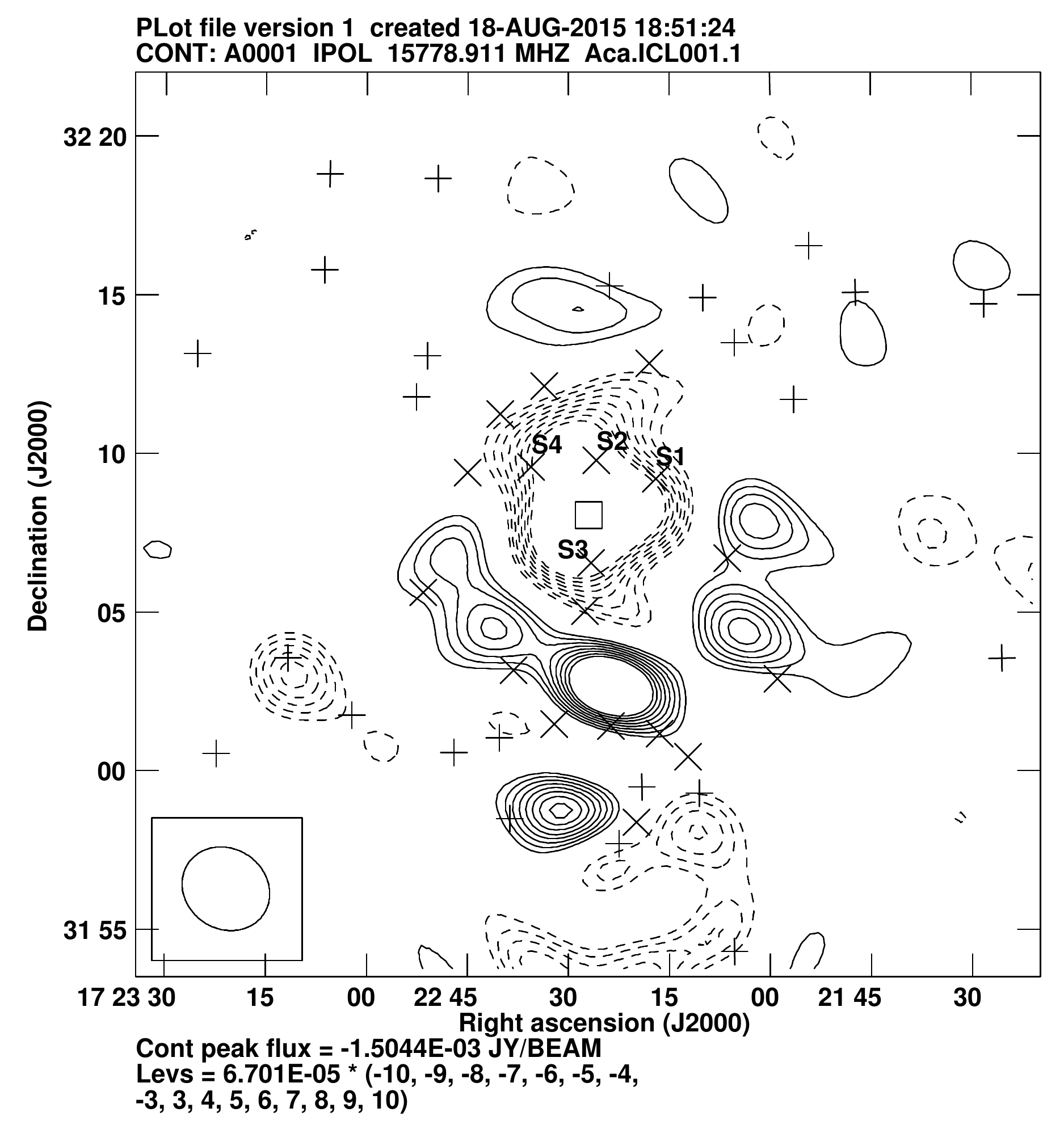}
\includegraphics[trim= 0mm 0mm 0mm -8mm, clip,width=88mm]{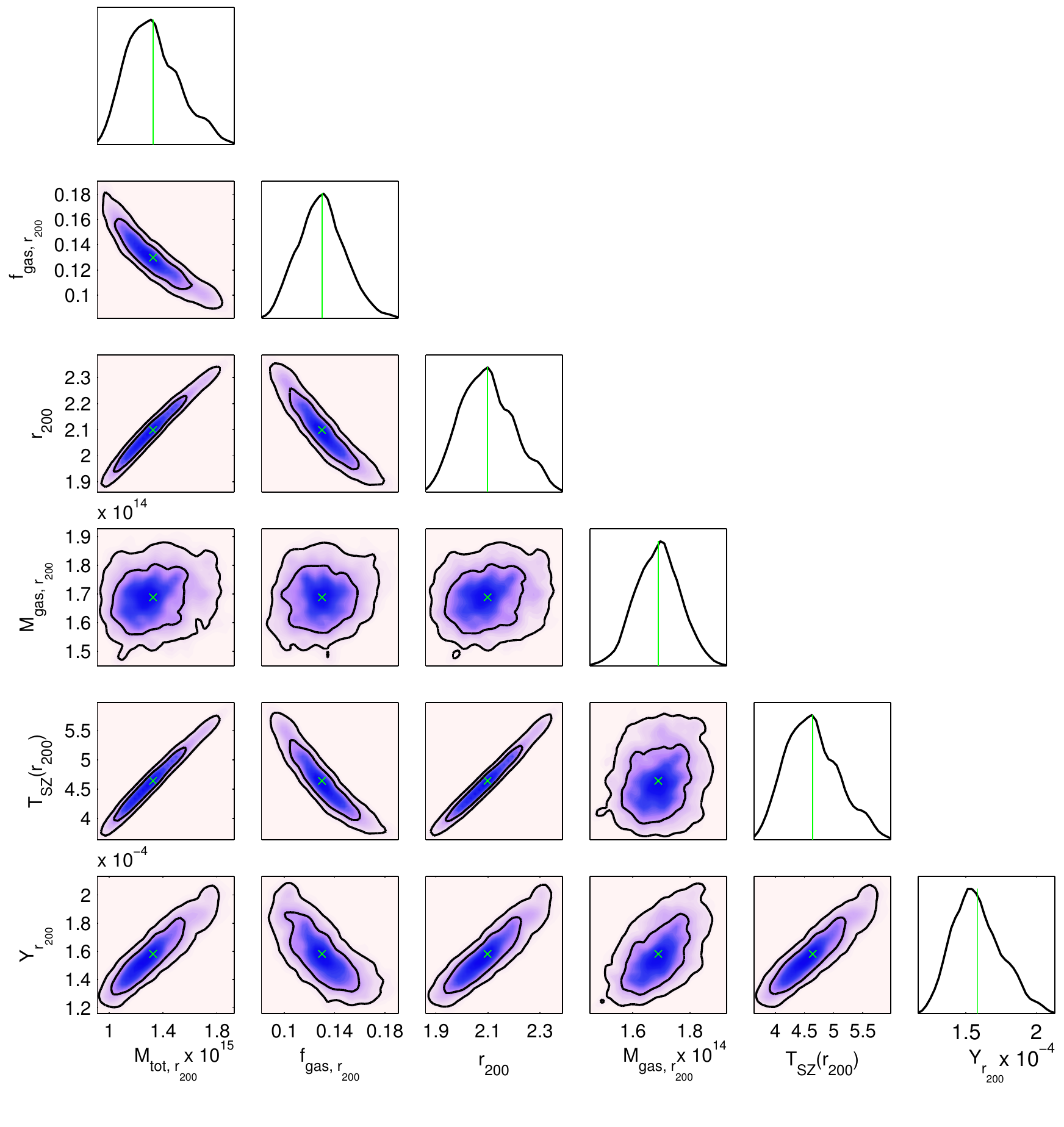}
\includegraphics[trim= 0mm 0mm 0mm -8mm, clip,width=88mm]{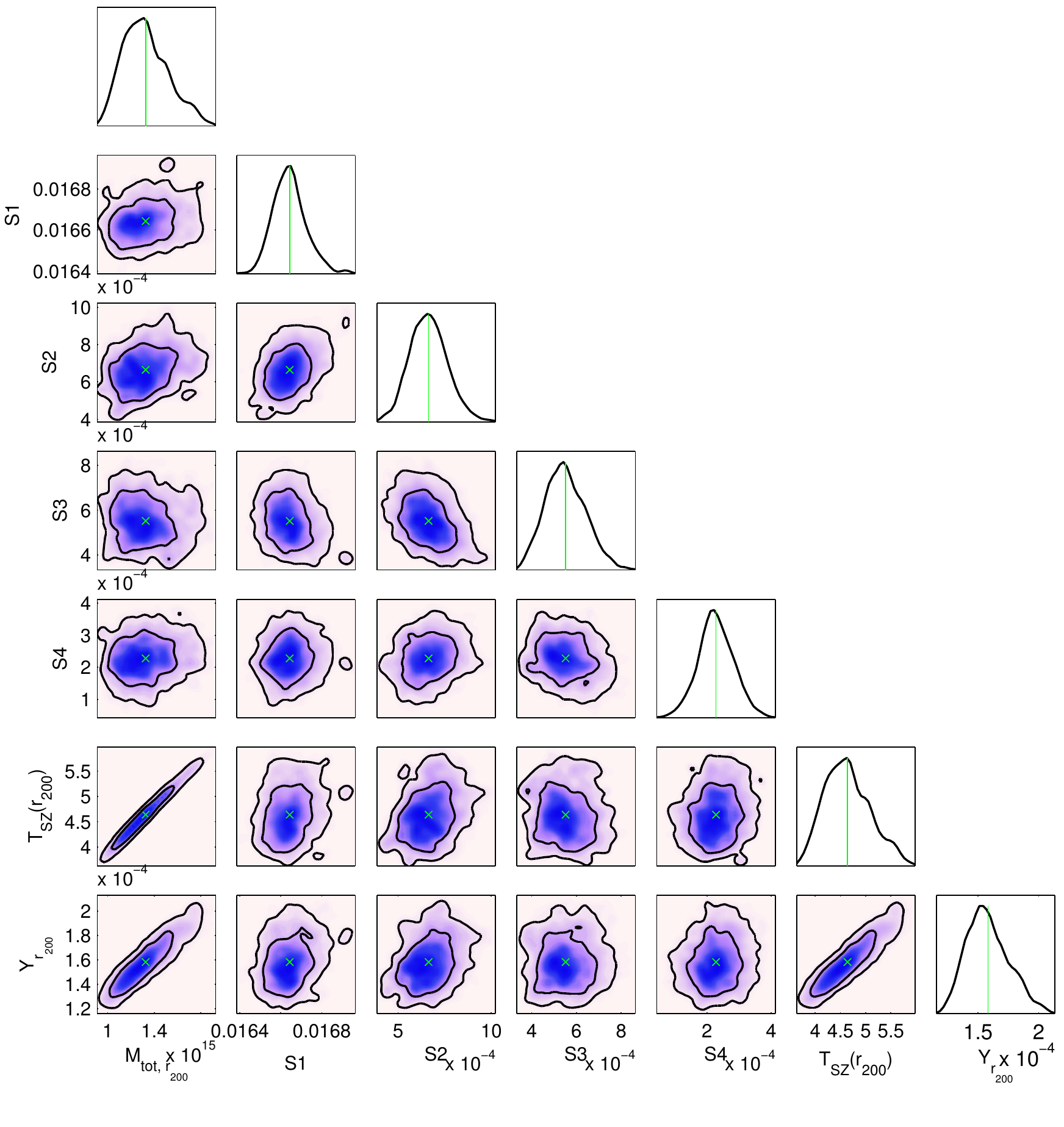}
\caption{\textbf{ A2261}. Top two panels: SA contour maps -- left shows the ($\sigma_{\rm SA} = 68\,\mu$Jy), right shows the source subtracted map ($\sigma_{\rm SA} = 67\,\mu$Jy, {\color{black}22$\sigma_{\rm SA}$ decrement}). See Figure \ref{fig:A611} caption for more details. Bottom two panels: \textsc{McAdam} fitted probability distributions -- left shows parameter probability distributions, right shows degeneracies between the fitted cluster parameter values and the flux of sources $\rm S1$, $\rm S2$, $\rm S3$, and $\rm S4$, labelled on the map in the top right panel. The green lines and crosses show the mean and the contour levels represent 68 per cent and 95 per cent confidence limits.
\label{fig:A2261}}
\end{figure*}

\begin{figure*}
\includegraphics[trim= 0mm 16.7mm 0mm 12mm, clip,width=88mm]{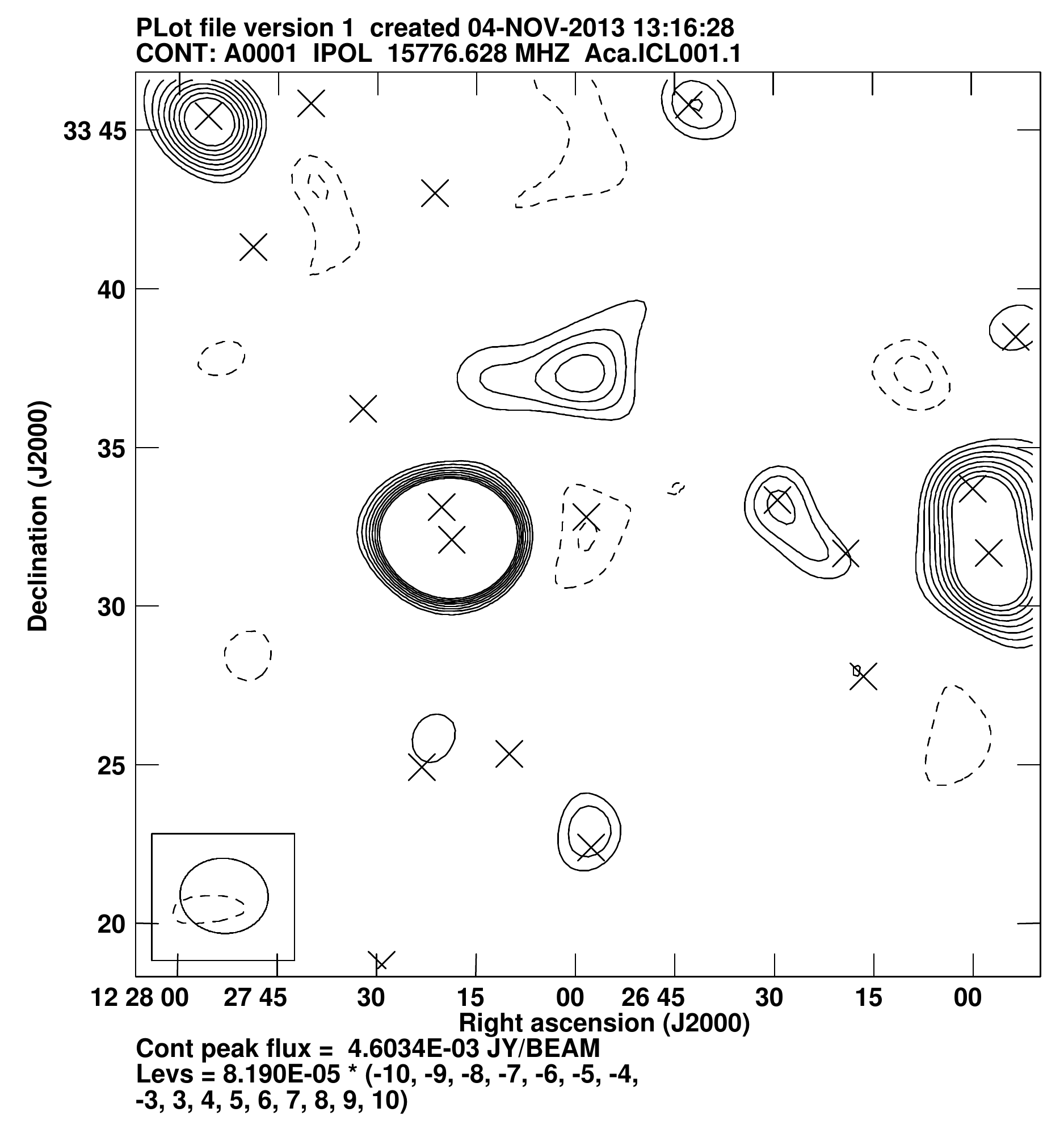}
\includegraphics[trim= 0mm 16.7mm 0mm 12mm, clip,width=88mm]{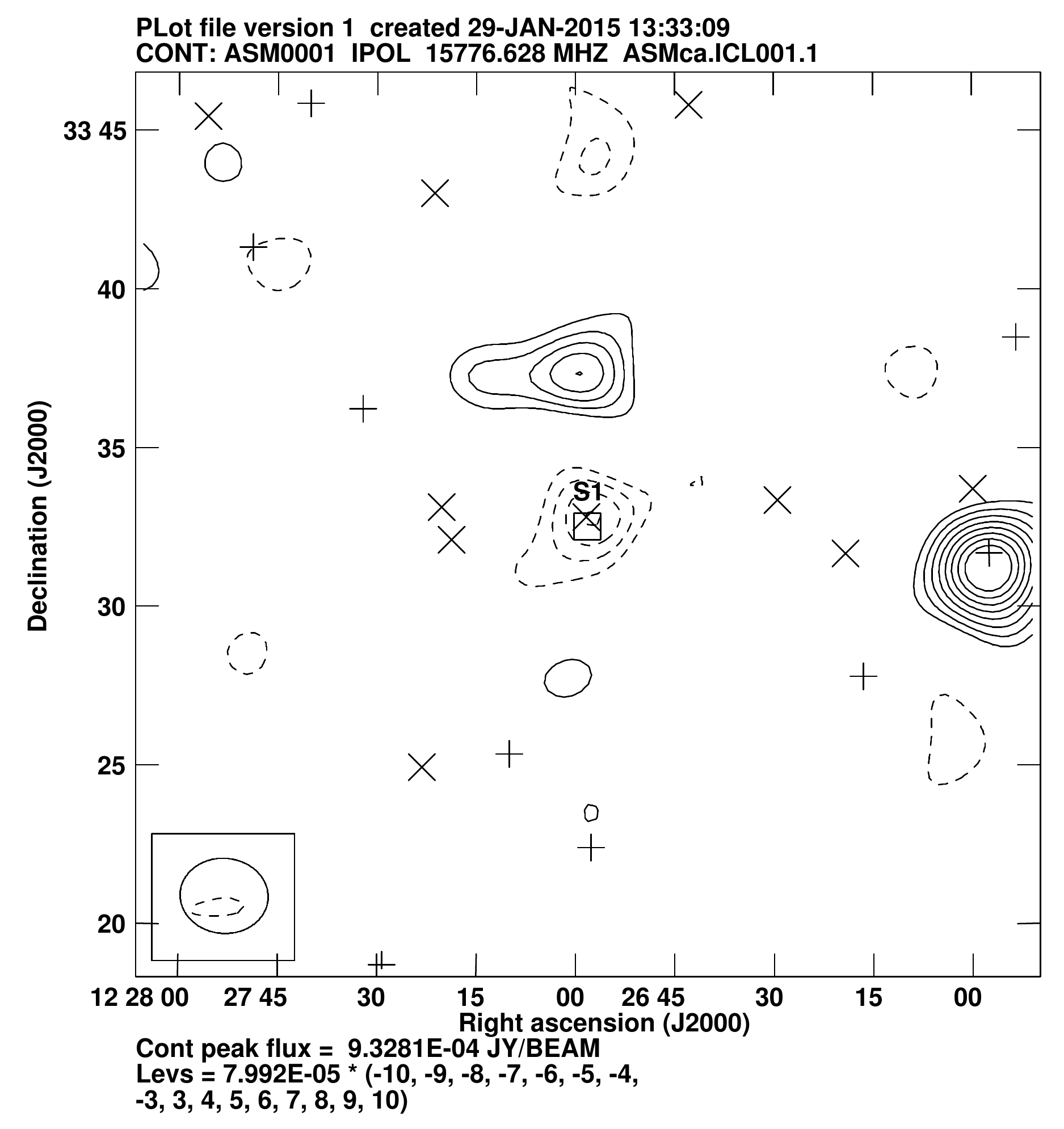}
\includegraphics[trim= 0mm 0mm 0mm -8mm, clip,width=88mm]{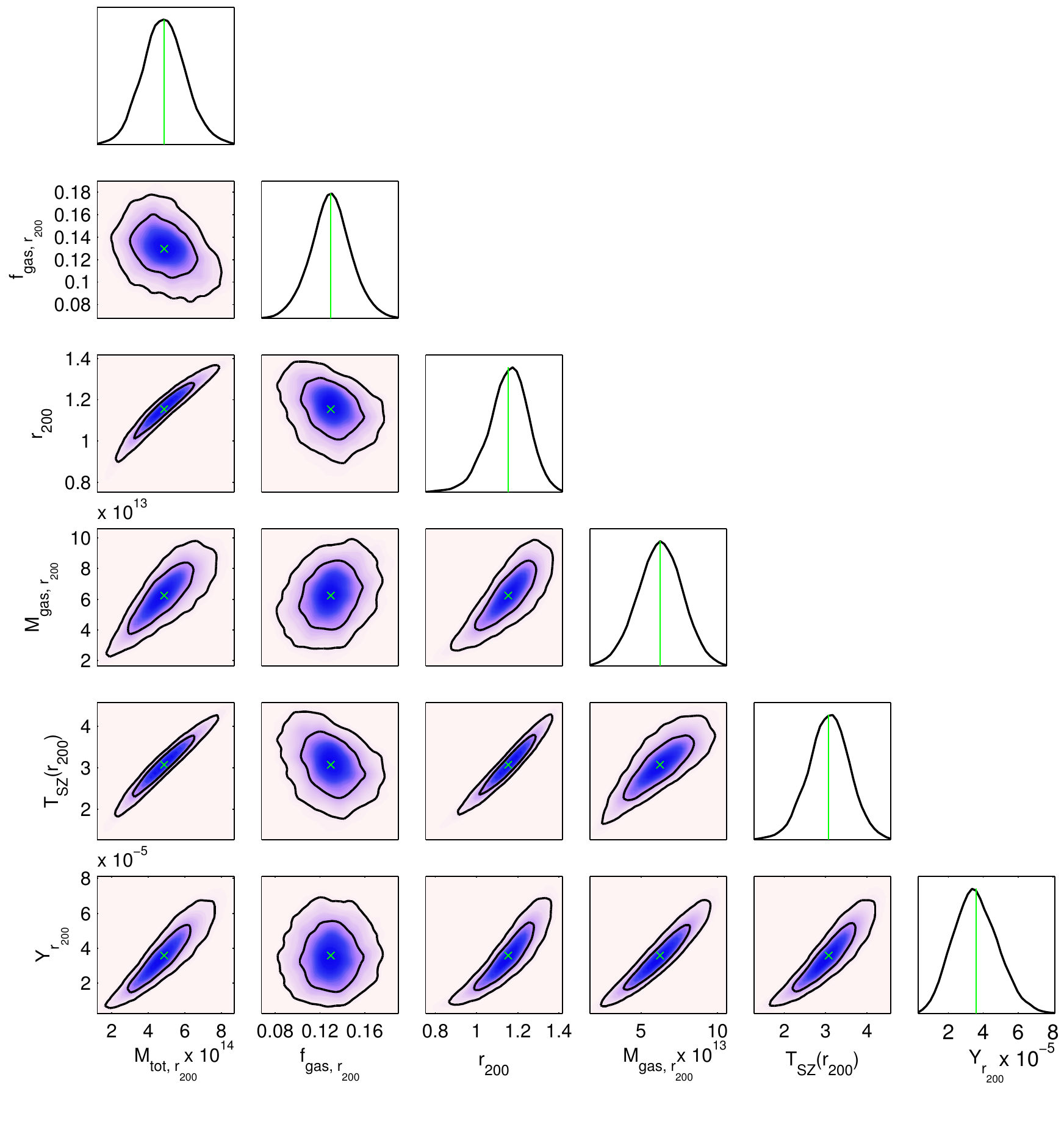}
\includegraphics[trim= 0mm 0mm 0mm -8mm, clip,width=88mm]{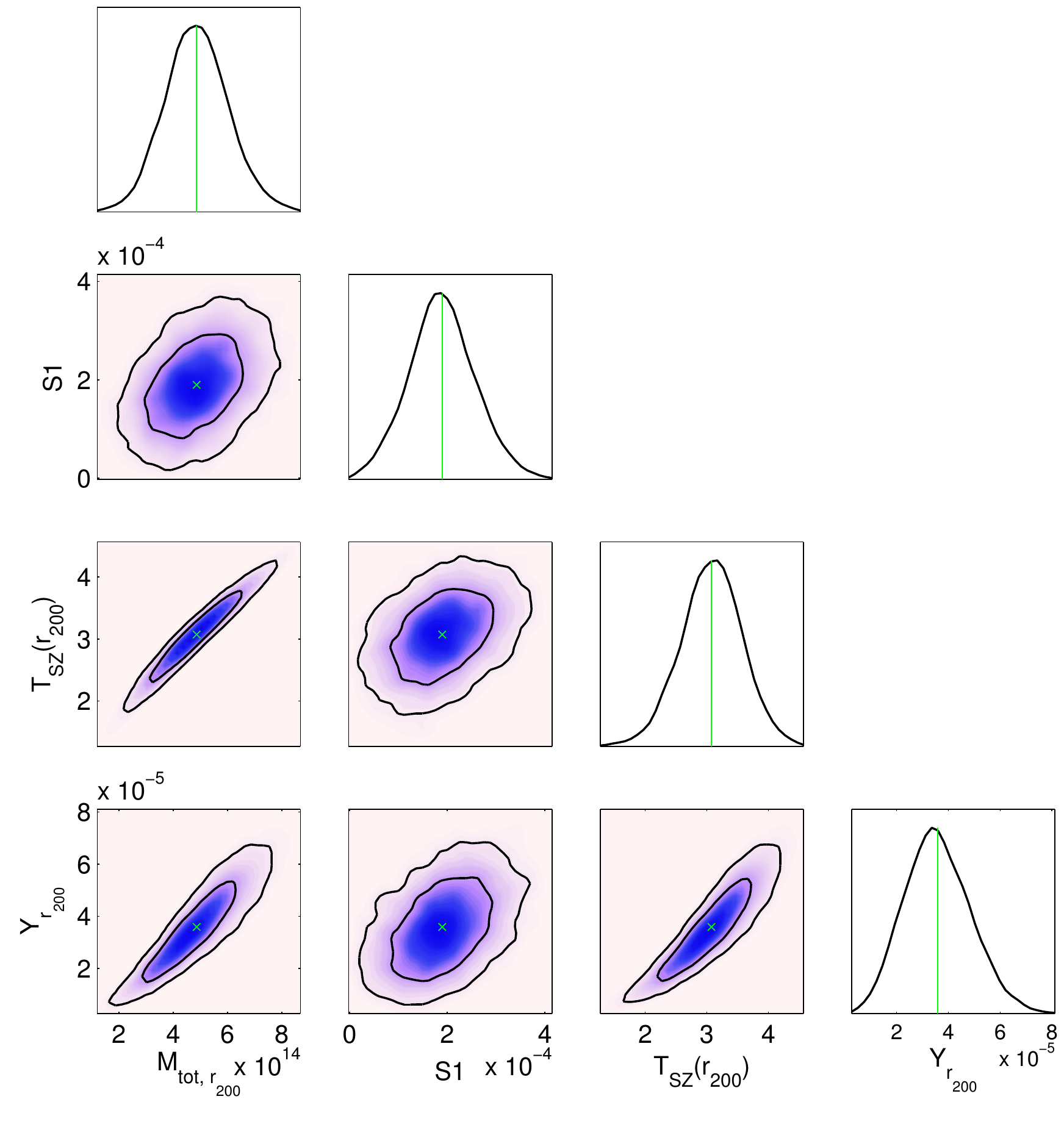}
\caption{\textbf{ CLJ1226+3332}. Top two panels: SA contour maps -- left shows the non-source-subtracted map ($\sigma_{\rm SA} = 82\,\mu$Jy), right shows the source subtracted map ($\sigma_{\rm SA} = 80\,\mu$Jy, {\color{black}6$\sigma_{\rm SA}$ decrement}). See Figure \ref{fig:A611} caption for more details. Bottom two panels: \textsc{McAdam} fitted probability distributions -- left shows parameter probability distributions, right shows degeneracies between the fitted cluster parameter values and the flux of source $\rm S1$, labelled on the map in the top right panel. The green lines and crosses show the mean and the contour levels represent 68 per cent and 95 per cent confidence limits. \label{fig:CLJ1226}}
\end{figure*}

\begin{figure*}
\includegraphics[trim= 0mm 16.7mm 0mm 12mm, clip,width=88mm]{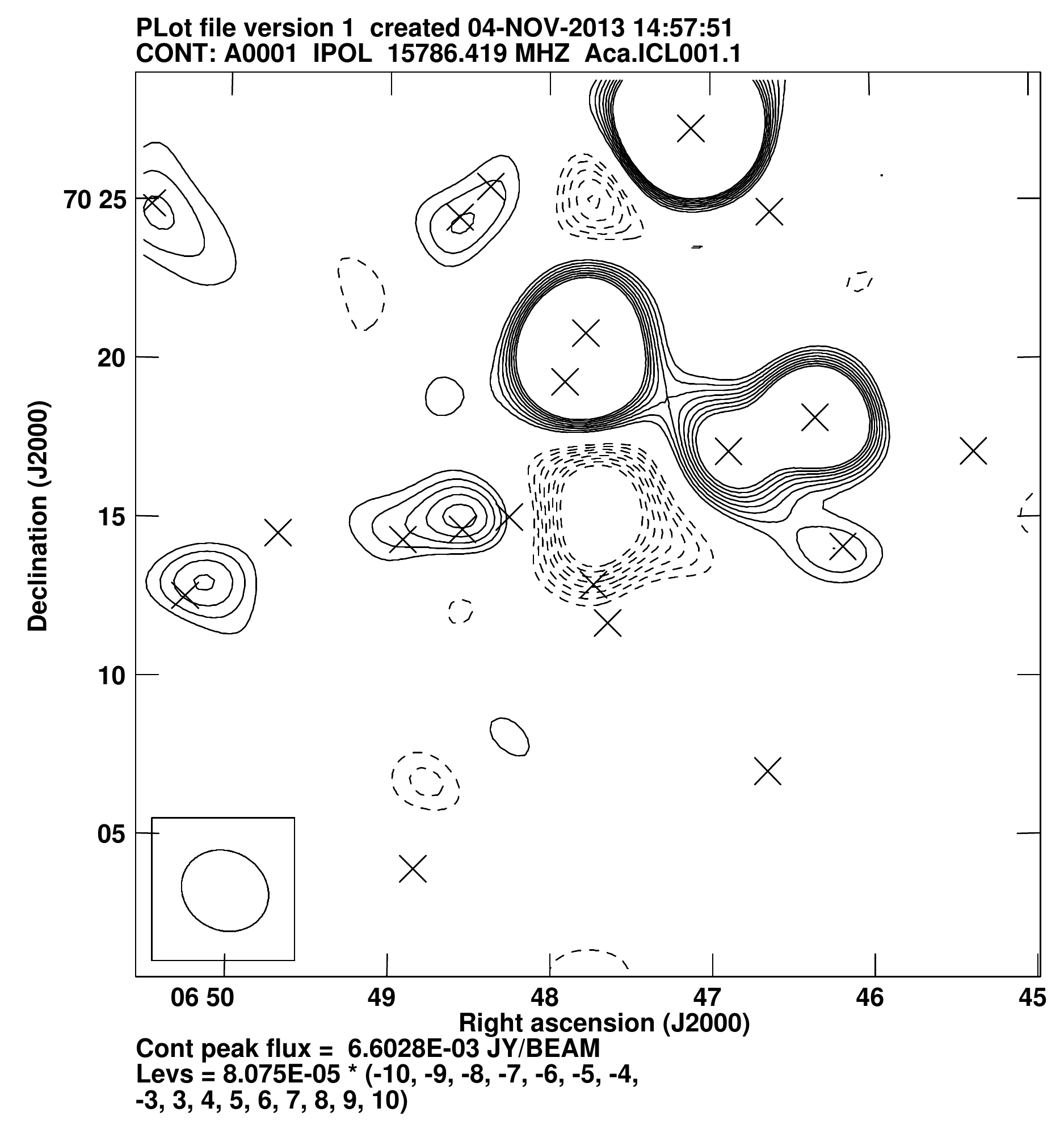}
\includegraphics[trim= 0mm 16.7mm 0mm 12mm, clip,width=88mm]{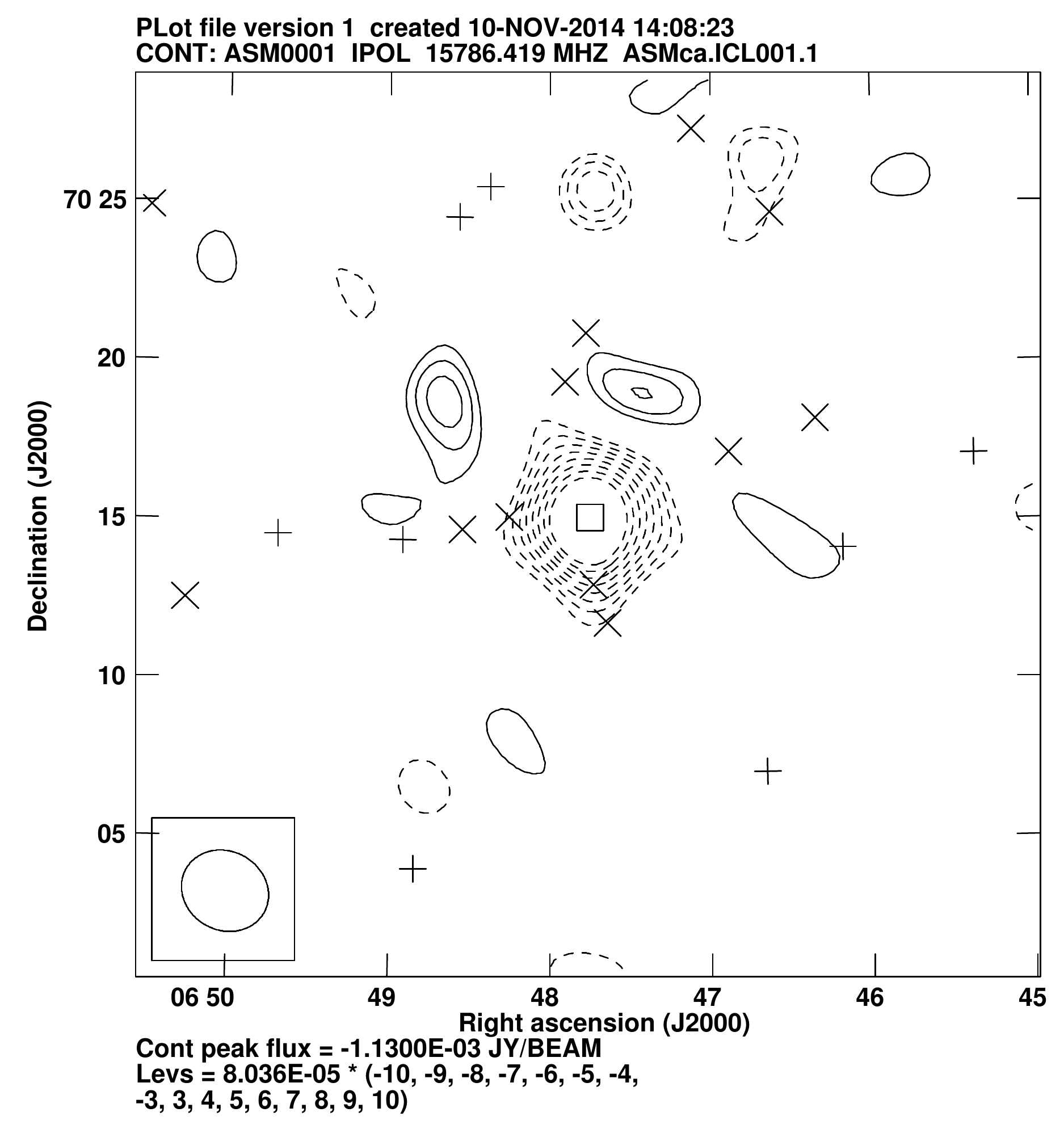}
\includegraphics[trim= 0mm -1mm 0mm -10mm, clip,width=100mm]{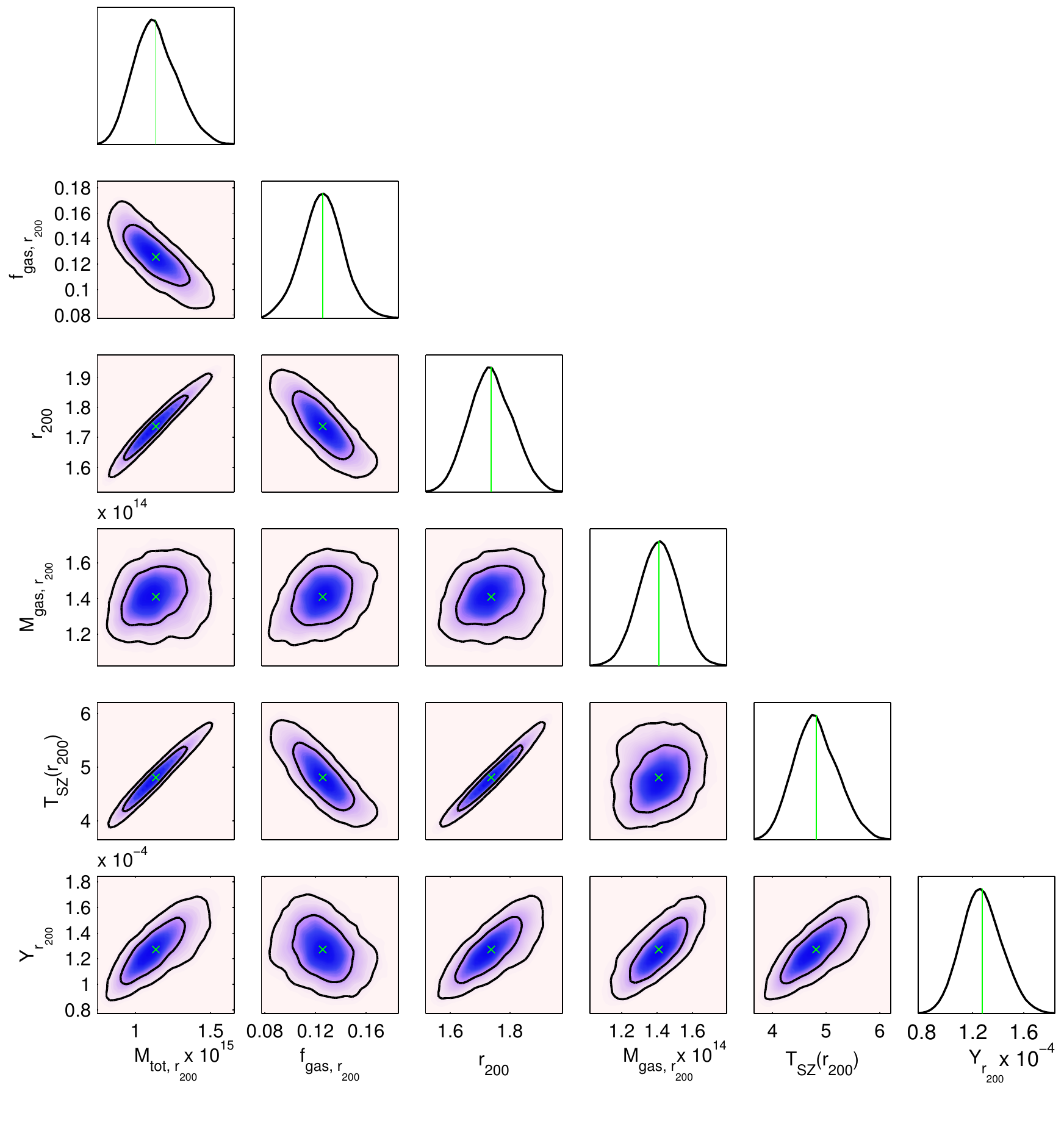}
\caption{\textbf{ MAJ0647+7015}. Top two panels: SA contour maps -- left shows the non-source-subtracted map ($\sigma_{\rm SA} = 81\,\mu$Jy), right shows the source subtracted map ($\sigma_{\rm SA} = 80\,\mu$Jy, {\color{black}14$\sigma_{\rm SA}$ decrement}). See Figure \ref{fig:A611} caption for more details. Bottom panel: \textsc{McAdam} fitted parameter probability distributions, the green lines and crosses show the mean and the contour levels represent 68 per cent and 95 per cent confidence limits. \label{fig:MAJ0647}}
\end{figure*}

\begin{figure*}
\includegraphics[trim= 0mm 16.7mm 0mm 12mm, clip,width=88mm]{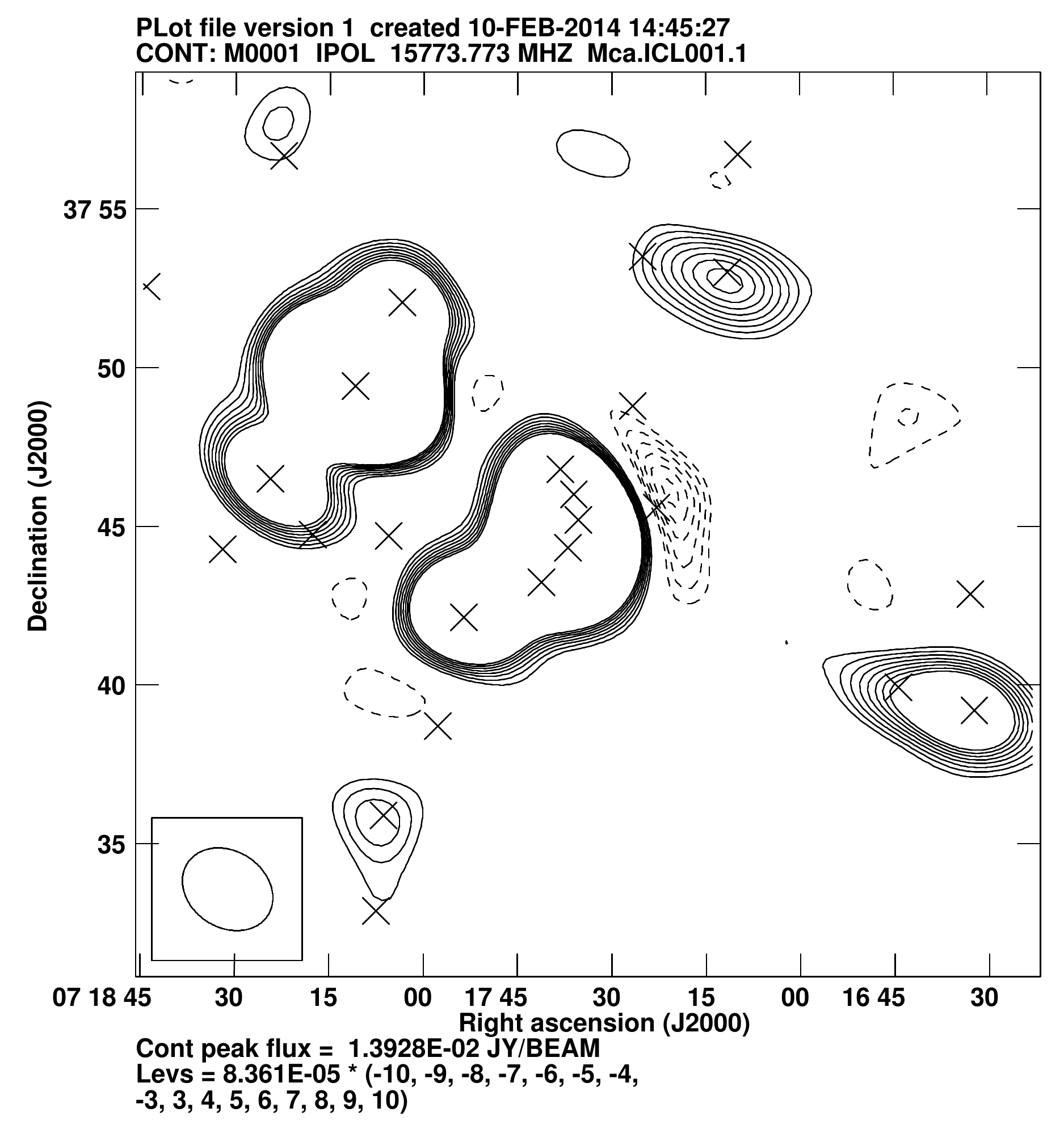}
\includegraphics[trim= 0mm 16.7mm 0mm 12mm, clip,width=88mm]{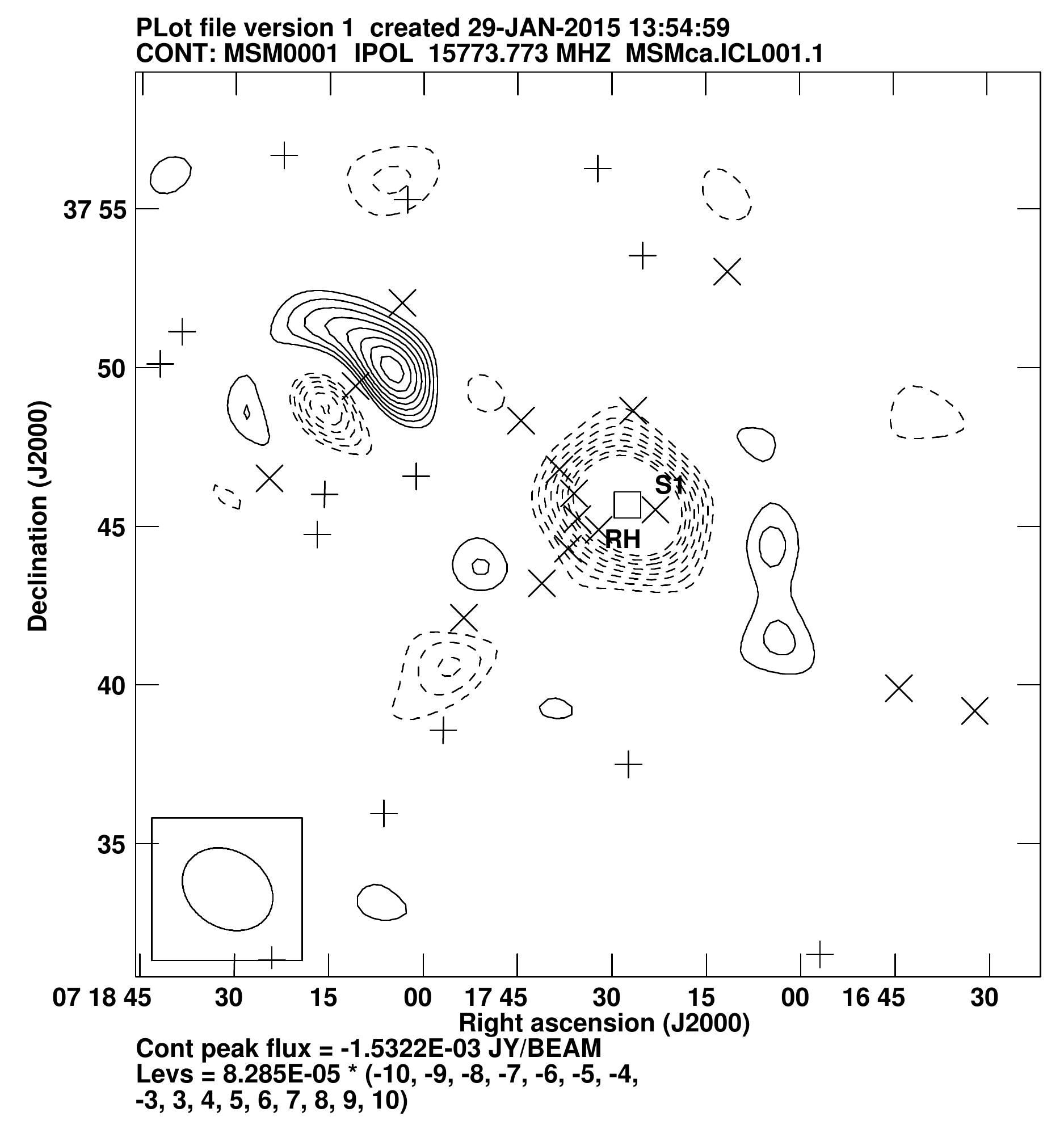}
\includegraphics[trim= 0mm 1mm 0mm -8mm, clip,width=88mm]{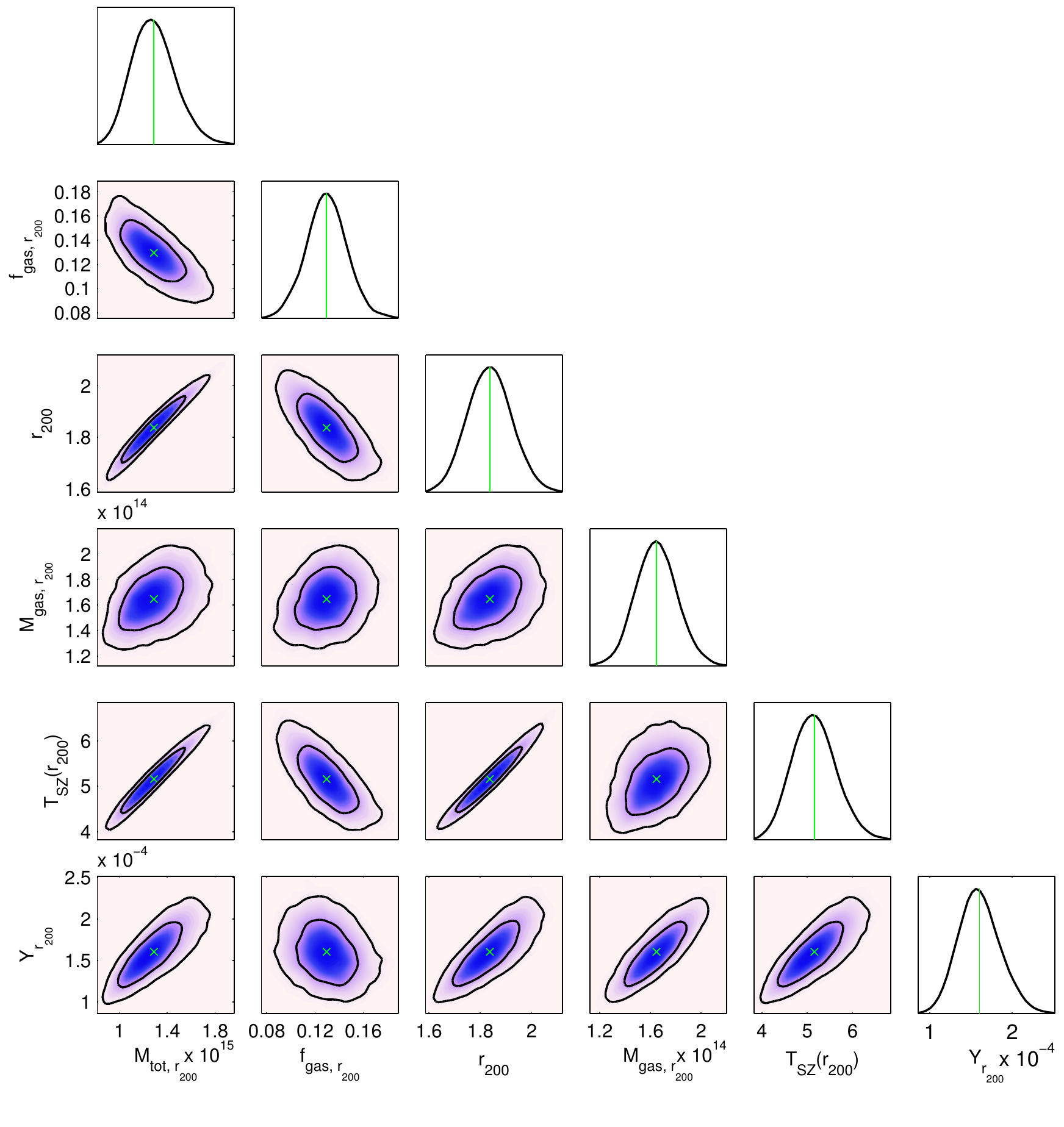}
\includegraphics[trim= 0mm 1mm 0mm -8mm, clip,width=88mm]{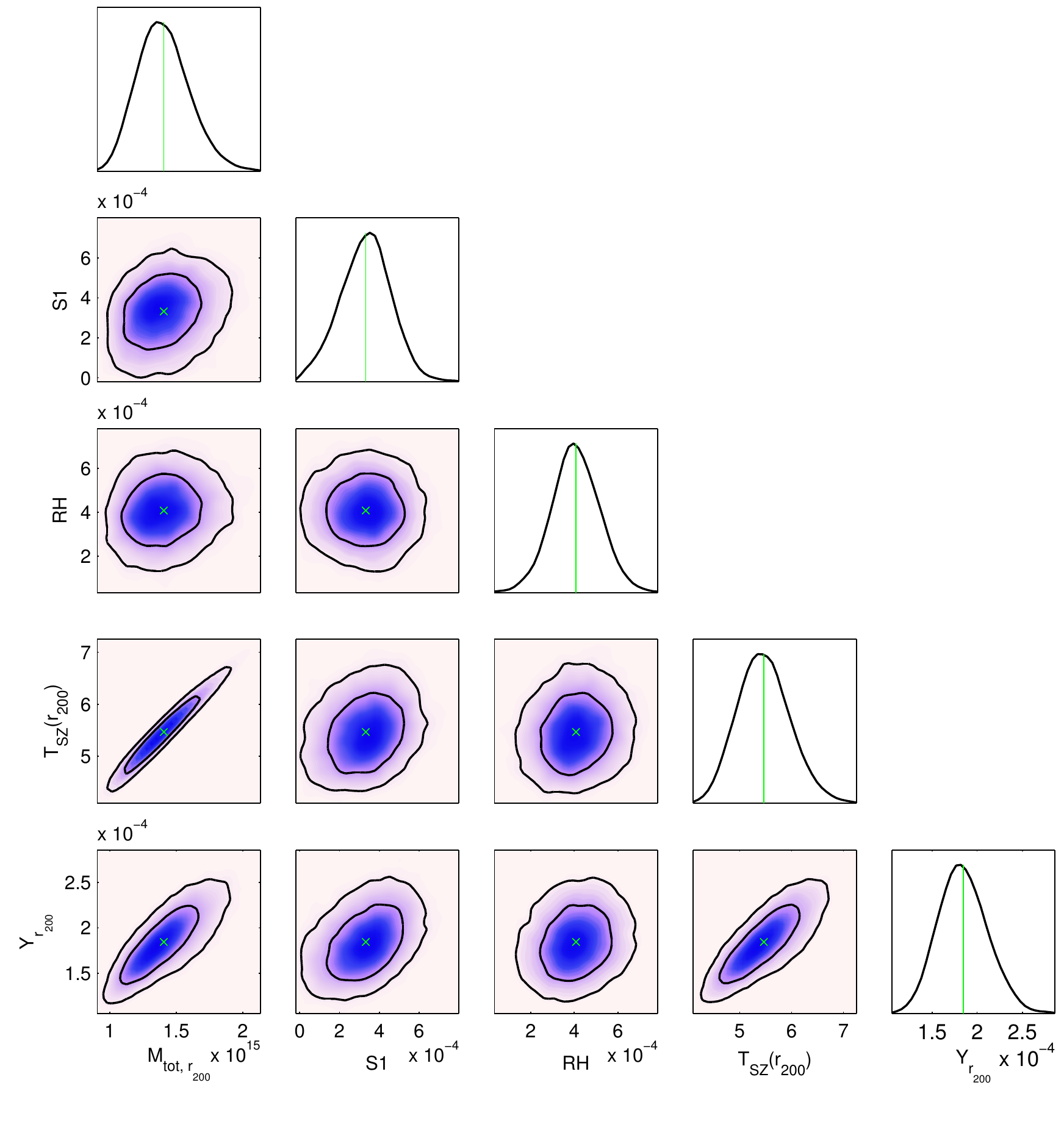}
\caption{\textbf{ MAJ0717+3745}. Top two panels: SA contour maps -- left shows the non-source-subtracted map ($\sigma_{\rm SA} = 84\,\mu$Jy), right shows the source subtracted map ($\sigma_{\rm SA} = 82\,\mu$Jy, {\color{black}18$\sigma_{\rm SA}$ decrement}). See Figure \ref{fig:A611} caption for more details. Bottom two panels: \textsc{McAdam} fitted probability distributions -- left shows parameter probability distributions, right shows degeneracies between the fitted cluster parameter values and the flux of source $\rm S1$ and possible radio halo remnant $\rm RH$, labelled on the map in the top right panel. The green lines and crosses show the mean and the contour levels represent 68 per cent and 95 per cent confidence limits. \label{fig:MAJ0717}}
\end{figure*}

\begin{figure*}
\includegraphics[trim= 0mm 16.7mm 0mm 12mm, clip,width=88mm]{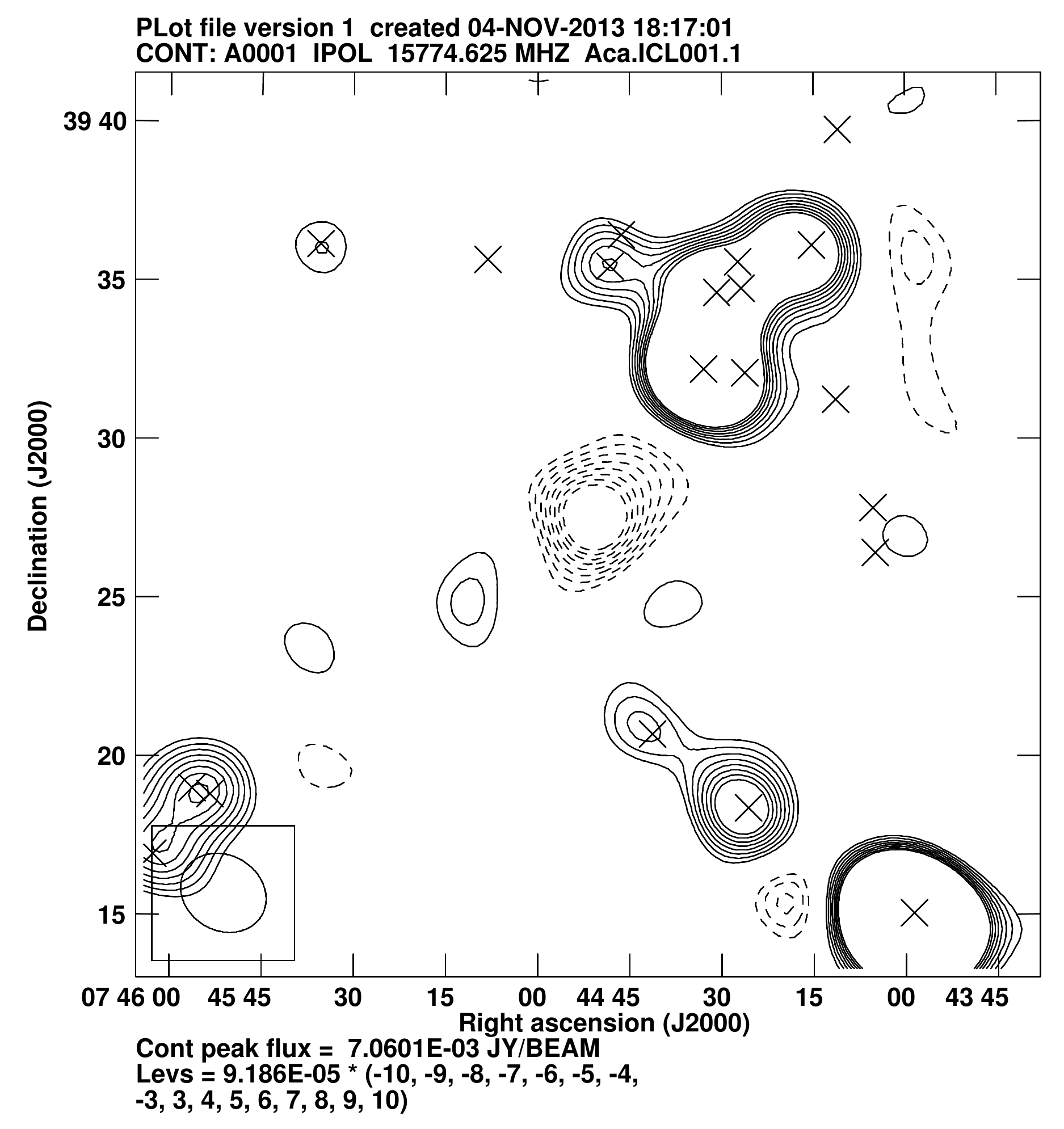}
\includegraphics[trim= 0mm 16.7mm 0mm 12mm, clip,width=88mm]{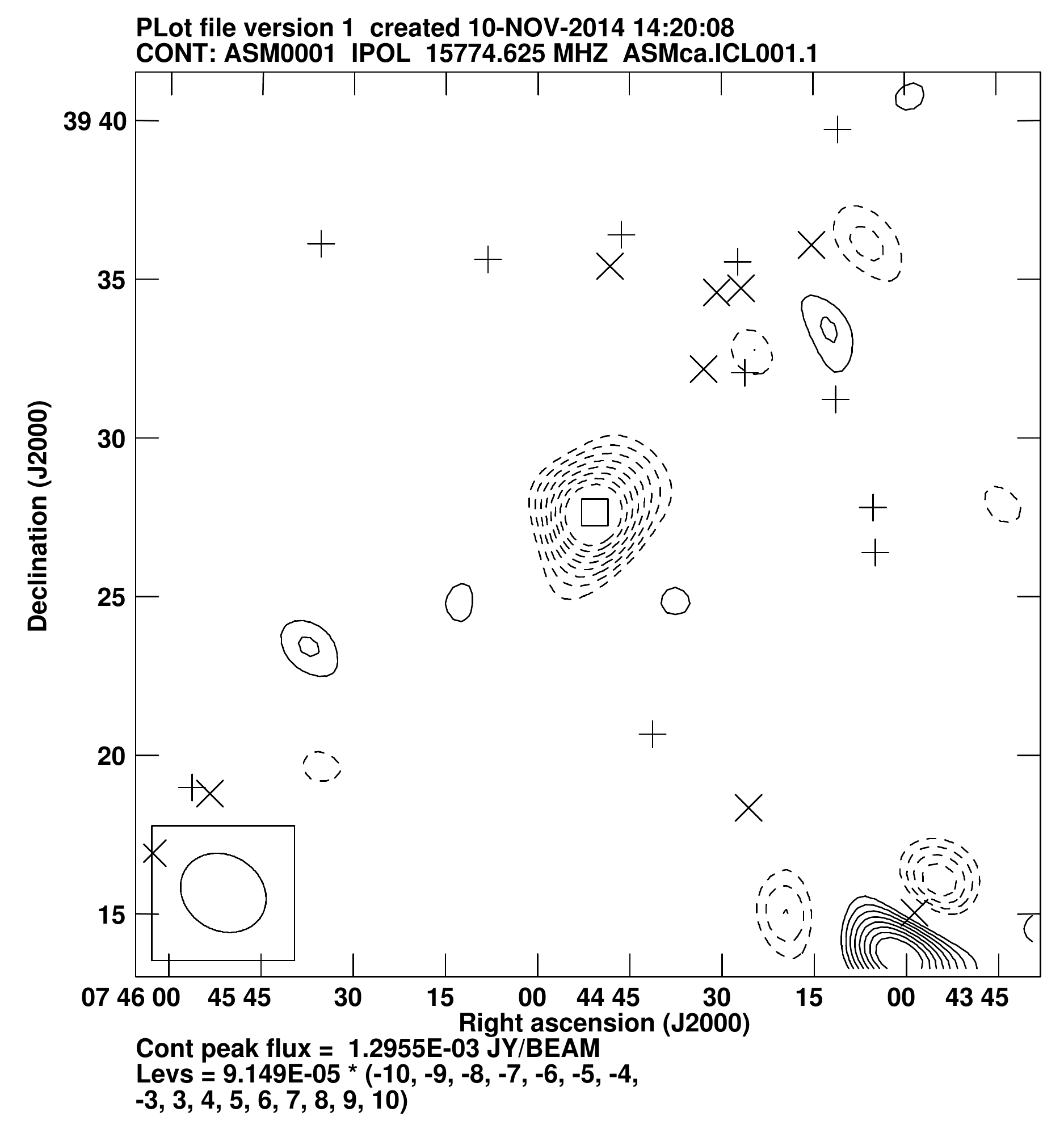}
\includegraphics[trim= 0mm 1mm 0mm -8mm, clip,width=100mm]{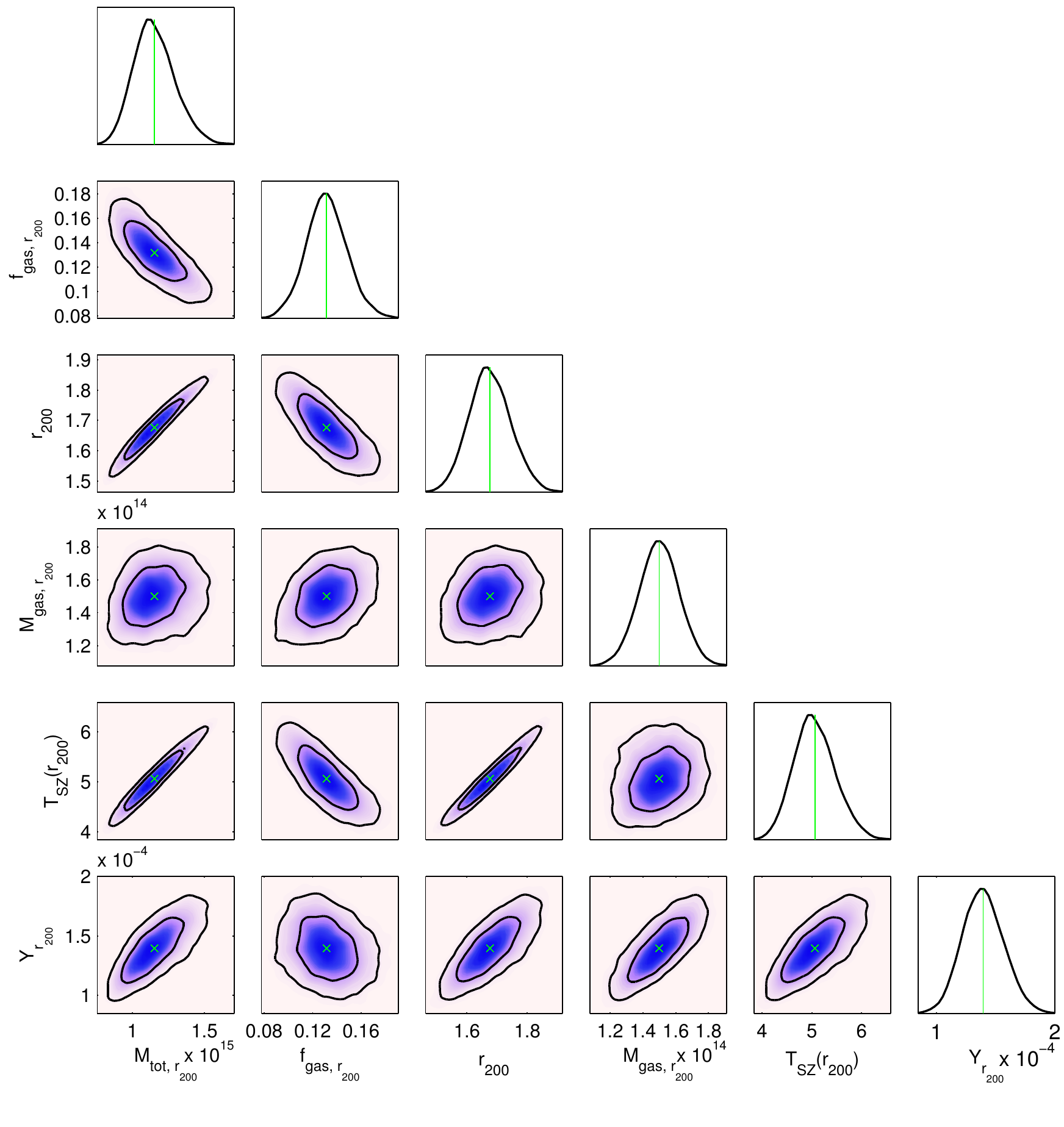}
\caption{\textbf{ MAJ0744+3927}. Top two panels: SA contour maps -- left shows the non-source-subtracted map ($\sigma_{\rm SA} = 92\,\mu$Jy), right shows the source subtracted map ($\sigma_{\rm SA} = 91\,\mu$Jy, {\color{black}13$\sigma_{\rm SA}$ decrement}). See Figure \ref{fig:A611} caption for more details. Bottom panel: \textsc{McAdam} fitted parameter probability distributions, the green lines and crosses show the mean and the contour levels represent 68 per cent and 95 per cent confidence limits. \label{fig:MAJ0744}}
\end{figure*}

\begin{figure*}
\includegraphics[trim= 0mm 16.7mm 0mm 12mm, clip,width=88mm]{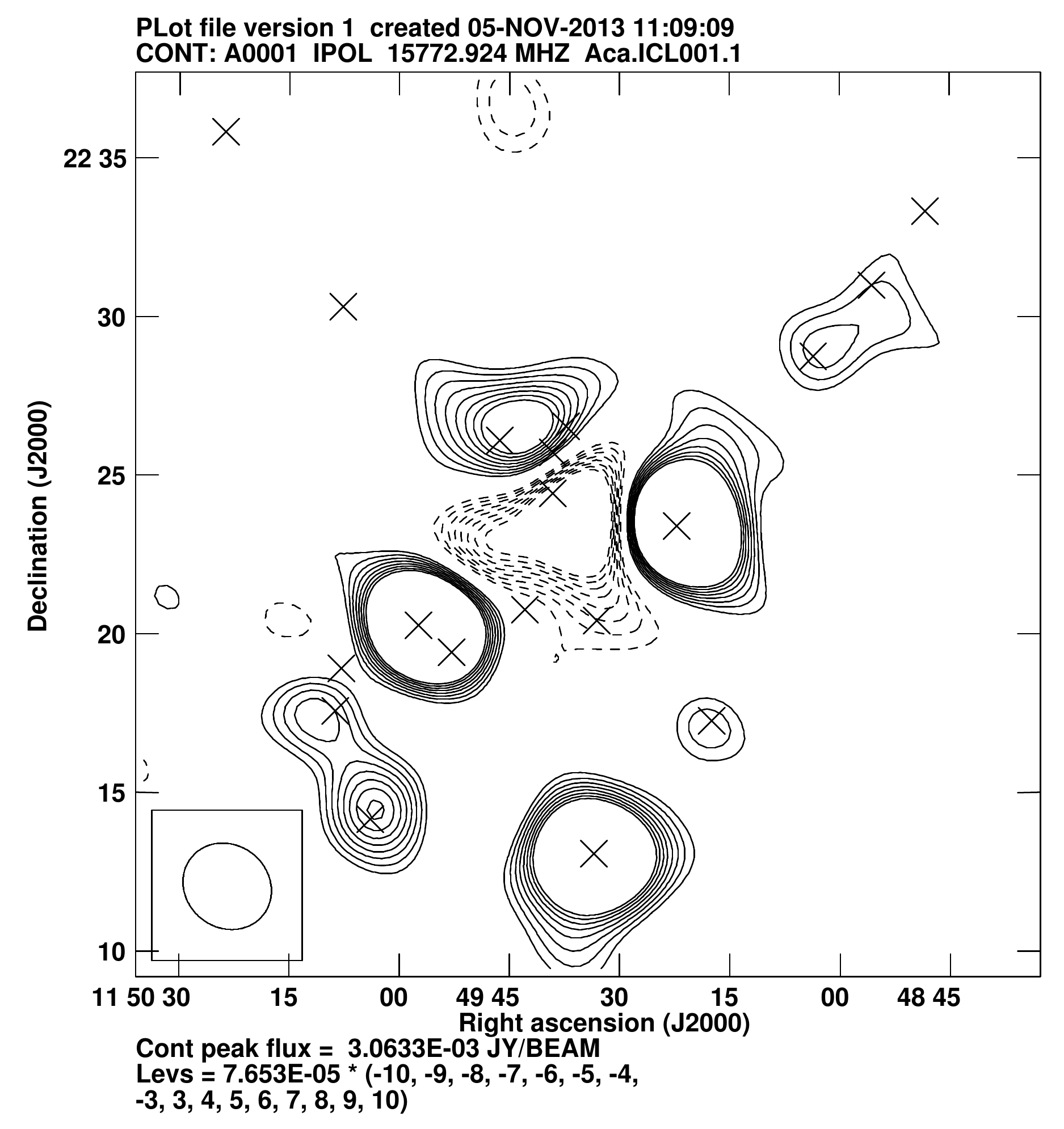}
\includegraphics[trim= 0mm 16.7mm 0mm 12mm, clip,width=88mm]{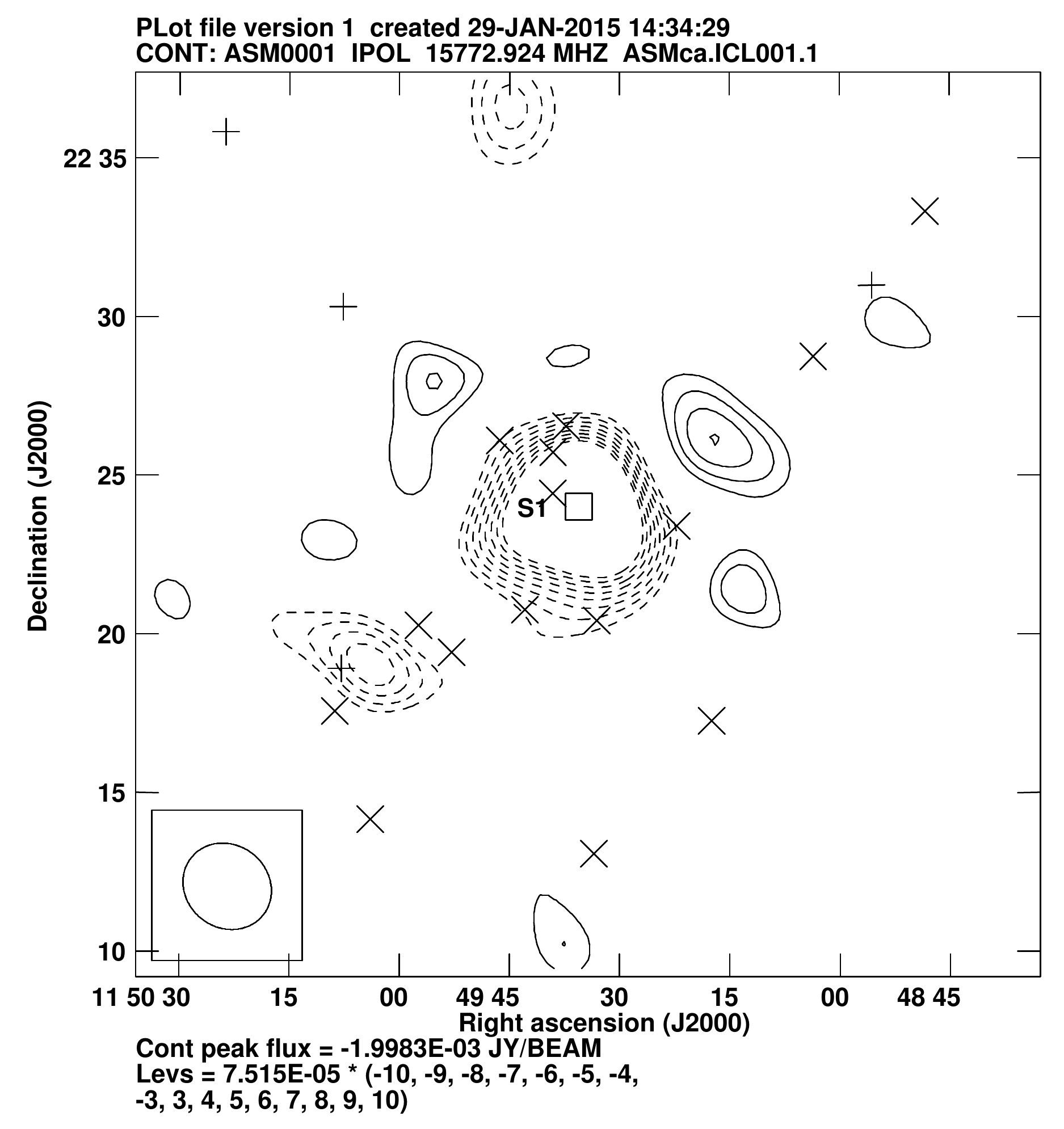}
\includegraphics[trim= 0mm 1mm 0mm -8mm, clip,width=88mm]{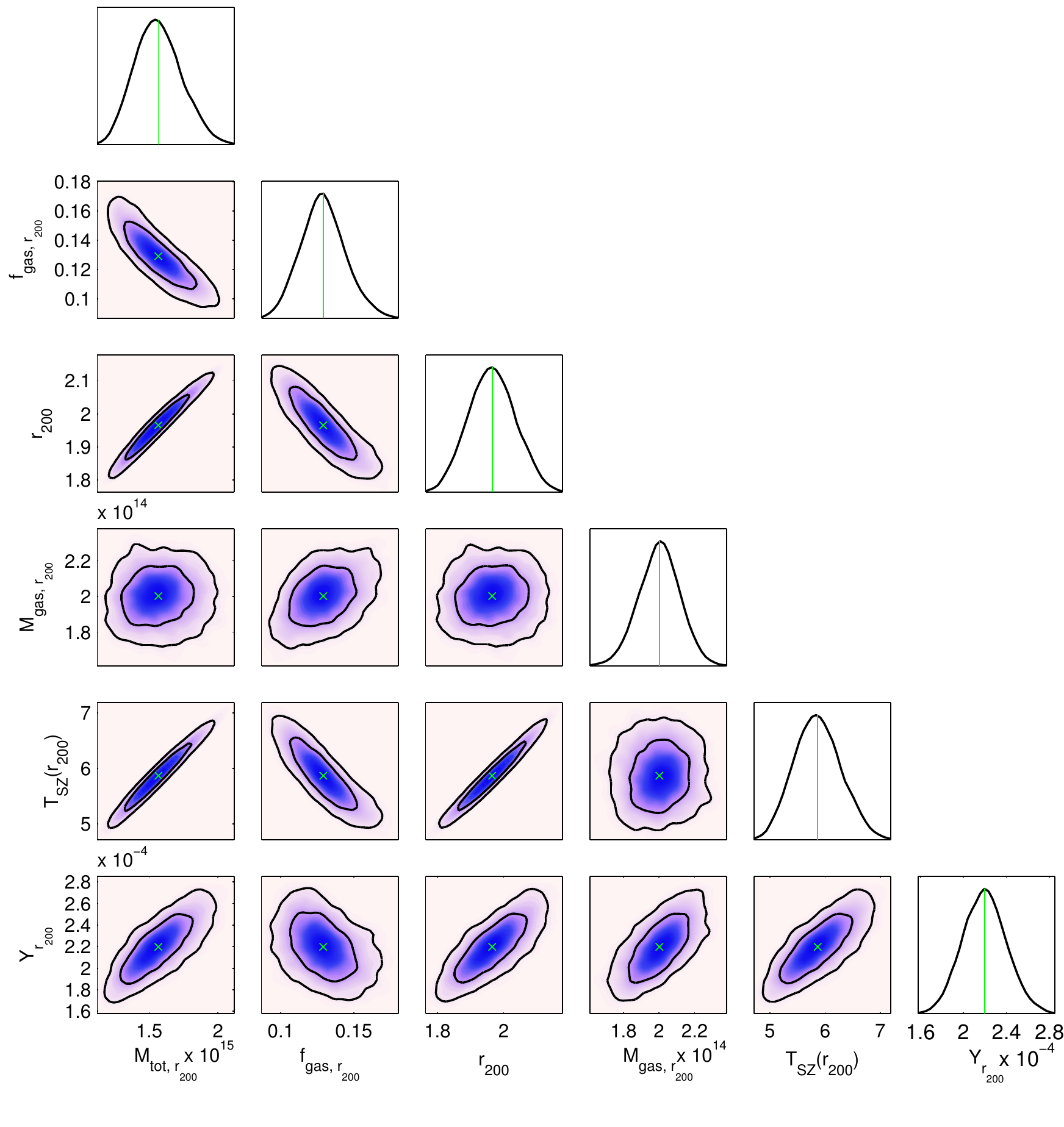}
\includegraphics[trim= 0mm 1mm 0mm -8mm, clip,width=88mm]{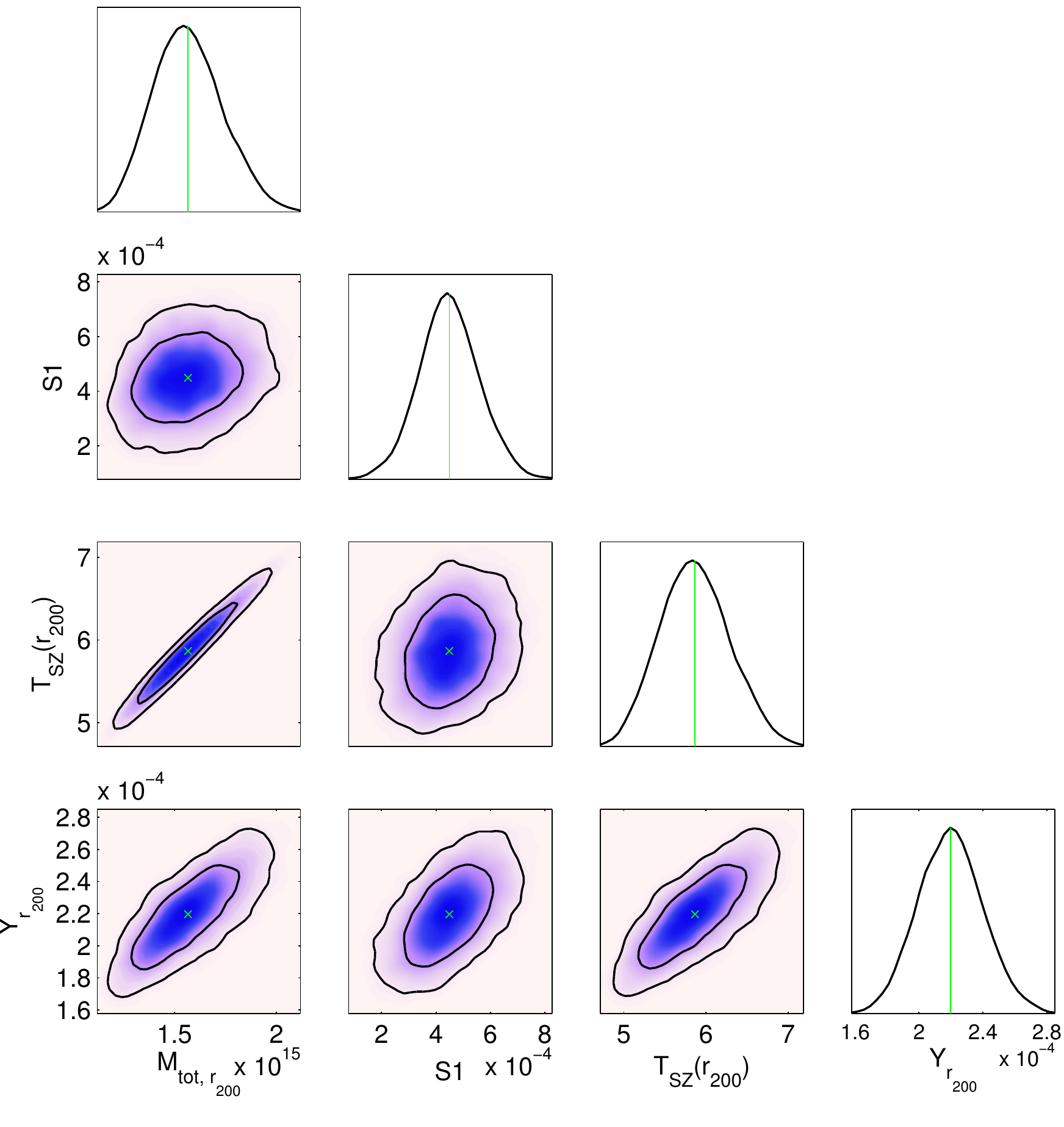}
\caption{\textbf{ MAJ1149+2223}. Top two panels: SA contour maps -- left shows the non-source-subtracted map ($\sigma_{\rm SA} = 77\,\mu$Jy), right shows the source subtracted map ($\sigma_{\rm SA} = 75\,\mu$Jy, {\color{black}27$\sigma_{\rm SA}$ decrement}). See Figure \ref{fig:A611} caption for more details. Bottom two panels: \textsc{McAdam} fitted probability distributions -- left shows parameter probability distributions, right shows degeneracies between the fitted cluster parameter values and the flux of source $\rm S1$, labelled on the map in the top right panel. The green lines and crosses show the mean and the contour levels represent 68 per cent and 95 per cent confidence limits. \label{fig:MAJ1149}}
\end{figure*}

\begin{figure*}
\includegraphics[trim= 0mm 16.7mm 0mm 12mm, clip,width=88mm]{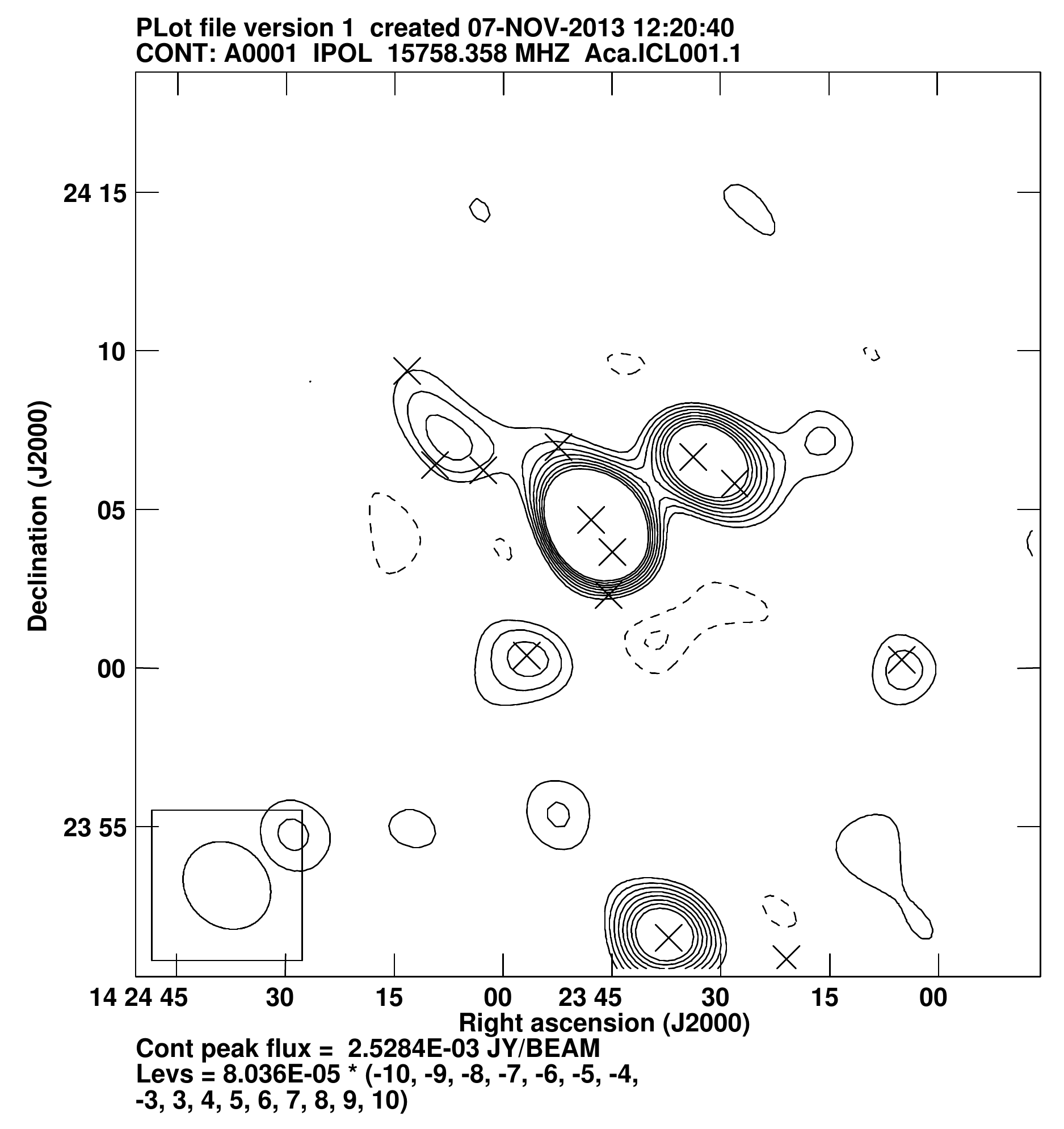}
\includegraphics[trim= 0mm 16.7mm 0mm 12mm, clip,width=88mm]{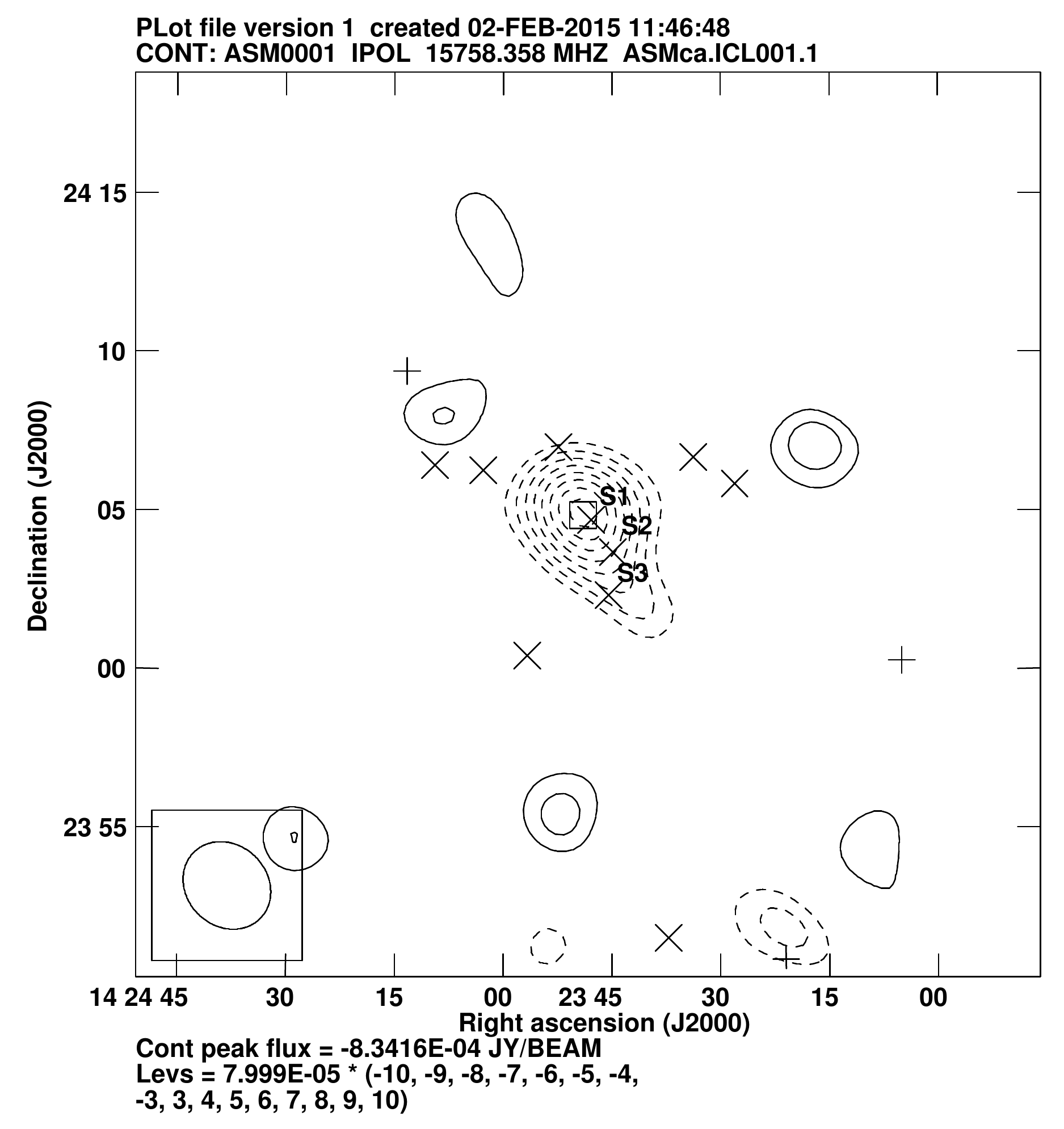}
\includegraphics[trim= 0mm 3mm 0mm -8mm, clip,width=88mm]{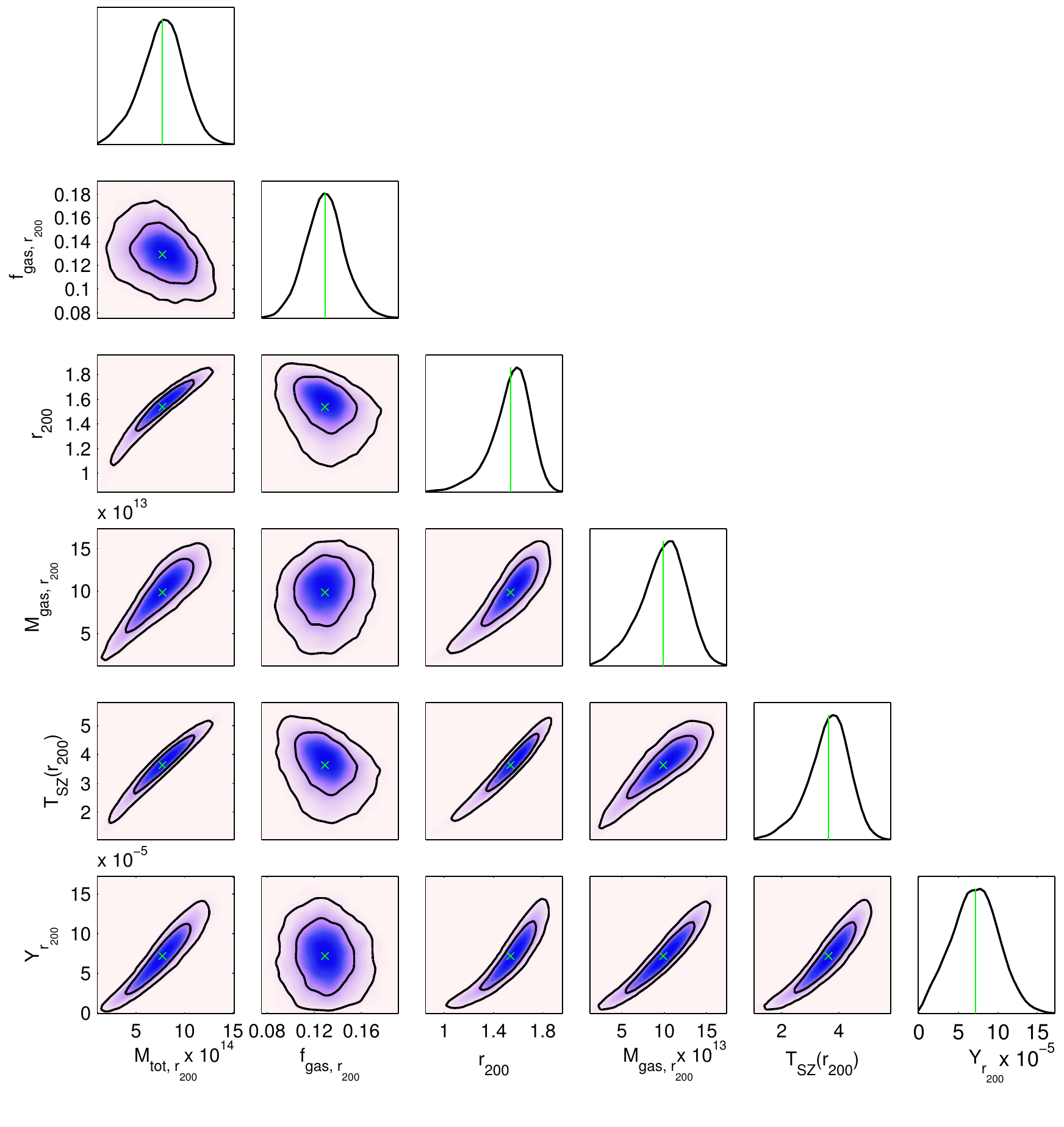}
\includegraphics[trim= 0mm 3mm 0mm -8mm, clip,width=88mm]{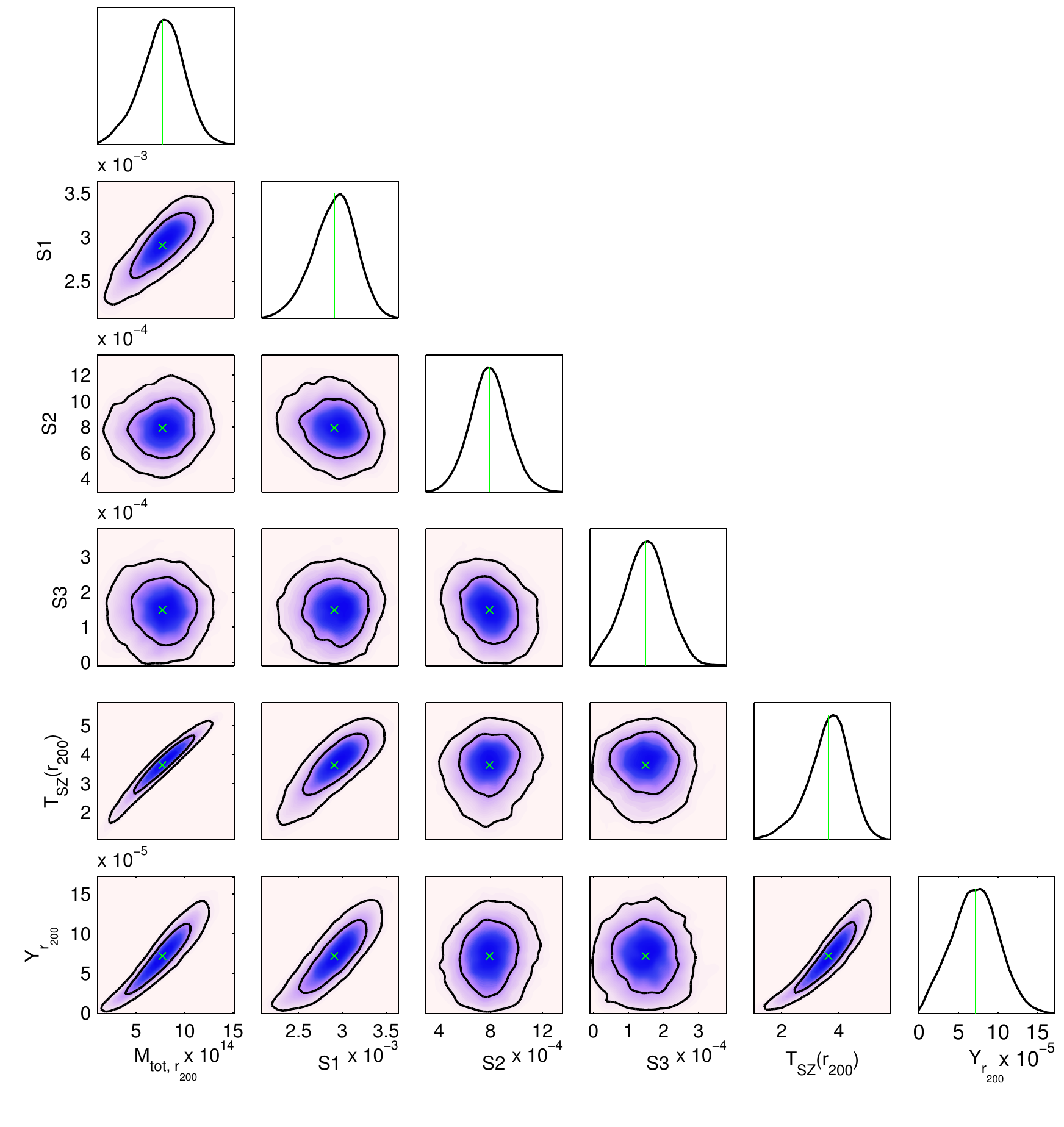}
\caption{\textbf{ MAJ1423+2404}. Top two panels: SA contour maps -- left shows the non-source-subtracted map ($\sigma_{\rm SA} = 80\,\mu$Jy), right shows the source subtracted map ($\sigma_{\rm SA} = 80\,\mu$Jy, {\color{black}10$\sigma_{\rm SA}$ decrement}). See Figure \ref{fig:A611} caption for more details. Bottom two panels: \textsc{McAdam} fitted probability distributions -- left shows parameter probability distributions, right shows degeneracies between the fitted cluster parameter values and the flux of sources $\rm S1$, $\rm S2$ and $\rm S3$, labelled on the map in the top right panel. The green lines and crosses show the mean and the contour levels represent 68 per cent and 95 per cent confidence limits. \label{fig:MAJ1423}}
\end{figure*}

\begin{figure*}
\includegraphics[trim= 0mm 16.7mm 0mm 12mm, clip,width=88mm]{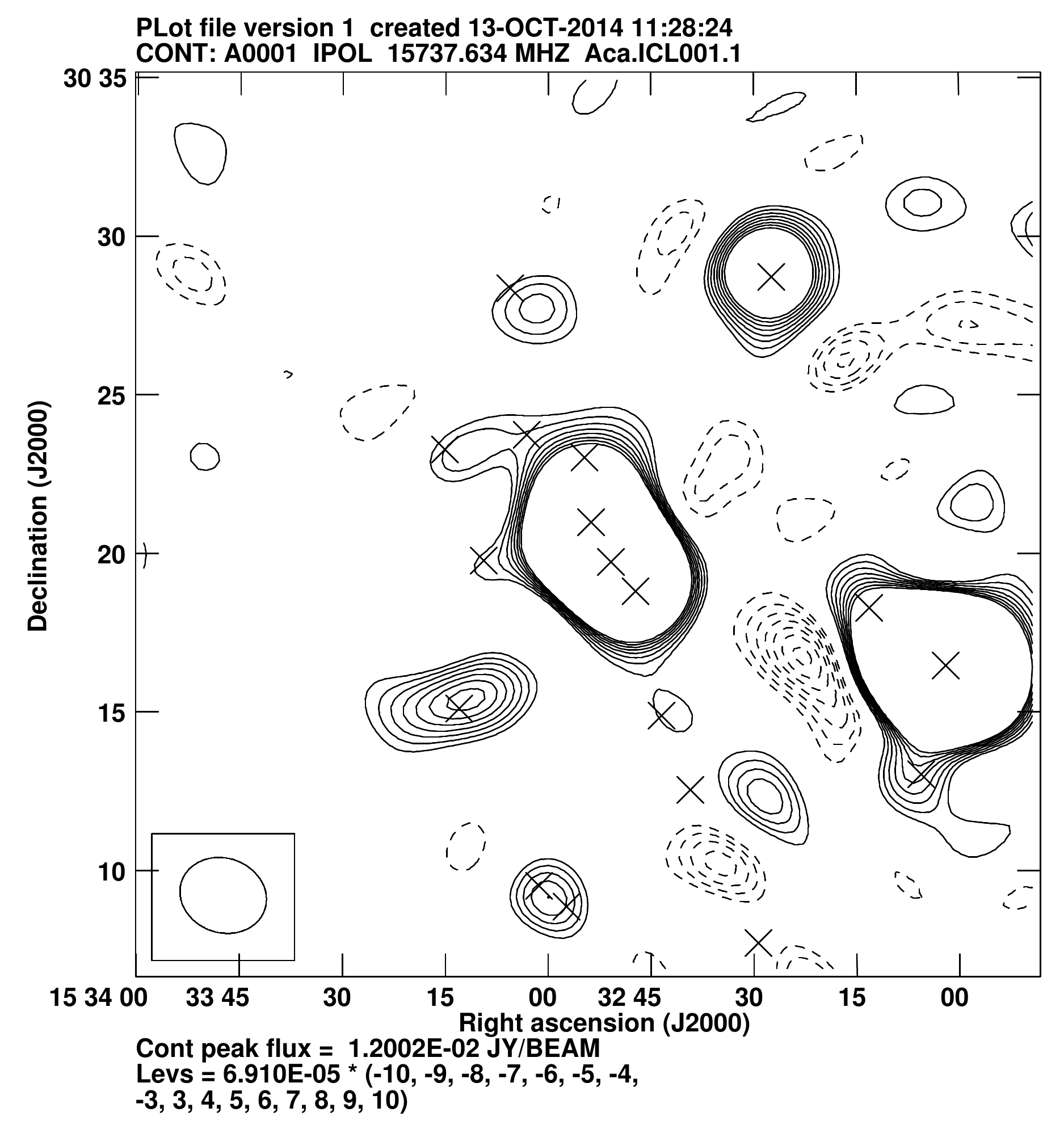}
\includegraphics[trim= 0mm 16.7mm 0mm 12mm, clip,width=88mm]{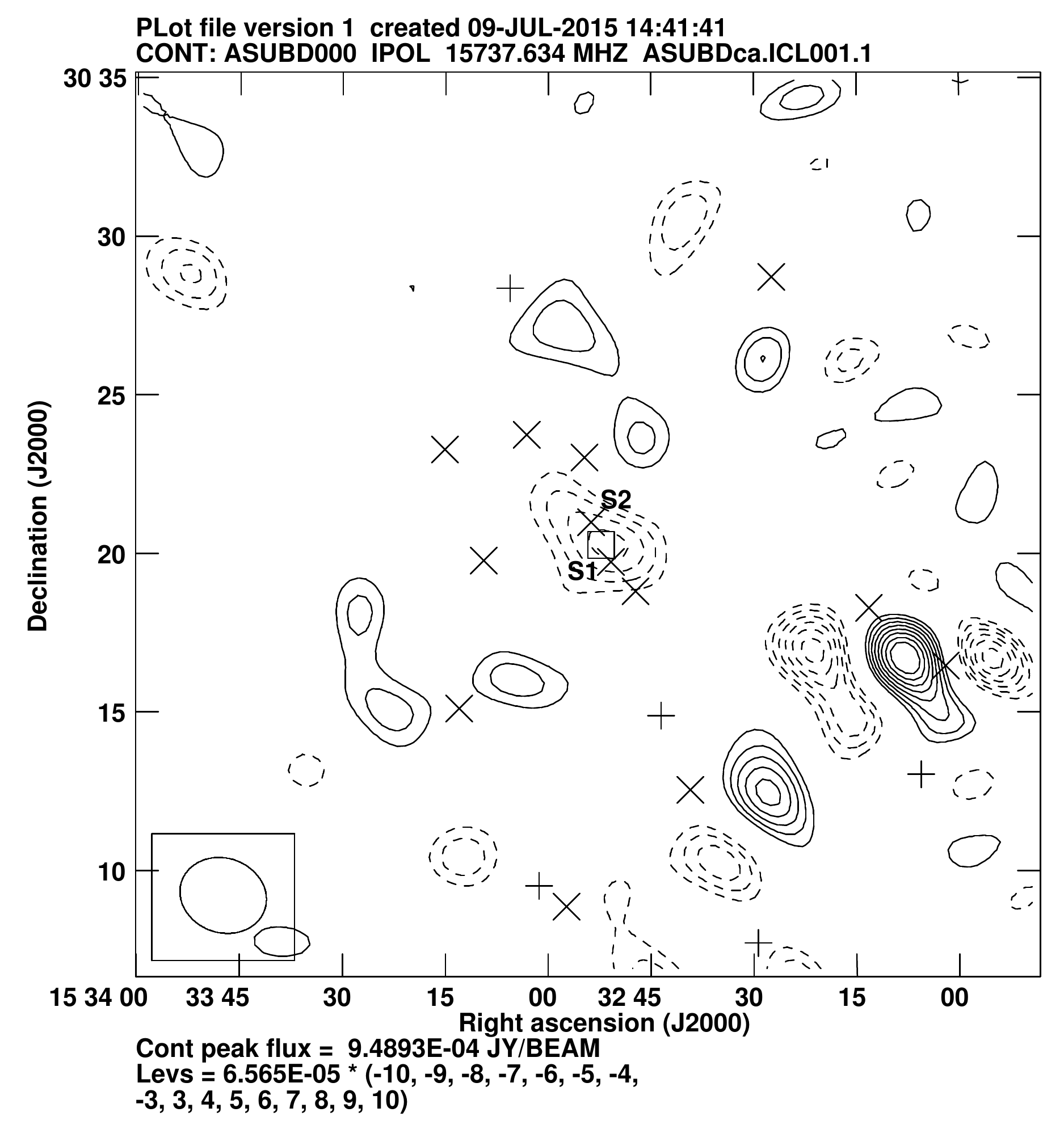}
\includegraphics[trim= 0mm 3mm 0mm -8mm, clip,width=88mm]{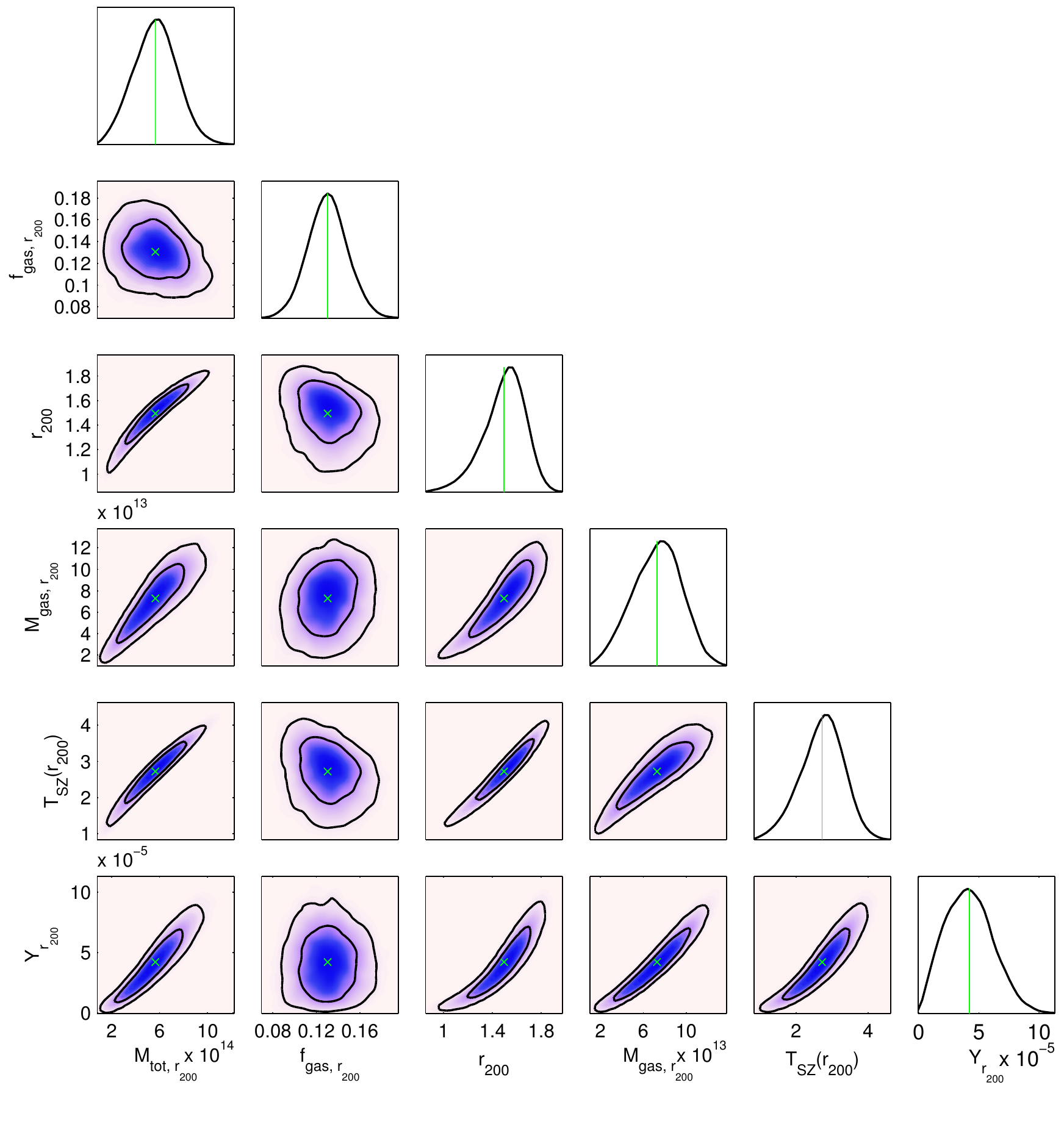}
\includegraphics[trim= 0mm 3mm 0mm -8mm, clip,width=88mm]{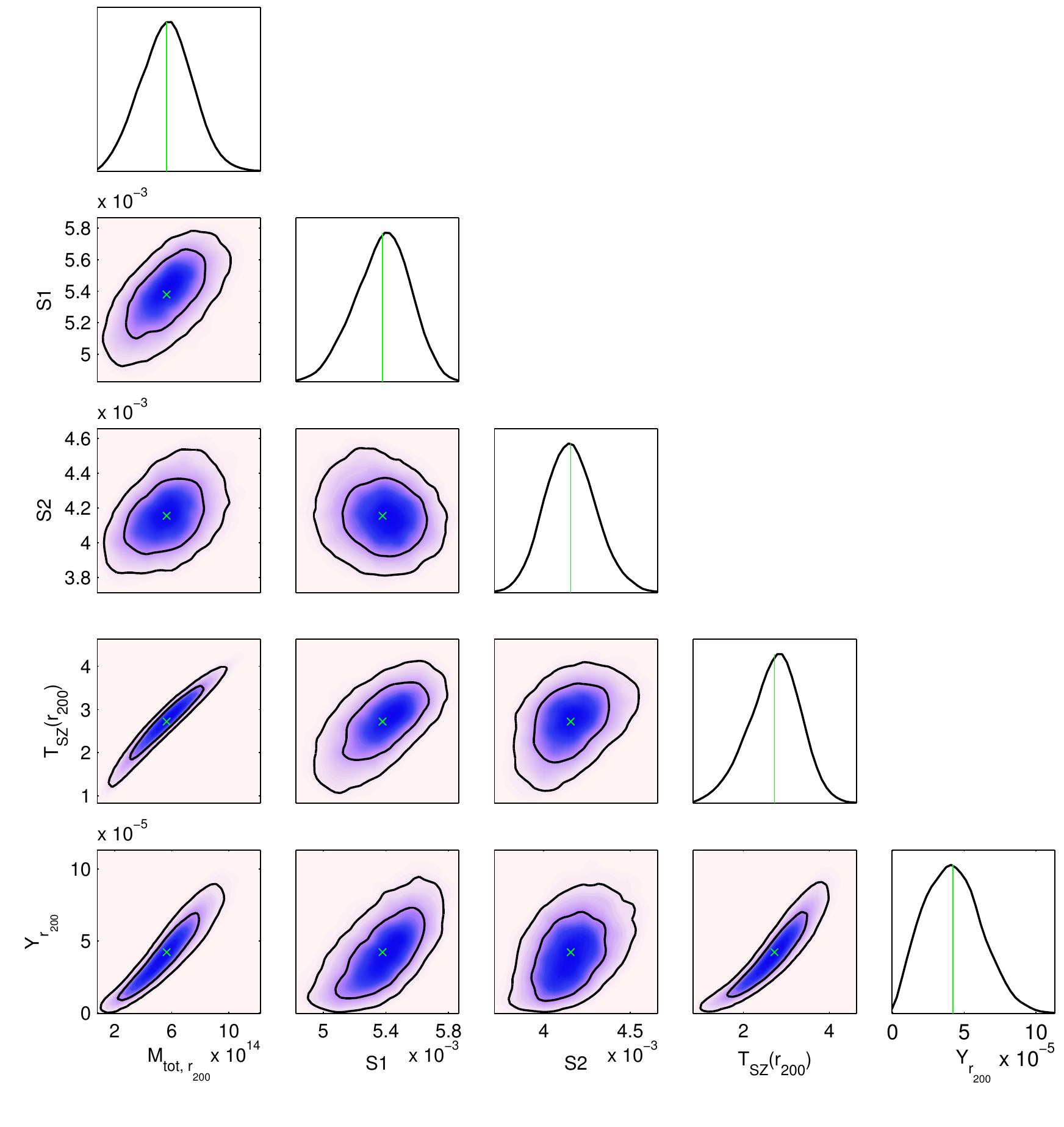}
\caption{\textbf{ RXJ1532+3021}. Top two panels: SA contour maps -- left shows the non-source-subtracted map ($\sigma_{\rm SA} = 69\,\mu$Jy), right shows the source subtracted map ($\sigma_{\rm SA} = 66\,\mu$Jy, {\color{black}6$\sigma_{\rm SA}$ decrement}). See Figure \ref{fig:A611} caption for more details. Bottom two panels: \textsc{McAdam} fitted probability distributions -- left shows parameter probability distributions, right shows degeneracies between the fitted cluster parameter values and the flux of sources $\rm S1$ and $\rm S2$, labelled on the map in the top right panel. The green lines and crosses show the mean and the contour levels represent 68 per cent and 95 per cent confidence limits. \label{fig:RXJ1532}}
\end{figure*}

In this section we present maps and probability distributions, and discuss each cluster individually. Tables \ref{tab:r200} and \ref{tab:r500} summarise some key \textsc{McAdam}-estimated parameter values. Table \ref{tab:r200} also includes cluster spectroscopic redshifts.
The naturally-weighted SA maps shown in Figures \ref{fig:A611} to \ref{fig:RXJ1532} are non-source-subtracted (the left of each Figure) and source-subtracted (the right of each Figure).
Contour levels correspond to $\pm$3$\sigma_{\rm SA}$ to $\pm$10$\sigma_{\rm SA}$, where solid contours indicate positive emission and dashed contours indicate decrement. The half-power contour of the synthesized beam for each map is displayed in the bottom left-hand corner. No primary beam correction or $uv$-taper has been applied.
In the source-subtracted maps the sources that are given Gaussian priors on flux density and spectral index are indicated by `$\times$' and those modeled with delta priors by `$+$'. The SZ fitted cluster centre position is marked with a $\square$.

On the parameter probability distributions, in Figures \ref{fig:A611} to \ref{fig:RXJ1532}, the green lines and crosses show the mean and the contour levels represent 68 per cent and 95 per cent confidence limits. The 2D marginalised posterior distributions show the degeneracies between parameter values and the correlations between uncertainties. We emphasise that derived parameters and their uncertainties are correlated -- this is important in Sections \ref{sc:scalingrel} and \ref{sec:xray}.
Where necessary we consider possible degeneracies between the fitted cluster parameter values and the fluxes of sources very close to the cluster centre; such degeneracy depends on the source position relative to the cluster centre and on the brightness of the source.

{\color{black}As stated in Section \ref{sec:priors}, SZ data provide little constraint on the value of $f_{{\rm gas},\,r_{200}}$.
This is demonstrated in the bottom right panel of Figure \ref{fig:A611} where we plot the 1D marginalised $f_{{\rm gas},\,r_{200}}$ posterior overlaid with its prior. For comparison, we also plot the 1D marginalised $M_{{\rm tot},\,r_{200}}$ posterior overlaid with its prior.}



\subsection{A611}
\label{sec:A611}
The SZ decrement is clearly visible in both un-subtracted and source-subtracted SA maps .
Source subtraction produces a map with very few residuals, and with no sources appearing at the cluster centre we expect the source environment has negligible effect on fitted cluster parameters.

\subsection{A1423}
\label{sec:A1423}
The complicated source environment has been subtracted successfully -- there are no sources $>$\,2\,mJy close to the cluster SZ decrement, which is visible in the source subtracted map at a high significance.
Residuals in the source-subtracted map suggest some extended emission in the field that is not visible to the higher resolution of the LA, but is far enough from the cluster centre to have little or no effect on the estimated parameter values. There is some degeneracy between the flux of the source closest to the cluster centre and the estimated parameter values; although the source flux is low, $\approx$\,0.5\,mJy, and the cluster is resolved, the analysis struggles to {\color{black}separate the relative contributions of} a source and a cluster of small angular size.

\subsection{A2261}
\label{sec:A2261}
This very massive cluster has a particularly complicated source environment with many sources up to 20\,mJy. A particularly deep, more-uniform noise level over the whole LA map was needed to characterise the sources sufficiently accurately in order to describe the source environment to \textsc{McAdam} without confusion due to side lobes.
In our initial analysis, the source environment was inaccurately described due to exclusion zones automatically placed by \textsc{Source-Find} around bright sources, within which other sources were not modelled. With the exclusion zones removed, the source environment was accurately modelled and the cluster analysed again, with the updated source information. The large number of sources -- and consequent high dimensionality -- necessitated the use of the next generation sampler \textsc{PolyChord} (\citealt{2015MNRAS.450L..61H} and \citealt{2015arXiv150600171H}) in the place of \textsc{MultiNest}.
There is very little difference between the parameter values estimated by the \textsc{MultiNest} and \textsc{PolyChord} analyses, showing that our analysis is robust against the presence of bright sources at $\approx$\,7\,arcmin from the cluster centre. Closer to the cluster centre there are other sources; for the four closest we find negligible degeneracies between the source fluxes and the SZ parameter values. This is particularly noteworthy for source $S_1$ which has a flux of nearly 17\,mJy but, as a point source, is easily distinguished in the analysis from the very extended cluster. Although there are significant residuals in the source-subtracted map they are far enough from the very large SZ decrememt that our parameter estimates are negligibly affected.

\subsection{CLJ1226+3332}
\label{sec:CLJ1226}
This is the highest redshift cluster in the CLASH sample at $z$\,=\,0.888 and also one of the least massive. Although the source environment is largely sparse, there is a low-significance source directly on the cluster X-ray centre. This is a particular concern for CLJ1226+3332 due to its high redshift and, therefore, small angular size: there is a larger degeneracy than is typical in the AMI-CLASH sub-sample, for a source of such low flux density ($\approx$\,0.18\,mJy) close to the cluster centre. Useful parameter constraints are still obtained, as is illustrated by Figure \ref{fig:CLJ1226}, bottom right, where the marginalised posteriors take full account of the degeneracies.

\subsection{MAJ0647+7015 -- strong-lensing selected}
\label{sec:MAJ0647}
The SZ decrement is clearly visible in both unsubtracted and subtracted SA maps. Our unsubtracted AMI maps show a source environment with no sources near the cluster centre and only sources of $<$\,4\,mJy towards the edge of the cluster decrement. The single bright source in the field is 21\,mJy but is $\approx$\,12\,arcmin from the cluster centre so has negligible effect on parameter estimation.

\subsection{MAJ0717+3745 -- strong-lensing selected}
\label{sec:MAJ0717}
This source environment requires special consideration. In the LA map we see a set of $>$\,5\,mJy point sources near the X-ray centre of the cluster.  Amongst these sources there is a slightly extended feature that the map-plane \textsc{Source-Find} cannot detect.
This cluster is known to host a radio halo (see e.g. \citealt{2011MNRAS.418.2754Z}, \citealt{2012A&ARv..20...54F}) and comparing this AMI feature to the 1.4-GHz WSRT map from \citet{2009AA...505..991V} we consider the possibility that this feature is the radio halo at AMI frequencies.
We have investigated how sensitive our parameter estimation is to the flux density of this feature by running the analysis twice with and without a source of 0.34\,mJy (measured from the LA map) at the peak position of the feature (labeled RH on the subtracted SA map).
The additional flux raises estimates of Y-values, masses and temperatures by less than 1$\sigma$ and the map decrement by less than 2$\sigma$. There is no degeneracy between parameter values and the additional flux and none associated with the sources closest to the cluster centre. {\color{black}The additional analysis is referred to as MAJ0717RH in the rest of this paper}.

\subsection{MAJ0744+3927}
\label{sec:MAJ0744}
We estimate MAJ0744+3927 to be among the more massive members of the sub-sample, and it has the second highest redshift in the complete CLASH sample. The source environment is not troublesome, with no sources detected near the cluster centre. The sources lying away from the cluster decrement are subtracted to leave very few residuals.

\subsection{MAJ1149+2223 -- strong-lensing selected}
\label{sec:MAJ1149}
Parameter estimation reveals this to be the most massive cluster in our AMI-CLASH sub-sample, with a large angular extent, despite its redshift. The radio source environment is fitted and subtracted well, leaving only a few residuals in the map which do not significantly affect the parameter estimation. We have checked the source closest to the cluster centre for degeneracy with the estimated cluster parameter values and find very little, as expected given the low flux density of the source and the large angular size of the cluster.

\subsection{MAJ1423+2404}
\label{sec:MAJ1423}
There are three detected sources lying in the direction of the cluster decrement. The flux densities estimated by \textsc{McAdam} are 2.9\,mJy, 0.8\,mJy, and 0.15\,mJy respectively. The only source close to the cluster centre that shows some degeneracy between the cluster parameter values and its flux density is $S_1$. The degeneracy is, as usual, taken into account in arriving at the marginalised posteriors.

\subsection{RXJ1532+3021}
\label{sec:RXJ1532}
There is a radio source at, and several close to, the map centre, all of which are modelled in \textsc{McAdam}. The source subtraction appears to have been successful with significant residuals only towards the edges of the SA map. We plot the degeneracy of the cluster parameter values with the flux estimates of the two sources closest to the cluster centre. Both sources show some degeneracy, the closer showing a significant amount.
As with the other resolved clusters showing degeneracy with central source fluxes, parameter estimates are still useful as the degeneracy is modelled in the analysis.


\section{Self-similar scaling relations}
\label{sc:scalingrel}

Assuming self-similarity, the theoretical scaling relations between cluster parameter values should describe the observational relationships, with the scatter dependent only on measurement uncertainties. The predicted $M$--$T$ relation is
\begin{equation}\label{eq:MT}
 M \propto T^{3/2},
\end{equation}
which arises from the potential $GM/R$ being $\propto T$ if all kinetic energy is in gas internal energy, and from $R \propto M^{1/3}$, where $M$ is the total mass within radius $R$. Similarly, the expected $Y$--$M$ relation is
\begin{equation}\label{eq:MT}
 Y \propto M^{5/3}.
\end{equation}
Studies of these scaling relations have been carried out using simulated and real X-ray and SZ data (see e.g. \citealt{2006ApJ...650..128K}, \citealt{2009ApJ...692.1033V}, \citealt{2009ApJ...692.1060V}, \citealt{2011ApJ...738...48A}).
\citet{2007MNRAS.380..437P} carry out simulations of two-body cluster mergers over a range of mass ratios and impact parameters (see Section \ref{sec:mergsim}) to investigate the effect of mergers on the scatter in scaling relations by tracking the merger evolution in scaling-relation planes. They find large changes in cluster observables predominantly \textit{along} the scaling-relation, with usually significantly less increase in displacement perpendicular to the direction of the relation.

Scaling relations are implicitly present in the model due, e.g., to the assumptions of HSE and pressure profile shape. The relationships between the sampling parameters (see Table \ref{table_priors}) and any other derived parameters (e.g. $Y$, $T$) are fixed by the model and can be described by power-laws in a similar way to typical scaling relations. We cannot therefore use the relationships between these sampling and derived parameters to investigate the dynamical state of a cluster.
In Section \ref{sec:MTYM} we investigate the correlations in our parameter estimation, {\color{black}focusing on} the scaling relation $Y$--$M$ {\color{black}which is used in further discussions in Section \ref{sec:xray}}.
\begin{figure}
\includegraphics[trim= -2mm 0mm 0mm 0mm, clip,width=85mm, height=71.5mm]{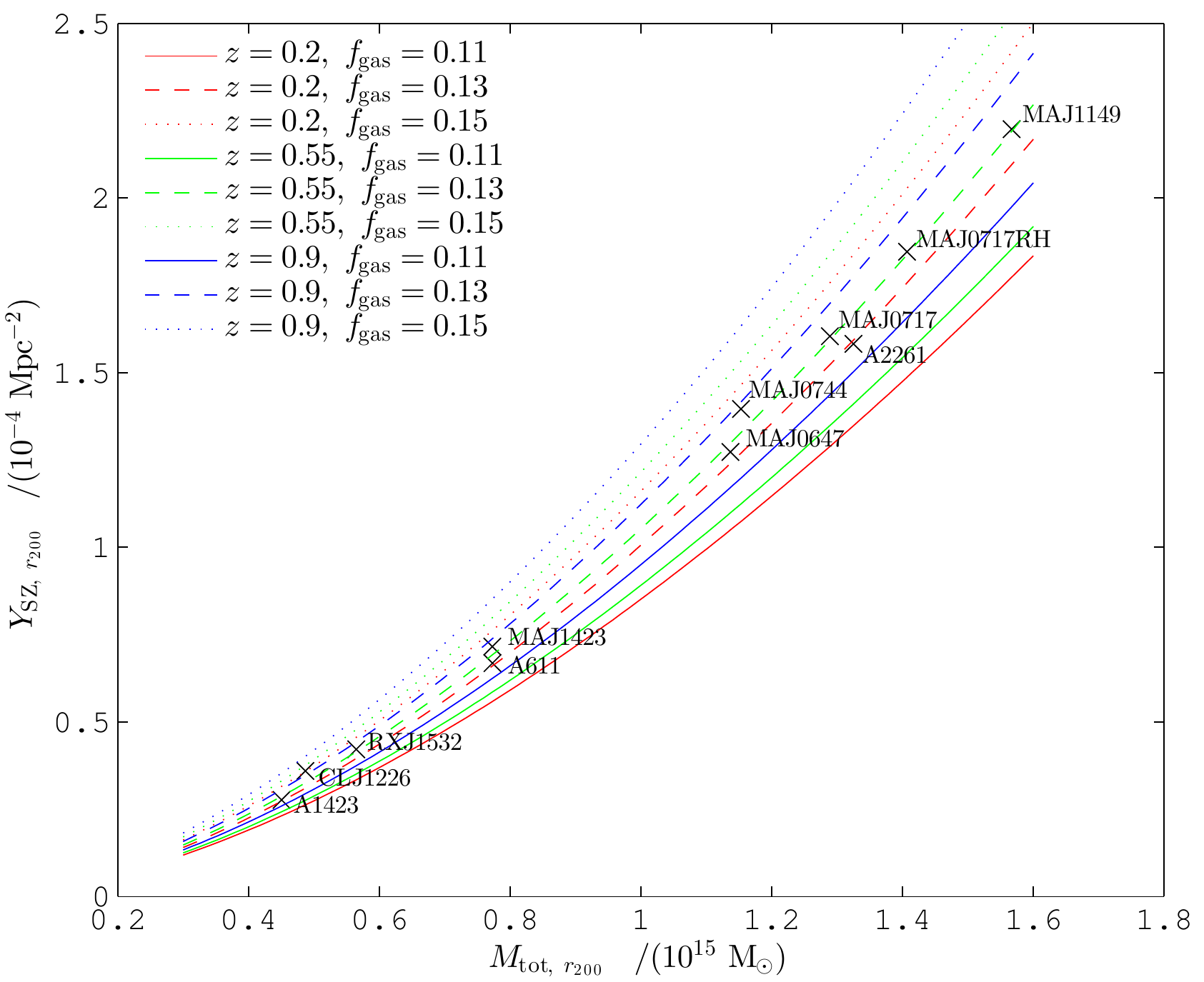}
\caption{Coloured lines showing $Y_{{\rm SZ},\,r_{200}}$ calculated for a range of $M_{{\rm tot},\,r_{200}}$ values and redshifts consistent with our sample ($3 \le M_{{\rm tot},\,r_{200}} < 16 \times 10^{14}$ \Msun\, and $z\,=\,0.2, 0.55$ and $0.9$). $f_{{\rm gas},\,r_{200}}$ varied to its $\pm 1\sigma$ prior values, i.e.\ 0.11, 0.13 and 0.15 -- over-plotted onto the McAdam estimated $Y_{\rm SZ}$ parameter within $r_{200}$ plotted against the estimated total mass within $r_{200}$ for each cluster. \label{fig:MTYMplot}}
\end{figure}

\subsection{$Y$--$M$}\label{sec:MTYM}

Deviations from these correlations occur because of differences in parameters in the analysis that are independent of the cluster shape parameters, such as the redshift and $f_{\rm gas,\,200}$.
Here, we use the model to demonstrate how these correlations would change if the analysis were sensitive to variations in $f_{\rm gas,\,200}$, $z$ and shape parameter $a$, properties that are all given restrictive priors in the analysis (see Section \ref{sec:priors}).

Figure \ref{fig:MTYMplot} shows theoretical predictions from our model (see Section \ref{sec:analysis}) for the $Y$--$M$ scaling relation, investigating the implicit relation; we use the model to calculate $Y_{{\rm SZ},\,r_{200}}$ given a range of $M_{{\rm tot},\,r_{200}}$ values and redshifts consistent with our sample, specifically 3$\times10^{14}$\,\Msun\,$\le\,M_{{\rm tot},\,r_{200}}$\,$<$\,16$\times10^{14}$\,\Msun~and $z$\,=\,0.2, 0.55 and 0.9. $f_{{\rm gas},\,r_{200}}$ is also varied to its $\pm 1\sigma$ prior values, giving 0.11, 0.13 and 0.15. We have found power-law best-fits $Y\,=\,AM^{\kappa}$, estimating $\kappa$ and $A$ with an orthogonal linear regression analysis \citep{1990ApJ...364..104I}, using the Case 3 implementation in the IDL script SIXLIN \footnote{Adapted from the \textsc{FORTRAN} script by \citet{1990ApJ...364..104I}. http://idlastro.gsfc.nasa.gov/ftp/pro/math/sixlin.pro}.
As expected, all fits are in good agreement with the self-similar prediction but there are dependencies of $A$ and $\kappa$ on $z$ and $f_{{\rm gas},\,r_{200}}$ (which is, of course, dependent on mass and redshift). Over-plotted in Figure \ref{fig:MTYMplot} are the AMI, \textsc{McAdam}-derived, values of $Y_{{\rm SZ},\,r_{200}}$ versus $M_{{\rm tot},\,r_{200}}$.
Fitted $\kappa$ values for $Y\,=\,AM^{\kappa}$ are not dependent on the value of $f_{{\rm gas},\,r_{200}}$ and fitted $A$ values vary by $\approx$25 per cent over the range of $f_{{\rm gas},\,r_{200}}$ values. SZ data provide no constraint on $f_{{\rm gas},\,r_{200}}$ so we use priors from X-ray and WMAP estimates of this property. Over the prior range the induced deviations of the fitted curves from self-similarity is small.
$\kappa$ and $A$ values are redshift-dependent as $\rho_{\rm crit}(z)$ (the critical density of the Universe at redshift z) and $c_{200}$ (the halo concentration parameter, which is a function of z \citep{2007MNRAS.381.1450N}), are used to derive cluster parameters from $M_{{\rm tot},\,r_{200}}$. We find $\kappa$ and $A$ values change by $<$0.5 per cent and 10 per cent respectively over the $z$ range used.

\begin{figure}
\includegraphics[trim= -2mm 0mm 0mm 0mm, clip,width=85mm, height=71.5mm]{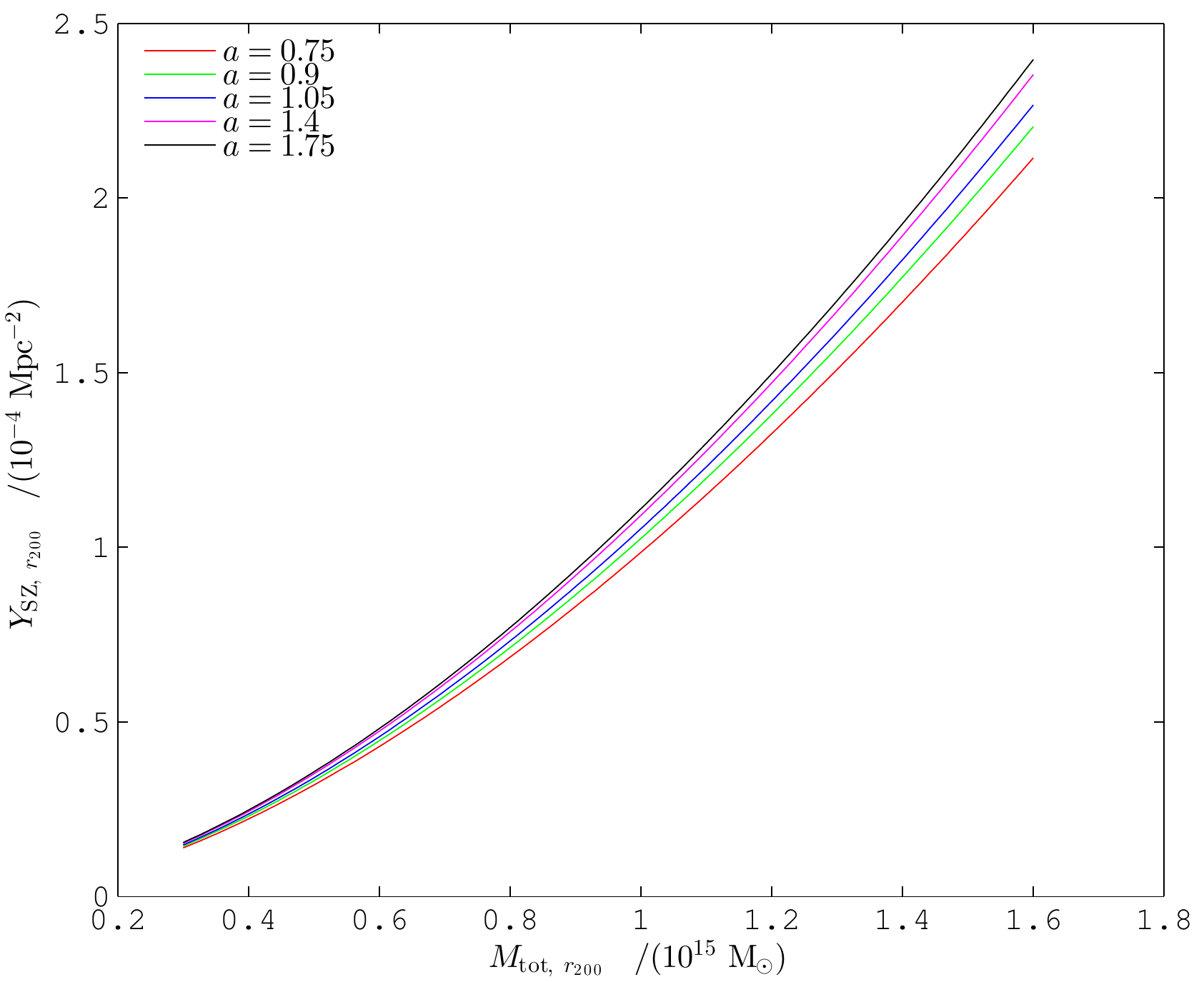}
\caption{$Y$--$M$ relations plotted using $M_{{\rm tot},\,r_{200}}$ and $Y_{{\rm SZ},\,r_{200}}$ values calculated with a fixed redshift and $f_{{\rm gas},\,r_{200}}$ of 0.55 and 0.13, respectively. Different coloured lines show different values of $a$ (the shape parameter describing the pressure profile shape around $r_{500}$) used in the calculation, as listed in the legend. \label{fig:MTYMaplot}}
\end{figure}

\citet{2010AA...517A..92A} (using $XMM$-$Newton$ data) and \citet{2013A&A...550A.131P} (using Planck SZ data), who find average pressure-profile shape parameters, both attribute deviations from the mean values to cluster dynamical state. \citet{2007MNRAS.380..437P} show large changes in cluster parameters, such as temperature, mass and luminosity, on Gyr timescales over the course of a merger. It is reasonable to expect the pressure profile, and the inferred shape parameters, to also vary substantially during a merger. This is in agreement with \citet{2013A&A...550A.128P} who fit a wide range of shape parameter values to five disturbed clusters (determined to be disturbed from X-ray morphology).

The assumed geometry and the chosen statistical methodology, in addition to cluster model (including pressure profile shape parameters) and assumptions of e.g. HSE, will also affect dynamical-state dependent parameter estimation. However, our analysis pipeline allows us to investigate the change with dynamical states due to a changes in $a$.
The shape parameter $a$ describes the slope of the pressure profile at radii best constrained by AMI SZ data.
We calculate $Y_{{\rm SZ},\,r_{200}}$ given the same range of $M_{{\rm tot},\,r_{200}}$ values. Redshift and $f_{{\rm gas},\,r_{200}}$ are fixed at 0.55 and 0.13 respectively, and $a$ is varied over a range representative of the deviations in $a$ found by Arnaud et al. and Planck Collaboration et al. II: between 0.75 and 1.75. As previously, we have found power-law best-fits for these $Y\,=\,AM^{\kappa}$ curves, plotted in Figure \ref{fig:MTYMaplot}. The $Y$--$M$ fits show very little change in $\kappa$ as $a$ is changed: less than a per cent over the range of $a$ values. The normalisation, $A$, changes by less than 15 per cent for $Y\,=\,AM^{\kappa}$. This variation in scaling relations with changing $a$ is indicative of the influence of mergers on scaling relation scatter: in agreement with Poole et al., the perpendicular scatter caused is small. As AMI data are sensitive to shape parameter $a$, we could relax the delta prior on $a$ and fit for it in the analysis. However, Figure \ref{fig:MTYMaplot} shows that this would not have much effect on the analysis.

\begin{figure*}
\includegraphics[trim= 0mm 0mm -2mm 0mm, clip,width=85mm,height=71.5mm]{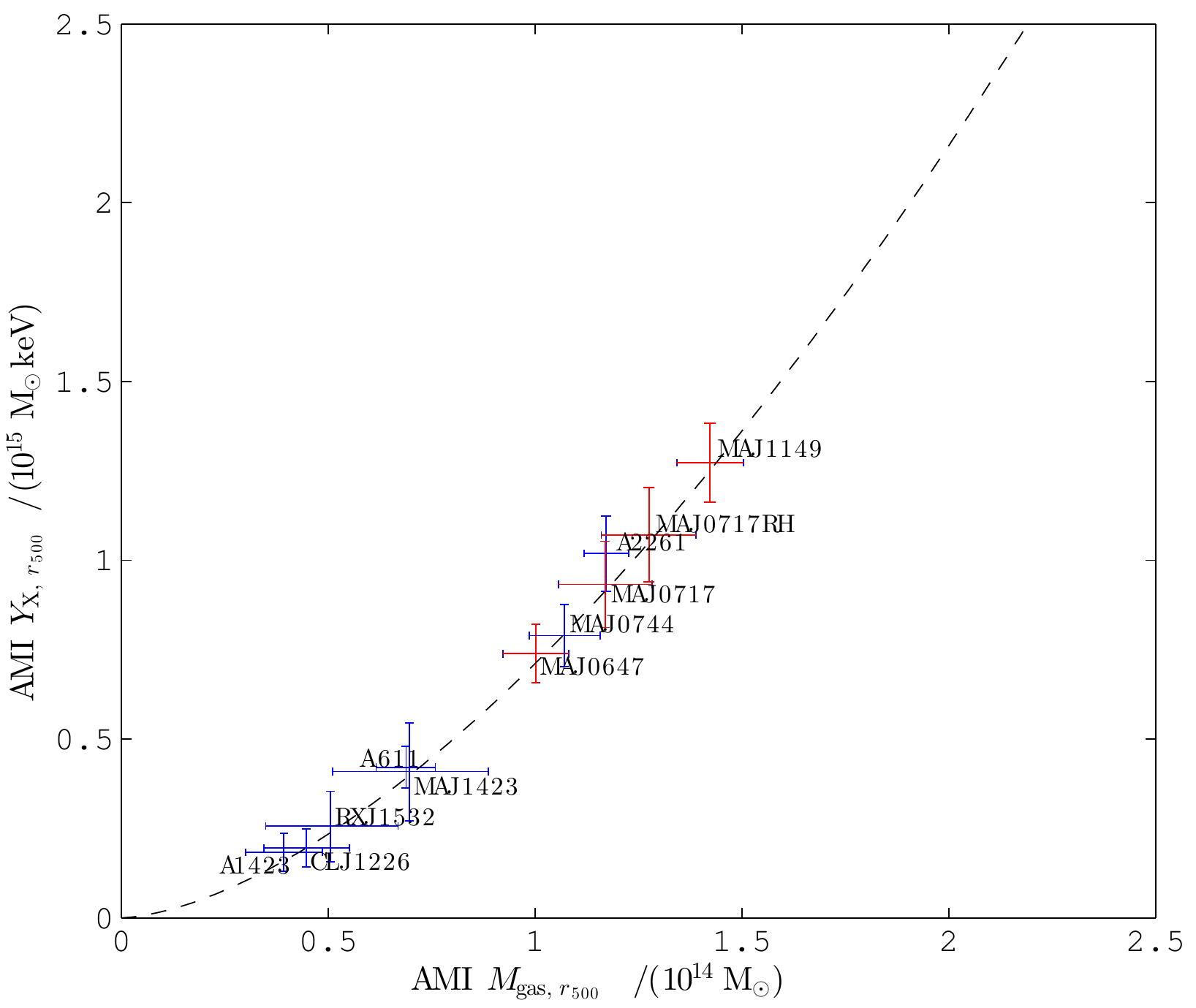}
\includegraphics[trim= -2mm 0mm 0mm 0mm, clip,width=85mm, height=71.5mm]{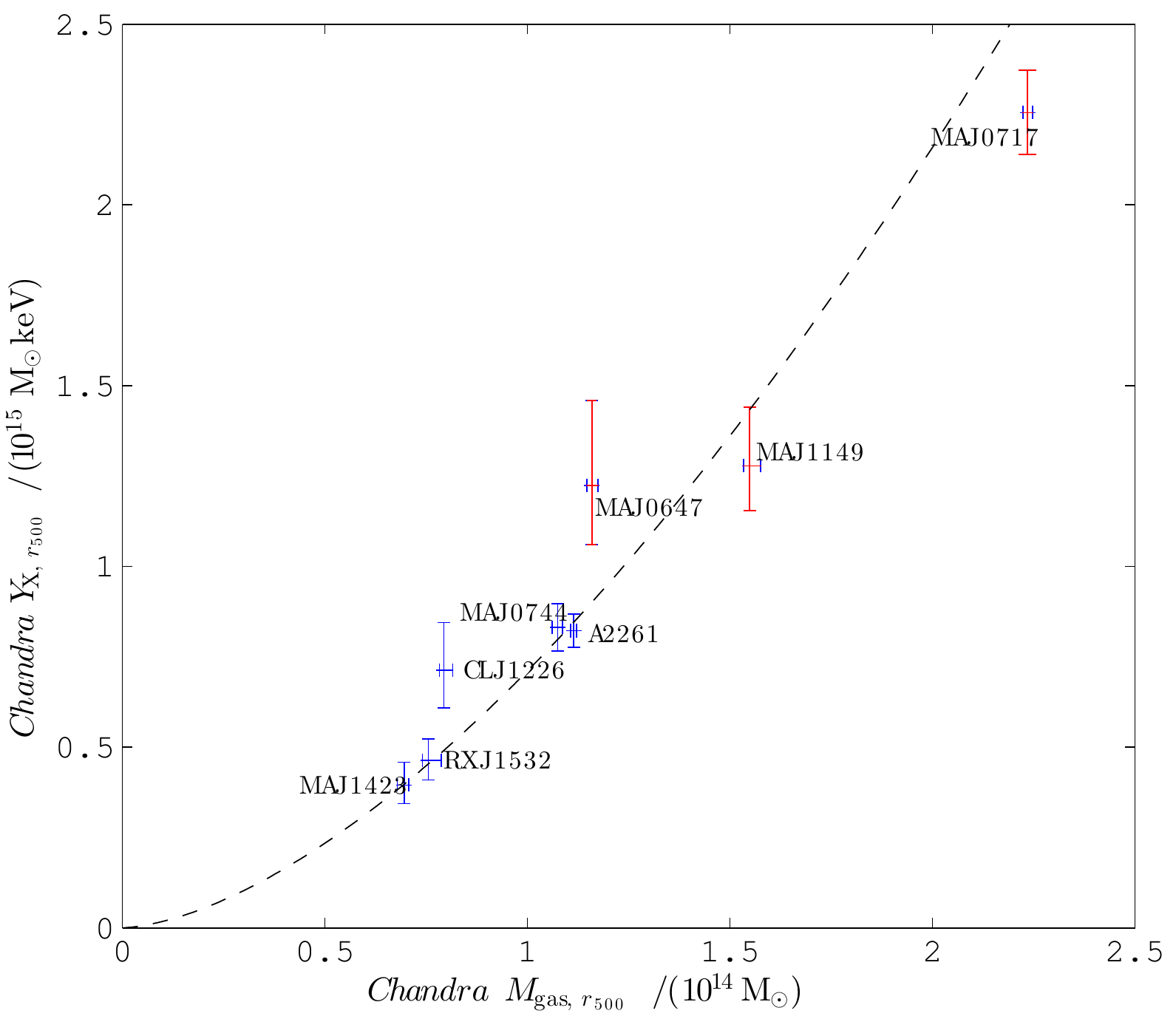}
\caption{Left (a): $Y_{\rm X}$ values calculated within $r_{500}$ from AMI estimated parameter values, with $T_{\rm SZ}$ weighted by $n_{\rm e}^2T^{1/2}$ {\color{black}(to imitate an X-ray-like weighting to $n_{\rm e}^2$ and $T^{1/2}$)}, plotted against the estimated gas mass within $r_{500}$ for each cluster. Error bars show 68 per cent confidence levels. Right (b): The X-ray estimated $Y_{\rm X}$ values within $r_{500}$ plotted against the estimated gas mass within $r_{500}$. Parameter values and errors from \citet{2008ApJS..174..117M}. Strong-lensing-selected sample members are displayed in red. Dashed lines show the $Y$--$M$ relation of $Y_{{\rm X},\,r_{500}}$ and $M_{{\rm gas},\,r_{500}}$ values calculated from our model with $z$\,=\,0.55, $f_{{\rm gas},\,r_{200}}$\,=\,0.13 and $a$\,=\,1.05. \label{fig:YM500plots}}
\end{figure*}


\subsection{$Y_{\rm SZ}$--$Y_{\rm X,\,AMI}$}

A tight relation between $Y_{\rm X}$ ($\equiv$ $M_{\rm gas}T$) from X-ray and $Y_{\rm SZ}$ (measured directly with an SZ telescope) is predicted from simulation \citep{2006ApJ...650..128K}.
We next check the tightness of $Y_{\rm SZ}$ (from AMI) versus $Y_{\rm X,\,AMI}$ (from AMI estimates of $M_{\rm gas}$ and $T_{\rm SZ,\,mean}$), to facilitate the comparison of $Y_{\rm X,\,AMI}$ with $Y_{{\rm X},\,Chandra}$ from X-ray studies of our clusters in Section \ref{sec:xray}. A very tight correlation is expected since both parameters are estimated from the same data, with the same model, and both measure the internal energy of the cluster gas.

The $Y_{\rm SZ}$--$Y_{\rm X}$ relation has previously been investigated for real and simulated data, such as \citet{2010AA...517A..92A}, \citet{2011ApJ...738...48A}, \citet{2012ApJ...760...67R}, and \citet{2014MNRAS.438...49R}, who find $Y_{\rm SZ} \propto {Y_{\rm X}}^{\kappa}$ where $\kappa$ is generally slightly lower than unity.

We derive $Y_{\rm X,\,AMI}$ from AMI SZ measurements by multiplying our AMI $M_{{\rm gas},\,r_{200}}$ values by our $T_{\rm SZ,\,mean}$ values, the mean SZ temperature found by taking the average of temperatures calculated with equation \ref{eq:Tgas} in the range (0.15\,$<$\,r\,$<$\,1)$r_{200}$.
We estimate $\kappa$ for the AMI-CLASH sub-sample using the orthogonal linear regression analysis (Case 3) used previously. As the uncertainties are highly correlated, we use an analysis that does not take error bars into account. We find $\kappa$\,=\,$0.978\pm0.009$ (where the error is evaluated from the scatter perpendicular to the fitted line) which shows the validity of assuming equivalence of $Y_{\rm SZ}$ and $Y_{\rm X}$ for our SZ parameter estimation and enables the use of $Y_{\rm X,\,AMI}$ in comparison with $Y_{{\rm X},\,Chandra}$ values.


\section{Cluster mergers and scaling relations}
\label{sec:xray}

{\color{black}Both X-ray and SZ studies find scaling relations that largely agree with self-similarity.} Scatter about these relations is small, and often attributed to the dynamical states of clusters in the sample used (see e.g. \citealt{2009ApJ...692.1033V}, \citealt{2011ApJ...738...48A}).
Due to the difference in dependence of SZ and X-ray measurement to $n_{\rm e}$ we expect scatter along scaling relations caused by mergers, predicted by \citet{2007MNRAS.380..437P}, to be larger for X-ray parameter values than SZ. We next compare AMI SZ parameter values with values from X-ray studies to investigate whether we can demonstrate this observationally.

In Figure \ref{fig:YM500plots} we plot two $Y$--$M$ scaling relations. On the left of Figure \ref{fig:YM500plots} we plot AMI parameter values, finding $Y_{\rm X,\,AMI}$ from $M_{{\rm gas},\,r_{500}}$ and the mean temperature in the range (0.15\,$<$\,r$\,<$\,1)$r_{500}$ given by equation \ref{eq:Tgas}.
On the right of Figure \ref{fig:YM500plots} we plot $Y_{{\rm X},\,Chandra}$--$M_{\rm gas}$ for eight of the AMI-CLASH sub-sample of ten using $Chandra$ X-ray estimated parameters from \citet{2008ApJS..174..117M} (A611 and A1423 are not included in their sample), who calculate $M_{\rm gas}$ internal to $r_{500}$ and find $Y_{{\rm X},\,Chandra}$ from $Y_{\rm X}\,=\,M_{\rm gas}T_{\rm X}$, where $T_{\rm X}$ is the mean X-ray temperature in the range (0.15\,$<$\,r\,$<$\,1)$r_{500}$.
AMI error bars are 68 per cent confidence limits; X-ray error bars are taken from Maughan et al..

SZ and X-ray measurements of a cluster are differently weighted with $n_{\rm e}$ and $T_{\rm e}$.
The SZ flux density depends linearly on $n_{\rm e}$ and $T_{\rm e}$ whereas the X-ray flux density depends on ${n_{\rm e}}^2$ and ${T_{\rm e}}^{1/2}$, where ${T_{\rm e}}^{1/2}$ is approximately proportional to the cooling function, $\Lambda (T)$, for the temperatures of the clusters considered here.
These different dependencies will cause parameter values estimated from X-ray to be more influenced by the higher density gas, relative to the SZ estimates.
When comparing SZ and X-ray temperatures we reduce this effect by introducing a weighting of ${n_{\rm e}}^2T^{1/2}$ to SZ-derived temperatures. This assumes $n_{\rm e}(r)$ accurately describes the cluster gas density with no additional density features such as shocking of cluster gas or fractionation. The $Y_{\rm X,\,AMI}$ values in Figure \ref{fig:YM500plots}(a) have been calculated with $T_{\rm SZ,\,mean}$ values weighted in this way.

The behaviour in Figure \ref{fig:YM500plots}\,(a) is, as expected, similar to the $Y$--$M$ scaling relation in Section \ref{sc:scalingrel}; in Figure \ref{fig:YM500plots}\,(b) the behaviour is consistent with the self-similar scaling relation but with a substantial scatter.
Although our AMI SZ data depends on dynamical state our model (see Section \ref{sec:model}) does not, as discussed in Section \ref{sec:MTYM}. The X-ray analysis of \citet{2008ApJS..174..117M} is dependent on cluster dynamical state, resulting in the scatter seen in Figure \ref{fig:YM500plots}\,(b), showing the sensitivity of X-ray measurement to mergers.
AMI values internal to $r_{500}$ produce an orthogonal-linear (Case 3) best-fit to $Y_{\rm X}\,\propto\,{M_{\rm gas}}^{\kappa}$ with $\kappa$\,=\,$1.594\pm0.025$, from which we have excluded A611 and A1423 in order to match the sample of Maughan et al..
Using X-ray parameter values from Maughan et al., the best-fit value of $\kappa$ is $1.425\pm0.091$.
Over-plotted onto Figure \ref{fig:YM500plots}\,(a) and (b) is the curve of $Y_{{\rm X},\,{\rm AMI},\,r_{500}}$ versus $M_{{\rm gas},\,r_{500}}$ values calculated from our model, as in Section \ref{sec:MTYM}, with $f_{{\rm gas},\,r_{200}}$\,=\,0.13, $a$\,=\,1.05 and $z$\,=\,0.55.

Both the discussions in the literature, cited in Table \ref{tab:litrev}, and studies showing the strong correlation between lensing strength and merger activity (see e.g. \citealt{2013ApJ...762L..30Z}), indicate that the three clusters selected for their lensing strength are significantly more disturbed than the remaining seven. In Figure \ref{fig:YM500plots}(b) these show slightly more deviation from the calculated curve relative to more relaxed systems, consistent with the small increase in perpendicular scatter caused by mergers found by \citet{2007MNRAS.380..437P}. CLJ1226+3332 also has a small deviation from the curve.

Despite the increase in perpendicular displacements caused by cluster dynamics, shown by Poole et al., X-ray scaling relations, like SZ relations, are seen to be consistent with self-similarity, (see e.g. \citealt{2009ApJ...692.1033V}); and indeed, this is what we see when plotting the X-ray parameters from Maughan et al., as illustrated by the over-plotted power-law on the $Y_{\rm X}$--$M_{\rm gas}$ plots in Figure \ref{fig:YM500plots}.

\begin{figure}
\includegraphics[trim= 1mm 0mm -16mm 0mm, clip,width=88mm]{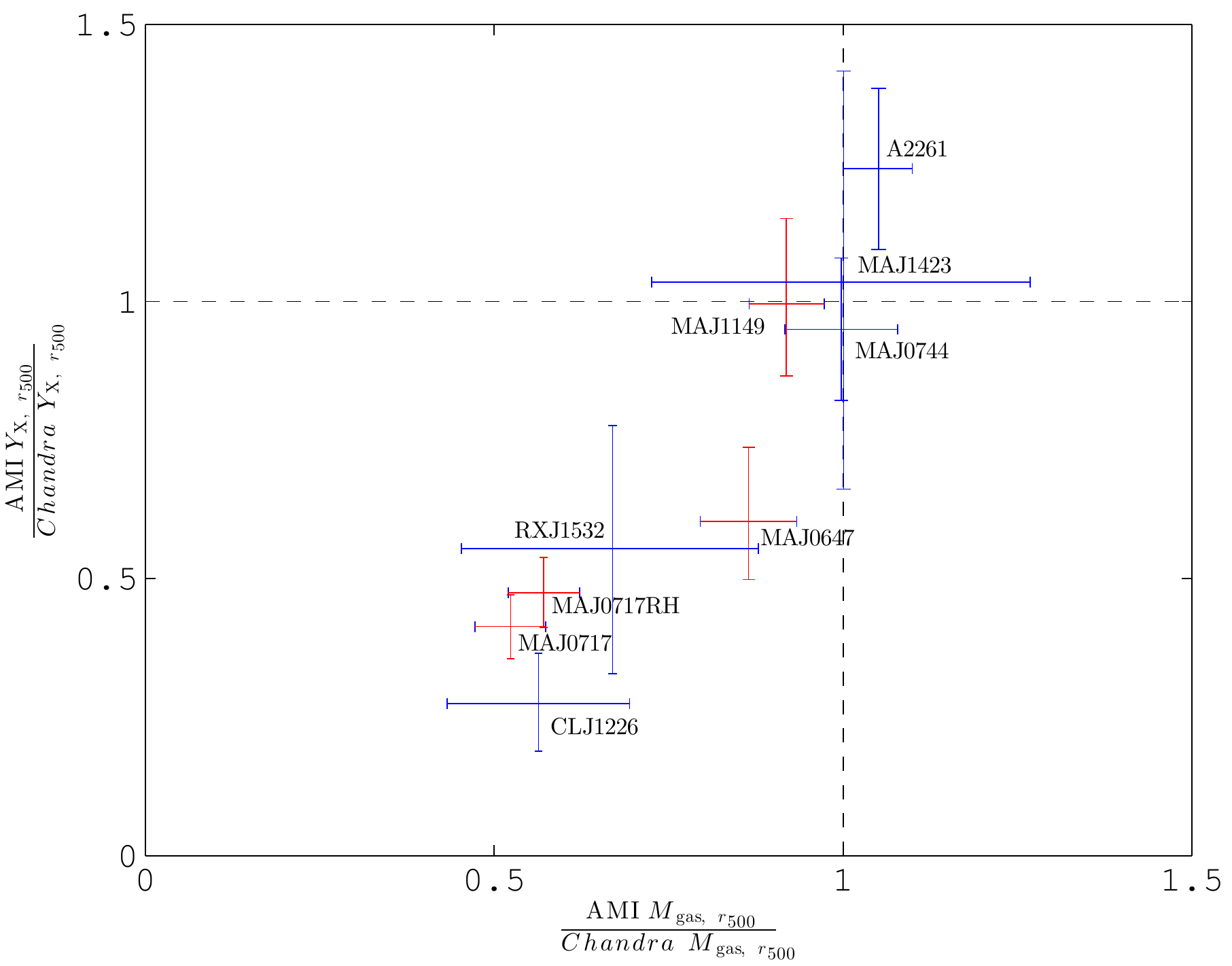}
\caption{The ratio of AMI and $Chandra$ $Y_{\rm X}$ within $r_{500}$ plotted against the ratio of AMI and $Chandra$ $M_{\rm gas}$ within $r_{500}$, to show discrepancies from deviations of the ratios from 1. $\frac{AMI~Y_{\rm X}}{Chandra~Y_{\rm X}}$\,=\,1 and $\frac{AMI~M_{\rm gas}}{Chandra~M_{\rm gas}}$\,=\,1 are plotted in dashed lines, the intersection of which indicates agreement of parameters. $Chandra$ parameter estimates are from \citet{2008ApJS..174..117M}. Errors are calculated from those in Figure \ref{fig:YM500plots}. Strong-lensing-selected sample members are displayed in red. \label{fig:YMratio}}
\end{figure}

However, the agreement of X-ray and SZ relations is not seen in individual cluster parameters on the $Y_{\rm X}$--$M_{\rm gas}$ plots: some cluster positions agree very well in $Y_{\rm X}$--$M_{\rm gas}$ space, with very little difference between SZ and X-ray estimated parameter values. Others disagree by 2$\times$ or more.
Figure \ref{fig:YMratio} shows the ratio of AMI and $Chandra$ $Y_{\rm X}$ values plotted against the ratio of AMI and $Chandra$ $M_{\rm gas}$ values and helps show the differences between the plots in Figure \ref{fig:YM500plots}. In Figure \ref{fig:YMratio}, deviations from 1 (marked by dashed lines in each dimension) show differences in the SZ-and X-ray-derived parameter values: the most significant deviations from 1:1 correspond to high values of $Chandra$ $Y_{\rm X}$ and $Chandra$ $M_{\rm gas}$.

Although the $\beta$-model, which is used by \citet{2008ApJS..174..117M}, has been shown to produce different estimates of parameter values to the GNFW model when used to analyse AMI SZ data (see \citealt{2013MNRAS.431..900S}), we do not expect this to produce the discrepancies seen in Figure \ref{fig:YMratio}. Recent studies have moved away from using the $\beta$-model, favouring the use of NFW and GNFW profiles which give a more detailed description of the changing gradient of the pressure profile over a wide range of $r$.
For the cluster masses and redshifts considered here, X-ray sensitivity is good out to $r_{500}$.
As a $\beta$-model gives a good general description of the pressure profile in a relaxed cluster in the range (0.15\,$<$\,r$\,<$\,1)$r_{500}$\footnote{Thus ignoring differences in descriptions of the cluster centre.}, estimates of $M_{\rm gas}$ within this range via a $\beta$-model, from good X-ray data, are likely to be robust.
A more detailed description of the pressure profile, provided by an NFW-like profile, is required to accurately infer $M_{\rm gas}$ in the range (0.15\,$<$\,r$\,<$\,1)$r_{500}$ from AMI SZ data that (for these clusters) have greatest sensitivity to scales $r_{500}$ to $r_{200}$. It is, therefore, reasonable to expect our AMI SZ parameter estimates to agree with those of Maughan et al. for relaxed, high signal-to-noise clusters.

The discrepancies in $Y_{\rm X}$, highlighted by Figure \ref{fig:YMratio}, show an effect that has not been removed by accounting for the difference in dependence of the SZ and X-ray signals to the density profile $n_{\rm e}(r)$ also present in $M_{\rm gas}$ estimates. This could result from additional density components such as shocking and/or fractionation in the cluster gas that are not described by $n_{\rm e}(r)$, and therefore not accounted for when adjusting the SZ temperature.
The apparent splitting of sample members into two populations in Figure \ref{fig:YMratio} indicates a systematic effect, whereas simple departures of clusters from relaxed morphology and HSE, assumed in the model, would cause random distributions of clusters about the 1:1 point.

\citet{2007MNRAS.380..437P} allow for, e.g., shocking, fractionation and radiative cooling, and show cluster observables evolving during a merger, following paths along the plotted scaling relations with small deviations perpendicular to the relations.
The difference in sensitivity of X-ray and SZ measurement to this merger activity allows us to investigate the large displacements along the scaling relation more clearly than looking at deviations of SZ or X-ray cluster parameters from a self-similar relation.

Displaying discrepancies between SZ and X-ray measurements as we do in Figure \ref{fig:YMratio} illustrates the discrepancies between SZ and X-ray temperature estimates in the context of the typical use of cluster parameter estimates. We next compare temperature estimates measured from SZ and X-ray directly.


\subsection{$T_{SZ}$ versus $T_X$}\label{sec:TxTsz}

\begin{figure*}
\includegraphics[trim= 0mm 0mm -4mm 0mm, clip,width=85mm,height=50mm]{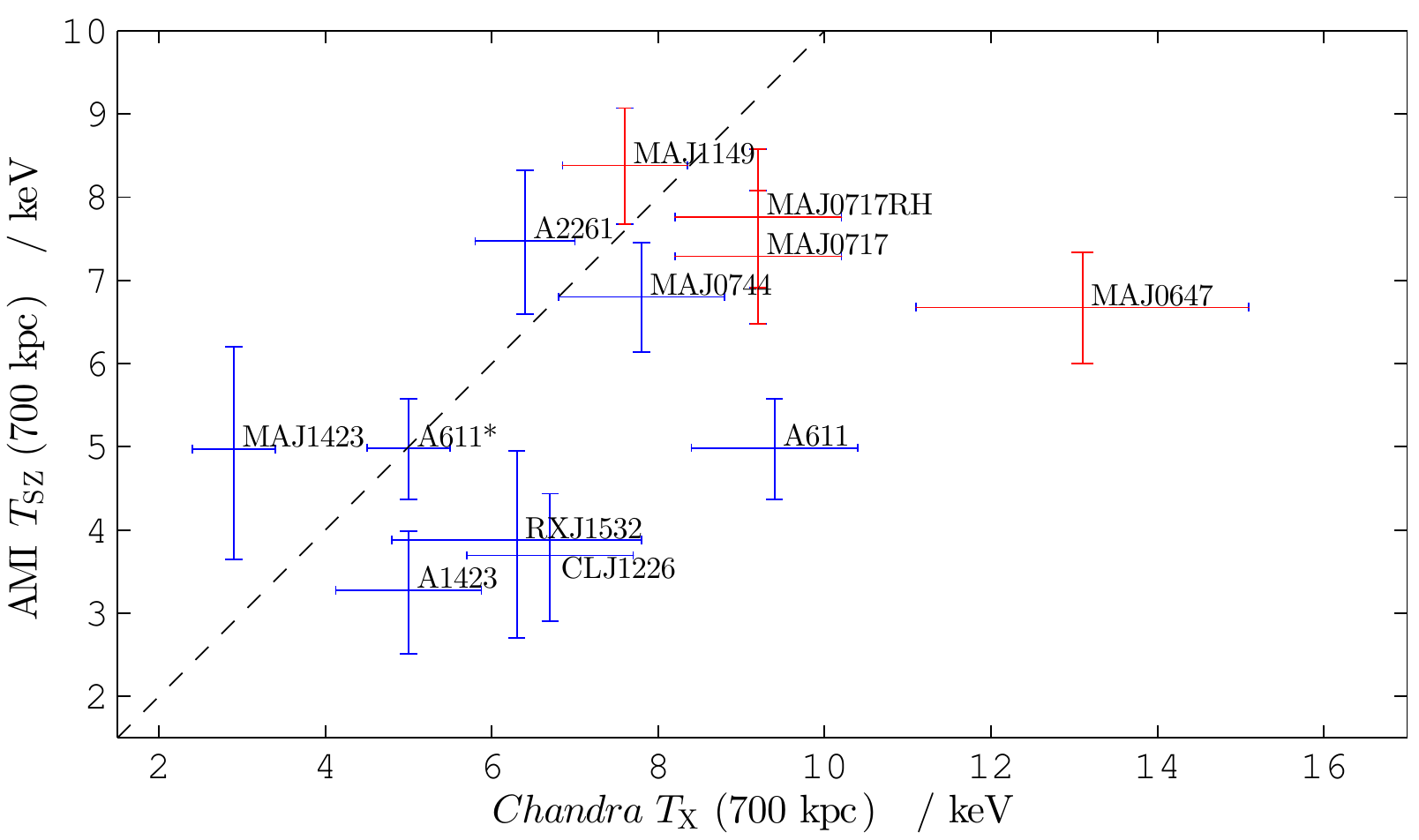}
\includegraphics[trim= -4mm 0mm 0mm 0mm, clip,width=85mm, height=50mm]{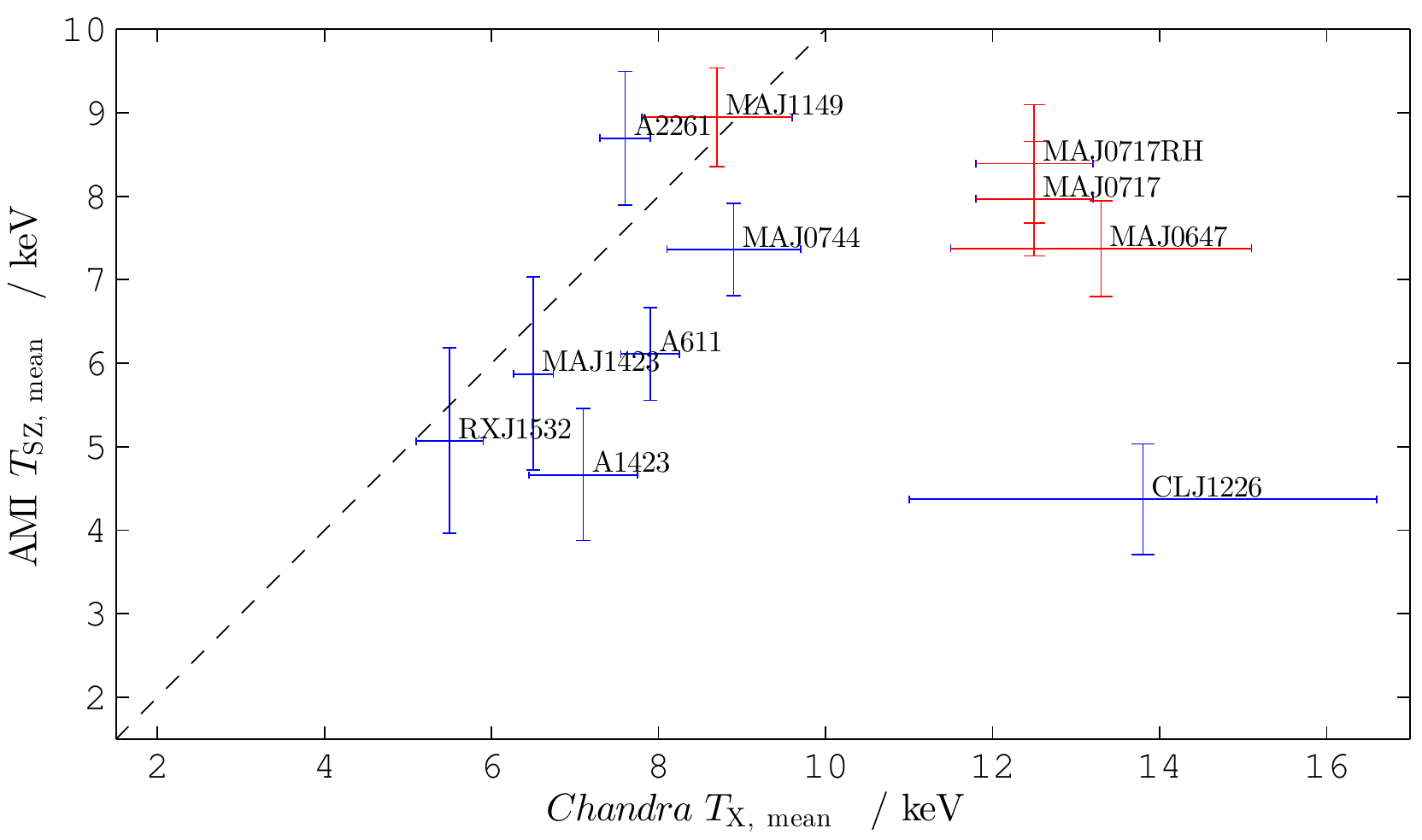}
\caption{SZ and X-ray temperatures for each cluster: Left\,(a) Shows AMI SZ temperatures at $r\,=\,700\,{\rm kpc}$ plotted against X-ray annularly averaged temperatures at $r\,=\,700\,{\rm kpc}$ from the ACCEPT database. Right\,(b) Shows mean AMI SZ temperatures plotted against X-ray mean temperatures from \citet{2012ApJS..199...25P}, both found in the range (0.15\,$<$\,r\,$<$\,1)$r_{500}$. Strong-lensing-selected sample members are displayed in red. Overplotted is the line of $T_{\rm SZ}\,=\,T_{\rm X}$. Error bars are 1$\sigma$. \label{fig:tempszx}}
\end{figure*}

\citet{2012MNRAS.425..162A} discuss using discrepancies in estimates of cluster temperature between SZ and X-ray to highlight mergers, probing boosts in cluster temperature through shocking and/or fractionation directly.
We follow Rodr\'{\i}guez-Gonz\'alvez et al. and plot the SZ temperature estimated by \textsc{McAdam} against the X-ray temperature from $Chandra$ data.
Temperature estimates are reported in the $Chandra$ ACCEPT database\footnote{\url{http://www.pa.msu.edu/astro/MC2/accept/}} \citep{2009ApJS..182...12C} in circular annuli which cover radii out to typically 500 to 1000\,kpc.
We calculate SZ temperatures at the highest radius that is reached by the $Chandra$ temperature profiles of all 10 AMI-CLASH clusters, $r$\,$\approx$\,700\,kpc, to obtain comparable temperature estimates at radii at which we can constrain parameters with SZ data. We plot $T_{\rm SZ}$ versus $T_{\rm X}$, at 700\,kpc in Figure \ref{fig:tempszx}\,(a). The ACCEPT temperature profile for A611 appears particularly uncertain, due to the automated way in which the analysis was carried out, necessitated by the size of this very useful database. A more detailed analysis of the same A611 $Chandra$ data is available in \citet{2011AA...528A..73D} producing a much lower temperature at high radius with smaller errors. Both 700-kpc X-ray temperature estimates are included in Figure \ref{fig:tempszx}\,(a).

Estimates of the \textit{average} X-ray temperature are found from Chandra data in \citet{2008ApJS..174..117M} (in the range (0.15\,$<$\,r\,$<$\,1)$r_{500}$) and \citet{2012ApJS..199...25P} (in the range 71.4\,kpc\,--\,714\,kpc from the centre of each cluster). There is good agreement between Maughan et al. and Postman et al. on the temperatures of 5 of the 8 clusters in both samples. There are larger differences between CLJ1226+3332, MAJ0647+7015 and MAJ0717+3745 X-ray temperatures but these are still consistent given the quoted error estimates.
We plot the mean AMI SZ temperature with the average cluster temperatures from Postman et al. in Figure \ref{fig:tempszx}(b). Comparing AMI SZ parameter values with three independent X-ray analyses addresses possible uncertainty regarding reduction and analysis methods.
Figures \ref{fig:YMratio} and \ref{fig:tempszx}\,(b) highlight the same two populations of clusters within our sample. Figure \ref{fig:tempszx}\,(a) is less demonstrative due to problems with annular averaging at high radius -- this is discussed further in Section \ref{sec:mergsim}.

\subsubsection{$XMM$--$Chandra$ $T_{\rm X}$ discrepancy}

In previous studies, X-ray estimated parameter values (such as gas mass) found using $Chandra$ data are often higher than those found from $XMM$-$Newton$ data: the discrepancies have been attributed to differences in temperature estimates between the instruments (see e.g. \citealt{2013ApJ...767..116M}). \citet{2015AA...575A..30S} find the discrepancy to be energy dependent, increasing with temperature to $\approx$\,29 per cent at 12\,keV (Earth-frame). \citet{2014ApJ...794..136D} use $XMM$ and $Chandra$ $T_{\rm X}$ profiles of the CLASH sample and find a radial dependence: at 100\,kpc they find no systematic difference in $XMM$ and $Chandra$ temperatures, but towards 1\,Mpc the discrepancy reaches $\approx$\,25 per cent.

We investigate the apparent separation of clusters into two populations in Figure \ref{fig:tempszx}\,(b) to see if an energy-dependent bias of $Chandra$ temperatures (from \citealt{2012ApJS..199...25P}) could be responsible. \citet{2012AA...545A..41B} find $XMM$ temperatures (in the range (0.15\,$<$\,r\,$<$\,1)$r_{500}$) for a sample of clusters including MAJ0744+3927, CLJ1226+3332 and MAJ0647+7015. From Figure \ref{fig:tempszx}\,(b) MAJ0647+7015 and CLJ1226+3332 have similar, very high, $T_{\rm X,\,mean}$ values. MAJ0744+3927 has a similar $T_{\rm SZ,\,mean}$ value to MAJ0647+7015 but a much lower $T_{\rm X,\,mean}$. In Table \ref{tab:xmmchan} we find differences between the temperatures measured for pairs of clusters by each instrument.
Given the error estimates quoted in Baldi et al. and Postman et al., we see that MAJ0647+7015 and CLJ1226+3332 $Chandra$ temperature estimates are not significantly biased high relative to MAJ0744+3927 compared with estimates from $XMM$. We emphasise that our clusters with high rest-frame temperatures tend to be at high redshift so that their measured temperatures are lower.

\citet{2012AA...545A..41B} find the $\approx$\,29 per cent discrepancy in $T_{\rm X}$ induces $\approx$\,15 per cent discrepancy in the mass estimate towards 12\,keV; this does not account for the nearly 50 per cent discrepancies in AMI and $Chandra$ mass estimates in Figure \ref{fig:YMratio}.

\begin{table}
\centering
\caption{Differences between temperature estimates for pairs of clusters in our sample. $T_{\rm X}$ values are core-excluded temperature estimates internal to $r_{500}$ from $Chandra$ observation \citep{2012ApJS..199...25P} and $XMM$ observation \citep{2012AA...545A..41B}. Uncertainties in $T_X$ estimates are the errors quoted in the respective papers; ``difference" significances have been obtained using the quadrature formula.}\label{tab:xmmchan}
\begin{tabular}{p{1.75cm}p{1.2cm}p{1.05cm}p{1.2cm}p{1.05cm}}
\hline
Cluster       & \multicolumn{2}{c}{$Chandra$}                     & \multicolumn{2}{c}{$XMM$}  \\\hline
              & $T_{\rm X}$ /\,keV & difference                    & $T_{\rm X}$ /\,keV       & difference   \\\hline
MAJ0647+7015  & $13.3\pm1.80$     & \multirow{2}{*}{2.23$\sigma$} & $9.30^{+0.45}_{-0.37}$  & \multirow{2}{*}{2.32$\sigma$} \\
MAJ0744+3927  & $8.9\pm0.80$      & \multirow{2}{*}{1.68$\sigma$} & $8.14\pm0.34$           & \multirow{2}{*}{2.49$\sigma$} \\
CLJ1226+3332  & $13.8\pm2.80$     &                               & $10.16^{+0.77}_{-0.73}$ &                               \\
\hline
\end{tabular}
\end{table}

\subsection{Comparison with simulations}\label{sec:mergsim}

Internal cluster dynamics during merger activity have been investigated through simulations covering a wide variety of merger scenarios and individual systems, see e.g. \citet{2010ApJ...717..908Z}, \citet{2011MNRAS.418..230V}, \citet{2012MNRAS.425L..76B} and \citet{2012ApJ...748...45M}. We focus on the comprehensive work of \citet{2006MNRAS.373..881P}, which follows the cluster merger process for three merger mass ratios for each of three merger impact parameters (see also \citealt{2001ApJ...561..621R}, \citealt{2002MNRAS.329..675R} and \citealt{2002ApJ...577..579R}). From example merger stages in the paper itself and the suite of simulation videos online, at \url{http://visav.phys.uvic.ca/~babul/Merger\_PaperI/}, this study provides simulations of both the X-ray and the SZ signals, ideal for the focus of this paper, as well as the X-ray temperature, gas surface density and entropy. While these simulations include the effects of radiative cooling, star formation and minimal feedback from supernovae, \citet{2006MNRAS.373..881P} make clear that they do not include factors such as feedback from AGN, magnetic fields, pressure due to cosmic rays, and conduction.

In the following, we compare the simulated X-ray morphology and temperature maps with real X-ray maps and temperature profiles on the $Chandra$ ACCEPT database.
For clusters that appear relaxed, or very close to relaxed, we class their dynamical state as $\gtrapprox$\,5\,Gyr since first pericentre (the first closest approach of the secondary cluster to the primary), as depicted by Poole et al.: we class A611, A1423, A2261 and MAJ1423+2404 in this way.

The X-ray morphologies of A611 and A1423 do not look similar to any of the simulated merger stages, with circular shapes and no visible concentrations of X-ray emission. Neither cluster is present in the \citet{2008ApJS..174..117M} sample, so are not included in Figure \ref{fig:YMratio}, but do not show significantly higher $T_{\rm X}$ values relative to $T_{\rm SZ}$ in Figure \ref{fig:tempszx}.

A2261 appears in both Figures \ref{fig:YMratio} and \ref{fig:tempszx}, also showing no significant difference in X-ray and SZ estimated parameter values. The X-ray morphology most closely resembles the Poole et al. simulations after 5\,Gyr apart from an area of isolated substructure lying approximately 0.7\,Mpc from the cluster centre (see e.g. Maughan et al.). Given the low X-ray temperature and agreement of SZ and X-ray parameter estimates, it could be the start of a very high mass-ratio merger, well before first pericentre.

MAJ1423+2404 also has a relaxed X-ray morphology and displays little discrepancy between X-ray-estimated and SZ-estimated $Y_{\rm X}$ vs $M_{\rm gas}$ in Figure \ref{fig:YMratio} and $T_{\rm SZ}$ vs $T_{\rm X}$ in Figure \ref{fig:tempszx}, also supporting a relaxed classification.

RXJ1532+3021 is reportedly a very strong cool-core cluster, supporting a relaxed classification. \citet{2013ApJ...777..163H} investigate the X-ray morphology using $XMM$-$Newton$ and deep $Chandra$ observations and reveal central substructure from substantial AGN activity and a cold front. The authors conclude that the origin of the cold front is either cool gas dragged out by the AGN outburst or turbulence from sloshing of the cool core induced by a minor merger.
Figure \ref{fig:tempszx} shows good agreement between $T_{\rm SZ}$ and $T_{\rm X}$ values but SZ- and X-ray-derived $Y_{\rm X}$ and $M_{\rm gas}$ values in Figure \ref{fig:YMratio} are more discrepant. This supports the theory of a low-level merger proposed by Hlavacek-Larrondo et al. which may be old enough to no longer exhibit the higher temperatures associated with mergers.

Although CLJ1226+3332 is classed by \citet{2008MNRAS.383..879A} and \citet{2012ApJS..199...25P} as relaxed given its X-ray morphology, differences in SZ and X-ray parameter values suggest otherwise. Figures \ref{fig:YMratio} and \ref{fig:tempszx} show large discrepancies between X-ray and SZ in $Y_{\rm X}$ and $M_{\rm gas}$ and in average temperature estimates. The average X-ray temperature and temperature profiles from $Chandra$ on the ACCEPT database, and $XMM$-$Newton$ in \citet{2007ApJ...659.1125M}, show very high temperatures for the low mass of CLJ1226+3332, suggesting an early-stage merger. In conjunction with its circular shape, slightly displaced X-ray peak brightness from the cluster centre and some substructure towards the south-west, CLJ1226+3332 is likely a recent, close to head-on, minor merger.
Maughan et al. and \citet{2009ApJ...691.1337J} also discuss the non-relaxed state of CLJ1226+3332.
Jee \& Tyson compare their weak-lensing mass reconstruction with the X-ray temperature map of Maughan et al.: they note the correlation of a high-temperature asymmetry with an area of low-luminosity substructure to the south-west of the cluster centre, concluding that the substructure has just passed through the primary cluster.

MAJ0647+7015 (strong-lensing selected) has high values of $T_{\rm X,\,mean}$ and $T_{\rm X}(700\,{\rm kpc})$, both of which are very discrepant from the SZ values in Figure \ref{fig:tempszx}. Figure \ref{fig:YMratio} shows large differences also between SZ and X-ray $Y_{\rm X}$ and $M_{\rm gas}$ values.
The X-ray morphology is non-circular with no central peak in X-ray surface brightness which, along with the high $T_{\rm X}$ values, indicate MAJ0647+7015 has recently undergone a close to head-on major merger.

MAJ0717+3745 has a complicated X-ray morphology that suggests multiple mergers: a triple merger, according to \citet{2012MNRAS.420.2120M}. Figure \ref{fig:YMratio} shows large differences between X-ray and SZ measurements of $Y_{\rm X}$ and $M_{\rm gas}$.
No significant discrepancy is evident between the SZ and X-ray temperatures at 700\,kpc, in Figure \ref{fig:tempszx}. This is likely due to effects of annular averaging over a complex X-ray temperature distribution that is poorly approximated by the model profile in both X-ray and SZ analyses. This is supported by a larger discrepancy in average temperature, for which annular averaging was not used.

Parameter estimation for MAJ0744+3927 suggests a relaxed state. Both $T_{\rm X,\,mean}$ and $T_{\rm X}(700\,{\rm kpc})$ are consistent with the values estimated from SZ, shown by Figure \ref{fig:tempszx}, and there is no significant discrepancy in $Y_{\rm X}$ and $M_{\rm gas}$ between SZ and X-ray in Figure \ref{fig:YMratio}. The ACCEPT X-ray temperature profile shows a low peak temperature for the mass of the cluster and some evidence for a cool-core. However, the X-ray morphology does not appear relaxed: there is a clear displacement in X-ray peak brightness from the centre and some extended clump features. These indicate an old merger returning to a relaxed state, supported by the small cool core.

In a strong-lensing investigation of MAJ1149+2223, \citet{2009ApJ...707L.163S} found the mass distribution to be best described as a main dark matter halo plus three, group-sized halos. \citet{2012MNRAS.420.2120M} suggest that the high velocity dispersion, measured by \citet{2007ApJ...661L..33E}, arises from merging along the line of sight.
There is, however, no significant discrepancy between X-ray--estimated and SZ--estimated parameter values in Figures \ref{fig:YMratio} and \ref{fig:tempszx}; we argue that there are two reasons that may explain this.
The first reason concerns the mass of the main halo compared with any of the masses of the group halos. Given the findings of Smith et al. and the low X-ray temperature, we expect the individual merging groups to be of low mass relative to the primary cluster: \citet{2006MNRAS.373..881P} show the X-ray temperatures reached in binary mergers of 1:1 and 3:1 are much higher than in 10:1 mergers, where the temperature is lower with larger impact parameter.
The lack of a high X-ray surface-brightness region in the ACCEPT $Chandra$ maps also indicates that mergers are of high impact-parameter.
The second reason concerns the complex shape of the system, evident in the ACCEPT images, for which annular averaging will be problematic. We expect the X-ray temperature profile to be biased low due to cooler, higher-radius gas being included in each annulus.

\section{Conclusions}

We have observed the eleven clusters in the CLASH sample that are accessible to AMI, and discard one due to a very bright source on the edge of the field of view. The remaining ten clusters have been analysed in a fully Bayesian way to give estimated parameter values from SZ measurement.

\begin{enumerate}

 \item Although 20 out of the 25 CLASH sample members were selected to be relaxed, we find disagreement in the literature on the dynamical states of many of our AMI-CLASH sub-sample, illustrating the difficulty in determining cluster dynamical state and identifying mergers.

 \item We investigate the correlations in our model and use it to calculate $Y$--$M$ curves, varying $f_{{\rm gas},\,r_{200}}$ and redshift. We discuss the effect these parameters have on the power-laws fitted to the curves: $Y$--$M$ is dependent on the cluster redshift and on the value of $f_{{\rm gas},\,r_{200}}$.

  \item X-ray studies have found sensitivity of the scatter about scaling relations to the dynamical states of the clusters included. Here, we investigate the sensitivity of our model correlations to dynamical state by introducing variations in cluster pressure profile shape parameters: a consequence of merger activity that can be induced in the analysis.
In our analysis these are fixed to the ``universal" values. Varying $a$, the shape parameter best constrained by AMI SZ data, over a large range induces only small changes in scaling relation power-laws.
This is consistent with \citet{2007MNRAS.380..437P} who track cluster observables of merging systems in the plane of scaling relations, finding very little perpendicular scatter but large variations along the scaling relation, over the course of a merger.

 \item  Due to the difference in dependence of SZ and X-ray measurement on $n_{\rm e}$, X-ray observables are much more sensitive to changes in mass, pressure, temperature and density during a merger. Discrepancies between parameter estimates derived from SZ and X-ray measurement can help identify clusters undergoing mergers. Denoting $M_{\rm gas}T$ by $Y_{\rm X}$ (for X-ray and SZ), we compare two $Y_{\rm X}$--$M_{\rm gas}$ scaling relations of the sub-sample members: one plotted using AMI SZ parameter values and the second using $Chandra$ X-ray parameters from \citet{2008ApJS..174..117M}. In addition to a difference in scatter we see an apparent ``movement" of sample members along the line of the relation, between SZ and X-ray. These discrepancies are visualised by plotting the ratios of AMI and $Chandra$ $Y_{\rm X}$ and $M_{\rm gas}$ values showing a population of our sub-sample for which the SZ and X-ray parameters agree well. Other clusters are discrepant by up to $\times$2, all towards higher X-ray $Y_{\rm X}$ and $M_{\rm gas}$ values \textit{along} the line of the relation.

  \item We also plot temperature estimates made from SZ and X-ray observation and find a similar split of clusters into two populations that are in agreement with those found from the ratios of AMI and $Chandra$ $Y_{\rm X}$ and $M_{\rm gas}$ values. This result comes from the comparison of AMI SZ parameter estimates with those from three independent X-ray analyses, addressing possible inconsistencies between methods.

 \item Dynamical state classifications in the literature of the ten clusters report: A611, MAJ1423+2404 and RXJ1532+3021 are relaxed; MAJ0647+7015, MAJ0717+3745 and MAJ1149+2223 are mergers; and A1423, A2261, CLJ1226+3332, and MAJ0744+3927 have mixed reports. Using discrepancies in SZ and X-ray parameter estimates and comparisons of \citet{2006MNRAS.373..881P} merger simulations with X-ray morphology we determine the dynamical states of the ten clusters in the sub-sample. We class A611, A1423, A2261 and MAJ1423+2404 as relaxed. MAJ0647+7015 and MAJ0717+3745, selected for their lensing strength, we find to be mergers. As expected from the mixed classifications in the literature, MAJ0744+3927 and CLJ1226+3332, although selected as relaxed, are also mergers. MAJ1149+2223, although reported in the literature as highly disturbed, shows no significant discrepancies between SZ and X-ray parameters. We conclude that low mass infalling cluster groups with high impact-parameters will cause less gas shocking and/or fractionation than lower mass-ratio mergers. We find evidence supporting the presence of an old, low level merger in RXJ1532+3021, postulated by \cite{2013ApJ...777..163H}.

\end{enumerate}

\section*{Acknowledgments}

We thank the anonymous reviewer for considered thoughts and useful comments that have improved the paper.
We also thank the staff of the Mullard Radio Astronomy Observatory for their invaluable assistance in the commissioning and operation of AMI, which is supported by Cambridge University.
WJH and CR are grateful for the support of STFC Studentships. CR also acknowledges the support of Cambridge University.
MO and YCP acknowledge support from Research Fellowships from Sidney Sussex College and Trinity College, Cambridge, respectively.
We thank Arif Babul for his assistance in accessing the Poole et al. online materials.
Much of this work was undertaken on the COSMOS Shared Memory system at DAMTP, Cambridge University, operated on behalf of the STFC DiRAC HPC Facility. This equipment is funded by BIS National E-infrastructure capital grant ST/J005673/1 and STFC grants ST/H008586/1, ST/K00333X/1.


\label{lastpage}
\end{document}